%% file: FR.tex
\documentclass[a4paper]{JHEP3} 

\usepackage{framed}
\usepackage{epsfig}
\usepackage{graphicx}
\graphicspath{{./}{figs/}}
\usepackage[latin1]{inputenc}
\usepackage{amssymb}
\usepackage{amsmath}
\usepackage{amsfonts}
\usepackage{epsfig}
\usepackage{slashed}
\usepackage{mathrsfs}
\usepackage{bbm}
\usepackage{feynarts}
\usepackage{subfigure}
\usepackage{multirow}
\usepackage{multicol}
\usepackage{axodraw}
\usepackage{url}

\def\btab#1\etab{\begin{tabular}{p{45mm}p{65mm}}#1\end{tabular}}
\def\btabx#1\etabx{\begin{tabular}{p{60mm}p{50mm}}#1\end{tabular}}
\def\btaby#1\etaby{\begin{tabular}{p{15mm}p{95mm}}#1\end{tabular}}
\def\bcen{\begin{center}}
\def\ecen{\end{center}}

\makeatletter
\def\url@leostyle{%
  \@ifundefined{selectfont}{\def\UrlFont{\sf}}{\def\UrlFont{\small\ttfamily}}}
\makeatother
\newcommand{\comment}[1]{}

\def\to{\rightarrow}

\def\Im{\mathrm{Im}}
\def\Re{\mathrm{Re}}

\def\d{\mathrm{d}}

\newcommand\fverb{\setbox\pippobox=\hbox\bgroup\verb}
\newcommand\fverbdo{\egroup\medskip\noindent%
                      \fbox{\unhbox\pippobox}\ }

\newcommand\fverbit{\egroup\item[\fbox{\unhbox\pippobox}]}
\newbox\pippobox
\newcommand{\ie}{{\it i.e.}}
\newcommand{\eg}{{\it e.g.}}

\newcommand{\pythia}{{\sc Pythia}}
\newcommand{\herwig}{{\sc Herwig}}
\newcommand{\sherpa}{{\sc Sherpa}}
\newcommand{\amegic}{{\sc Amegic++}}
\newcommand{\apacic}{{\sc Apacic++}}
\newcommand{\mgme}{{\sc MadGraph/MadEvent}}
\newcommand{\alpgen}{{\sc Alpgen}}
\newcommand{\helac}{{\sc Helac}}
\newcommand{\comix}{{\sc Comix}}
\newcommand{\madgraph}{{\sc MadGraph}}

\newcommand{\feynrules}{{\sc FeynRules}}

\newcommand{\whizard}{{\sc Whizard}}
\newcommand{\calchep}{{\sc CalcHep}}
\newcommand{\comphep}{{\sc CompHep}}
\newcommand{\lanhep}{{\sc LanHep}}
\newcommand{\feynarts}{{\sc FeynArts}}
\newcommand{\formcalc}{{\sc FormCalc}}

\newcommand{\be}{\begin{equation}}
\newcommand{\ee}{\end{equation}}
\newcommand{\ba}{\begin{eqnarray}}
\newcommand{\ea}{\end{eqnarray}}
\newcommand{\bt}{\begin{tabular}}
\newcommand{\et}{\end{tabular}}
\newcommand{\bfig}{\begin{figure}}
\newcommand{\efig}{\end{figure}}

\newcommand{\lagran}{\begin{cal}L\end{cal}}
\newcommand{\lag}{\lagran}
\newcommand{\GeV}{\mathrm{\;GeV}}

\newcommand{\mathematica}{{\sc Mathematica}}
\def\beq{\begin{equation}}
\def\eeq{\end{equation}}
\def\bsp#1\esp{\begin{split}#1\end{split}}
\newcommand{\doub}[2]{\left(\begin{array}{c} #1 \\ #2\end{array}\right)}
\newcommand{\doublet}[2]{\bpm#1\\#2\epm}
\newcommand{\del}{\partial}

\def\1{\mathchoice{\rm 1\\\
mskip-4.2mu l}{\rm 1\mskip-4.2mu l}%
{\rm 1\mskip-4.6mu l}{\rm 1\mskip-5.2mu l}} 

\def\bpm{\begin{pmatrix}}
\def\epm{\end{pmatrix}}
\def\del{\partial}

\def\bsp#1\esp{\begin{split}#1\end{split}}
\def\hc{{\rm h.c.}}

\def\cw{{\cos\theta_w}}
\def\cwd{{\cos^2\theta_w}}

\def\ca{{\cos\alpha}}
\def\sa{{\sin\alpha}}
\def\cb{{\cos\beta}}
\def\sb{{\sin\beta}}
\def\tb{{\tan\beta}}

\title{A comprehensive approach to new physics simulations}

\author{Neil Christensen$^{(1)}$, Priscila de Aquino$^{(2,3)}$, Celine Degrande$^{(2)}$, Claude Duhr$^{(2)}$, Benjamin Fuks$^{(4)}$, Michel Herquet$^{(5)}$, Fabio Maltoni$^{(2)}$, Steffen Schumann$^{(6)}$\\
$^{(1)}$ Department of Physics and Astronomy, Michigan State University, East Lansing, MI 48824, USA \\
$^{(2)}$ Center for Particle Physics and Phenomenology, Universit\'e Catholique de Louvain, B-1348 Louvain-la-Neuve, Belgium \\
$^{(3)}$ Instituut voor Theoretische Fysica, Katholieke Universiteit Leuven, Celestijnenlaan 200D, B-3001 Leuven, Belgium \\
$^{(4)}$ Institut Pluridisciplinaire Hubert Curien, Universit\'e de Strasbourg, IN2P3-CNRS, BP28, F-67037 Strasbourg Cedex 2, France\\
$^{(5)}$ Nikhef Theory Group, Kruislaan 409, 1098 SJ Amsterdam, The Netherlands \\
$^{(6)}$ Institut f\"ur Theoretische Physik, Universit\"at Heidelberg, Philosophenweg 16, D-69120, Heidelberg, Germany \\
}

\abstract{We describe a framework to develop, implement and validate any perturbative Lagrangian-based particle physics model for further theoretical, phenomenological and experimental studies. The starting point is \feynrules, a \mathematica\ package that allows to generate Feynman rules for any Lagrangian and then, through dedicated interfaces, automatically pass the corresponding relevant information to any supported Monte Carlo event generator. We prove the power, robustness and flexibility of this approach by presenting a few examples of new physics models (the Hidden Abelian Higgs Model, the general Two-Higgs-Doublet Model, the most general Minimal Supersymmetric Standard Model, the Minimal Higgsless Model, Universal and Large Extra Dimensions, and QCD-inspired effective Lagrangians) and their implementation/validation in \feynarts/\formcalc, \calchep, \mgme, and \sherpa. }

\preprint{CP3-09-24\\
HD-THEP-09-11\\
IPHC-PHENO-09-01\\
MSUHEP-090612\\
NIKHEF-2009-009}


\begin{document}

\input{intro}

\input{hidden}

\input{feynrules}

\section{Interfaces}
\label{app:Interfaces}

In this section we provide a concise description of the \feynrules~interfaces to several matrix-element generators and symbolic tools available to perform simulations/calculations from Lagrangian-based theories. The most  important features of the general structure of the new physics models in the codes and their limitations are emphasized. Complete description of the options and more technical details can be found in the \feynrules~user's manual, available on the \feynrules~website.  Interfaces to other codes, once available, will be included in the main release of the package and documented in the user's manual.

\input{feynarts}

\input{calchep}

\input{madgraph}

\input{sherpa}

\section{Models}
\label{sec:models}

In this section we briefly present the implementation of the Standard Model and other several important New Physics models in \feynrules.  Our aim is to show that very complete and sophisticated implementations are possible and that \feynrules\ offers a very natural
and convenient framework where models can be first developed (from the theoretical point of view) and then tested/constrained against experimental data. Since the main focus is the implementation procedure, the actual model descriptions, as well as the information about values of parameters, are kept to the minimum. More exhaustive information about each of the following models, all of which being publicly available, can be found on the \feynrules\ website.

\input{sm}

\input{2hdm}

\input{mssm1}

\input{3-site}

\input{ued1}

\input{effectivel}

\section{Validation of the implemented models}\
\label{sec:validation}
In this section, we first review some generic features of the various validation procedures used to assess the model robustness and then move to the results of the validation process for the Beyond the Standard Model theories presented in the previous section.
\input{validation}

\input{outlook}

\acknowledgments{We are particularly thankful to the organizers and participants of the MC4BSM series of workshops for the many stimulating discussions that have helped in shaping the approach presented in this paper. Several of us are grateful to the members of their respective MC development teams (\calchep, \mgme\ and \sherpa) and to Thomas Hahn for \feynarts/\formcalc\ for their support and help in developing the interfaces to \feynrules. The authors are also grateful to C.~Grojean for valuable discussions on the SILH model. C.~Degrande and C.~Duhr are research fellows of the ``Fonds National de la Recherche Scientifique'', Belgium. This work was partially supported by the Institut Interuniversitaire des Sciences Nucl\'eaires and by the Belgian Federal Office for Scientific, Technical and Cultural Affairs through the Interuniversity Attraction Pole P6/11. It is also part of the research program of the ``Stichting voor Fundamenteel Onderzoek der Materie (FOM)", which is financially supported by the ``Nederlandse organisatie voor Wetenschappelijke Onderzoek (NWO)". N.D.C. is supported by the U.S. Department of Energy under Grant No. DE-FG02-06ER41418. The authors of \feynrules are grateful to R.~Franceschini and K.~Kannike for their valuables comments on the code, and especially  to M.~Wiebusch for implementing the covariant derivative.}

\appendix


\input{FRconvention}

\section{Validation tables}
\label{app:val}
In this Appendix we report the main results of our work, \ie, the validation tables. In general, the tables list quantities (such as decay widths or cross sections) that have no direct phenomenological interest but they are physical, easily reproducible and provide an exhaustive check of the (complex) values of all the couplings of the model. In several instances, other powerful checks (such as gauge invariance, unitarity cancellation at high energy, and so on) have been performed that are not presented here. When possible, comparisons between the so-called ``stock implementation", \ie\ implementations already available in the Monte Carlo tools, have been made as well as comparisons between different Monte Carlo's also in different gauges. All numbers quoted in this section are expressed in pico-barn and correspond to a collision in the center-of-mass frame. In all cases agreement to better than 1\% was obtained.

\input{sm2}

\input{2hdm2}

\input{mssm2}
\input{3-site2}
\input{ued2}

\clearpage

\bibliography{physics}

\end{document}

%% file: intro.tex
\section{Introduction}


At the Large Hadron Collider (LHC) discoveries most probably will not be an easy task. 
The typical final states produced at this proton-proton collider running at very
high energies will involve a large number of jets, heavy-flavor
quarks, as well as leptons and missing energy, 
providing an overwhelming background to many new physics
searches. Complex signal final state signatures will then need a very
careful understanding of the detector and an accurate modeling of the
data themselves. In this process, Monte Carlo (MC) simulations will
play a key role in describing control data sets and devising robust
search strategies.


Already the first step, \ie, establishing ``an excess over the Standard Model (SM)
background'', might be very difficult, depending on the type of
signature involved~\cite{Mangano:2008ag}. At this stage, matrix-element-based MC (which give reliable predictions for shapes and can
still be tuned to some extent to the data) will be used to describe
backgrounds and possibly candidates signals. For some specific
signals, an accurate prediction of the background normalization and
shapes, validated via control samples, could be also needed. At the same time, 
accurate measurements and comparisons with the best theoretical predictions 
(\eg, at the next-to-next-to-leading order, resummation calculations, \ldots) of a set of
standard-candle observables will also be mandatory to claim a good
understanding and control of physics and detector effects. Very
accurate predictions, possibly including even weak corrections, and a
reliable estimate of errors (such as those introduced by the parton distribution functions)
will then be needed.  

Once the presence of excess(es) is confirmed,
model building activities will be triggered, following both top-down
and bottom-up approaches. In each case, tools that are able to make
predictions for wide classes of Beyond the Standard Model (BSM) physics, as well
as those that help in building up an effective field theory from the
data (such as the so called OSET method~\cite{ArkaniHamed:2007fw}),
could be employed.  Finally, as a theoretically consistent picture
arises, measurements of the parameters (masses, spin, charges) will be
carried out. In this case it will be necessary to have at least next-to-leading order (NLO)
predictions (\ie, a reliable normalization) for the signal
processes. As our knowledge about the detector and the newly
discovered physics scenario gets stronger, more accurate
determinations will be possible and sophisticated analyses tools could
be employed, on the very same lines as current top quark analyses at
the Tevatron collider, \eg, see Ref.~\cite{Bellettini:2008zz}.


As schematically outlined above, Monte Carlo simulations will play a
key, though different role at each stage of the exploration of
the TeV scale, \ie, the discovery and identification of BSM Physics,
and the measurement of its properties.  The realization of the need
for better simulation tools for the LHC has spurred an intense
activity over the last years, that has resulted in several important
advances in the field.


At the matrix-element level, these include the development of general
purpose event generators, such as
\comphep/\calchep~\cite{Pukhov:1999gg,  Boos:2004kh,Pukhov:2004ca},
\mgme~ \cite{Stelzer:1994ta,Maltoni:2002qb,Alwall:2007st,Alwall:2008pm} ,
\sherpa~\cite{Gleisberg:2003xi,Gleisberg:2008ta} and 
\whizard~\cite{Kilian:2001qz}, as
well as high efficiency multiparton generators which go beyond the
usual Feynman diagram techniques, such as \alpgen~\cite{Mangano:2002ea},
\helac~\cite{Papadopoulos:2006mh} and \comix~\cite{Gleisberg:2008fv}. As a result, the problem of
generating automatically leading-order matrix elements (and then cross
sections and events) for a very large class of renormalizable
processes has been solved. Quite amazingly, enormous progress has
also been achieved very recently in the automatization of NLO
computations. First the generation of the real corrections with
the appropriate subtractions has be achieved in an automatic
way~\cite{Gleisberg:2007md,Seymour:2008mu,Hasegawa:2008ae,Frederix:2008hu,Czakon:2009ss}.  Then
several new algorithms for calculating loop amplitudes numerically
have been proposed (see, \eg, Ref.~\cite{Zanderighi:2008na} for a review)
and some of them successfully applied to the computation of SM
processes of physical interest~\cite{Ellis:2009zw,Berger:2009zg,vanHameren:2009dr}.


An accurate simulation of a hadronic collision requires a careful
integration of the matrix-element hard process, with the full parton
showering and hadronization infrastructure, as efficiently provided by \pythia~\cite{Sjostrand:2006za, Sjostrand:2007gs},
\herwig~\cite{Corcella:2000bw, Bahr:2008pv} and \sherpa. Here also, significant progress has been made in
the development of matching algorithms such as that by Catani, Krauss,
Kuhn and Webber
(CKKW)~\cite{Catani:2001cc,Krauss:2002up,Mrenna:2003if}, Mangano
(MLM)~\cite{Mangano:2006rw} and
others~\cite{Lonnblad:2001iq,Lavesson:2005xu,Hoeche:2009rj}, in their
comparison~\cite{Hoche:2006ph,Alwall:2007fs} and application to
SM~\cite{Mangano:2006rw, Krauss:2004bs} and BSM~\cite{Alwall:2008qv}
processes.  A breakthrough in merging fixed order calculations and parton showers
was achieved by Frixione, Webber and
Nason~\cite{Frixione:2002ik,Frixione:2003ei}, who showed how to
correctly interface an NLO computation with a parton shower to avoid
double counting and delivered the first event generator at NLO,
MC@NLO. More recently, a new method along the same lines, dubbed
POWHEG, has been proposed~\cite{Nason:2004rx} and applied to Drell-Yan
and Higgs
production~\cite{Alioli:2008gx,Alioli:2008tz,Hamilton:2008pd,Hamilton:2009za}.


The progress in the field of Monte Carlo tools outlined above shows
that we are, or will be soon, able to simulate all the relevant SM
processes at the LHC with an unprecedented level of accuracy.  It is
therefore worth considering the status of the predictions for physics
Beyond the Standard Model.  Quite interestingly, the challenges in
this case are of a quite different nature. The main reason is that presently
there is not a leading candidate for BSM, but instead a plethora
of models have been suggested, based on very different ideas in continuous evolution. The implementation of complex  BSM models 
in existing general purpose event generators like those enumerated above remains a 
long, often painstaking and error-prone, process.  The derivation of the numerous
Feynman rules to describe the new interactions, and their implementation 
in codes following conventions is a very uninteresting and time consuming activity. In addition, the validation of a
given implementation often relies on a comparison of the obtained
analytical and numerical results with those available in the
literature. Again, due to presence of various conventions, the
restricted number of public results and the lack of a dedicated
framework, such a comparison is often done manually, in a partial and not systematic way. 
Finally, besides a handful of officially endorsed and
publicly distributed BSM models (\eg, the Minimal Supersymmetric Standard Model), many implementations
remain private or only used by a restricted set of theorists and/or
phenomenologists, and never get integrated into the official chain of simulation
tools used by experimental collaborations. Instead, various ``home-made'' or ``hacked''
versions of existing MC softwares are commonly used for specific BSM
process studies, leading to problems in the validation, traceability and
maintenance procedures.



In this work we address the problem of having an efficient framework
where any new physics model can be developed and its phenomenology can
be tested against data. A first step in the direction of deriving Feynman rules automatically
starting from a model Lagrangian has been made in the 
context of the \comphep/\calchep\ event generator with the \lanhep\ 
package~\cite{Semenov:1998eb}. Our aim is to go beyond this scheme and 
create a general and flexible environment where communication between theorists and 
experimentalists in both directions is fast and robust. The desiderata for the new physics
phenomenological framework linking theory to experiment and vice versa which we
provide a solution for are:

\begin{enumerate}

\item General and flexible environment, where any perturbative Lagrangian-based
      model can be developed and implemented.

\item Modular structure with interfaces to several multi-purpose MC's
      and computational tools.

\item Robust, easy-to-validate and easy-to-maintain.

\item Integrable in the experimental software frameworks.

\item Full traceability of event samples.

\item Both top-down and bottom-up approaches are natural.

\end{enumerate}

This paper is organized in five main sections and various
appendices. In the first section, by discussing a simple example, we
expose the strategy which we propose to address the challenges that model
builders, phenomenologists and experimentalists have to face to study
the phenomenology of a new physics model. In
Section~\ref{sec:feynrules}, we briefly recall how the \feynrules\
package works and present some of the new features recently
implemented. Section~\ref{app:Interfaces} contains a brief description of the various interfaces already available.
Section~\ref{sec:models} contains the information about
the models that have already been implemented. In Section~\ref{sec:validation} we present our strategy to validate BSM model implementations, and illustrate our procedure on the already implemented models. Finally, in Section~\ref{sec:outlook}, we discuss the
outlook of our work.  In the appendices we collect some technical
information as well as a few representative validation tables, which constitute
the quantitative results of this paper.

%% file: hidden.tex
\section{A simple example: from the Standard Model to the Hidden Abelian Higgs Model}
\label{sec:hidden}

From the phenomenological point of view, we can distinguish two classes of BSM models. The first class of models consists of straightforward extensions of the SM, obtained by adding one (or more) new particles and interactions to the SM Lagrangian. In this bottom-up approach, one generally starts from the SM, and adds a set of new operators according to some (new) symmetry. The second class of models are obtained in a top-down approach, where the fundamental Lagrangian is determined by the underlying global and local symmetries, and the SM is only recovered in some specific limit.
 
In this section we describe in detail our framework to develop, test, implement and validate any perturbative Lagrangian-based particle physics model. We concentrate on the case of the bottom-up models, and show how our framework allows to easily extend the SM  and to go in a straightforward way from the model building to the study of the phenomenology. We will comment on the top-down models in subsequent sections. 

\subsection{The model}
As an illustration, we use the Hidden Abelian Higgs (HAH) Model, described in Ref.~\cite{Wells:2008xg}. This model can be seen as the simplest way to consistently add a new gauge interaction to the SM. More specifically, we consider an $SU(3)_C\times SU(2)_L\times U(1)_Y\times U(1)_X$ gauge theory where all SM particles are singlets under the new gauge group $U(1)_X$. A new Higgs field $\phi$ is added that is also a singlet under the SM gauge group and breaks the $U(1)_X$ symmetry when it acquires its vacuum expectation value (vev), $\langle\phi\rangle = {\xi/\sqrt{2}}$ . The most general Lagrangian describing this model is given by
\beq\label{eq:LSMHidden}
\lagran_{\rm HAH} = \lagran_{\rm HAH, Gauge} + \lagran_{\rm HAH, Fermions} + \lagran_{\rm HAH, Higgs}+ \lagran_{\rm HAH, Yukawa},
\eeq
with
\beq\bsp\label{eq:Lhidden}
\lagran_{\rm HAH, Gauge} =&\, -{1\over 4}\,G_{\mu\nu}^a\,G^{\mu\nu}_a -{1\over 4}\,W_{\mu\nu}^i\,W^{\mu\nu}_i  -{1\over 4}\,B_{\mu\nu}\,B^{\mu\nu}  -{1\over 4}\,X_{\mu\nu}\,X^{\mu\nu} + {\chi\over 2}\,B_{\mu\nu}\,X^{\mu\nu},\\
\lagran_{\rm HAH, Higgs} =&\, D_\mu\Phi^\dagger\,D^\mu\Phi +D_\mu\phi^\dagger\,D^\mu\phi +\mu_\Phi^2\,\Phi^\dagger\Phi +\mu_\phi^2\,\phi^\dagger\phi - \lambda\,(\Phi^\dagger\Phi)^2 \\
&\,- \rho\,(\phi^\dagger\phi)^2 - \kappa\,(\Phi^\dagger\Phi)\,(\phi^\dagger\phi),
\esp
\eeq
where $\Phi$ denotes the SM Higgs field. The covariant derivative reads
\beq\label{eq:covdel}
D_\mu = \del_\mu - ig_s\,T^a\,G_\mu^a - ig\,{\vec\sigma\over2}\cdot\vec W_\mu\, - ig'\,Y\,B_\mu- ig_X\,X\,X_\mu,
\eeq
and we define the field strength tensors,
\beq \bsp\label{eq:fieldstrength}
  X_{\mu\nu}   =& \del_\mu X_\nu-\del_\nu X_\mu,\\
  B_{\mu\nu}   =& \del_\mu B_\nu-\del_\nu B_\mu,\\
  W_{\mu\nu}^i =& \del_\mu W_\nu^i-\del_\nu W_\mu^i + g^\prime \,\epsilon^{ijk}\,W_\mu^j\,W_\nu^k,\\
  G_{\mu\nu}^a =& \del_\mu g_\nu^a-\del_\nu g_\mu^a + g_s\, f^{abc}\,g_\mu^b\,g_\nu^c.~
\esp\eeq $g_s$, $g$, $g'$ and $g_X$ denote the four coupling constants associated with the $SU(3)_C\times SU(2)_L\times U(1)_Y\times U(1)_X$ gauge groups while $T$, $\sigma^i$, $Y$ and $X$ are the corresponding generators and $\epsilon^{ijk}$ represents the totally antisymmetric tensor. 
We do not write explicitly the terms in the  Lagrangian describing the matter sector of the theory as they are identical to those of the SM described in detail in Section~\ref{sec:SM},
\beq
\lag_{\rm HAH, Fermions} = \lag_{\rm SM, Fermions} {\rm ~~and~~}\lag_{\rm HAH, Yukawa} = \lag_{\rm SM, Yukawa}.
\eeq
The kinetic mixing term in $\lagran_{\rm HAH, Gauge}$ induces a mixing between the two $U(1)$ gauge fields and thus a coupling between the matter fermions and the new gauge boson. The kinetic terms for the $U(1)$ fields can be diagonalized via a $GL(2,\mathbb{R})$ rotation,
\beq
\doub{\tilde X_\mu}{\tilde B_\mu} = \left(\begin{array}{cc} \sqrt{1-\chi^2} & 0\\ -\chi & 1\end{array}\right)\,\doub{X_\mu}{B_\mu}.
\eeq
After this field redefinition, the gauge sector takes the diagonal form
\beq
\lagran_{\rm HAH, Gauge} = -{1\over 4}\,G_{\mu\nu}^a\,G^{\mu\nu}_a -{1\over 4}\,W_{\mu\nu}^i\,W^{\mu\nu}_i  -{1\over 4}\,\tilde B_{\mu\nu}\,\tilde B^{\mu\nu}  -{1\over 4}\,\tilde X_{\mu\nu}\,\tilde X^{\mu\nu},
\eeq
but the covariant derivative now contains an additional term coupling the field $\tilde X_\mu$ to the hypercharge,
\beq
D_\mu = \del_\mu - ig_s\,T^a\,G_\mu^a - ig\,{\vec\sigma\over2}\cdot\vec W_\mu\, - ig'\,Y\,\tilde B_\mu- i(g_X\,X + g'\,\eta\,Y)\,\tilde X_\mu,
\eeq
with $\eta = \chi/\sqrt{1-\chi^2}$.
When the Higgs fields acquire their vev\footnote{We work in unitary gauge.},
\beq
\Phi = {1\over\sqrt{2}}\,\doub{0}{H+v} {\rm ~~and~~} \phi = {1\over\sqrt{2}}\,(h+\xi),
\eeq
the gauge symmetry gets broken to $SU(3)_C\times U(1)_{EM}$, and we obtain a non-diagonal mass matrix for the neutral weak gauge bosons, which can be  diagonalized by an $O(3)$ rotation,
\beq
\left(\begin{array}{c} \tilde B\\ W^3\\ \tilde X\end{array}\right) = \left(\begin{array}{ccc} c_w & -s_w\,c_\alpha & s_w\,s_\alpha \\ s_w & c_w\,c_\alpha & -c_w\,s_\alpha \\ 0& s_\alpha & c_\alpha\end{array}\right)\,\left(\begin{array}{c} A\\ Z\\ Z'\end{array}\right),
\eeq
where the mixing angles are given by
\beq
s_w \equiv \sin\theta_w = {g'\over \sqrt{g^2+g'^2}} {\rm ~~and~~}  \tan 2\theta_\alpha = {-2s_w\eta\over 1-s_w^2\eta^2-\Delta_Z},
\eeq
with $\Delta_Z=M_X^2/M_{Z_0}^2$, where $M_X$ and $M_{Z_0}$ denote the masses of the gauge bosons before the kinetic mixing,
\beq
M_X^2 = \xi^2\,q_X^2\,g_X^2 {\rm ~~and~~} M_{Z_0}^2 = (g^2+g'^2)\,v^2/4,
\eeq
and $q_X$ denotes the $U(1)_X$ charge carried by $\phi$. The photon  remains massless while the two other states acquire a mass given by
\beq\label{eq:Zpmass}
M_{Z,Z'}^2 = {M_{Z_0}^2\over 2}\,\left[(1+s_w^2\eta^2+\Delta_Z) \pm \sqrt{(1-s_w^2\eta^2-\Delta_Z)^2+4s_w^2\eta^2}\right].
\eeq

As a result of  electroweak symmetry breaking, non-diagonal mass terms for the Higgs fields appear that can be diagonalized via an orthogonal transformation,
\beq
\doub{H}{h} = \left(\begin{array}{cc} c_h & s_h\\ -s_h & c_h\end{array}\right)\, \doub{h_1}{h_2},
\eeq
where the mixing angles and mass eigenvalues are given by
\beq\bsp
\tan 2\theta_h &\,= {\kappa\,v\,\xi\over\rho\,\xi^2-\lambda\,v^2},\\
M^2_{1,2} &\,= \lambda\,v^2+\rho\,\xi^2\mp \sqrt{(\lambda\,v^2+\rho\,\xi^2)^2 + \kappa^2\,v^2\,\xi^2}.
\esp
\eeq

Once the Lagrangian is written down and diagonalized in terms of mass eigenstates, one can easily identify the minimal set of parameters which the model depends on. Not all the parameters introduced above are independent, as most of them are related by some algebraic relations, \emph{e.g.}, the relation between the mass eigenvalues of the gauge bosons, Eq.~(\ref{eq:Zpmass}), and the fundamental parameters appearing in the Lagrangian. Our choice of independent input parameters is given in Table~\ref{tab:hiddenparams}. All other parameters appearing in $\lag_{\rm HAH, Gauge}$ and $\lag_{\rm HAH, Higgs}$ can be reexpressed in terms of these parameters. Note, however, that there are strong experimental constraints from LEP on the masses and the couplings of additional neutral gauge bosons to fermions, which need to be taken into account when building the model. In Ref.~\cite{Wells:2008xg} it was pointed out that $\eta=0.01$ is still allowed. In general, in order to determine a benchmark point that takes into account the direct and indirect experimental constraints,  it is required to perform (loop) computations for several physical observables. We will comment more on this in the next subsection.
\begin{table}[!t]
\begin{center}
\begin{tabular}{lcr}
\hline
Parameter & Symbol & Value\\
\hline\hline
$U(1)_X$ coupling constant  & $\alpha_X$ & 1/127.9\\
Kinetic mixing parameter & $\eta$ & 0.01\\
$Z$ pole mass & $M_Z$ & 91.188 GeV \\
$Z'$ pole mass & $M_{Z'}$ & 400.0 GeV \\
SM Higgs quartic coupling & $\lambda$ & 0.42568\\
Abelian Higgs quartic coupling & $\rho$ & 0.010142\\
Abelian/SM Higgs interaction & $\kappa$ & 0.0977392\\
\hline
\end{tabular}
\caption{\label{tab:hiddenparams} Input parameters for the Hidden Abelian Higgs model. Other SM input parameters are not shown.}
\end{center}
\end{table}

Let us note that, although Eq.~(\ref{eq:Lhidden}) is a very simple extension of the SM, from a more technical point of view an implementation of the HAH model in a matrix-element generator is already not trivial. In this case, it is not sufficient to start from the existing SM implementation and just add the vertices contained in $\lagran_{\rm HAH, Higgs}$, because mixing in the gauge and scalar sectors implies that all SM vertices involving a Higgs boson and/or a $Z$ boson need to be modified. For example, although there is no direct coupling between the abelian Higgs field $\phi$ and the matter fermions, all the Yukawa couplings receive contributions from the two mass eigenstates $h_1$ and $h_2$, weighted by the mixing angle $\theta_h$, resulting in an almost complete rewriting of the SM implementation. In the next subsection we will describe how this difficulty can be easily overcome and the phenomenology of the Hidden Abelian Higgs model studied.

\subsection{From model building to phenomenology}
The starting point of our approach is \feynrules~(see Section~\ref{sec:feynrules}). Since in this case we are interested in a simple extension of the SM, it is very easy to start from the \feynrules~implementation of the SM which is included in the distribution of the package and to extend the model file by including the new particles and parameters, as well as the HAH model Lagrangian of Eq.~\eqref{eq:Lhidden}. Note that, at variance with the direct implementation into a matrix-element generator where we need to implement the vertices one at the time, we can work in \feynrules\ with a Lagrangian written in terms of the gauge eigenstates and only perform the rotation to the mass eigenbasis as a second step. This implies that it is not necessary to modify $\lagran_{\rm HAH, Fermions}$ since the new fields only enter through $\lagran_{\rm HAH, Gauge}$ and $\lagran_{\rm HAH, Higgs}$.

Several functions are included in \feynrules~to perform sanity checks on the Lagrangian (\emph{e.g.}, hermiticity). The diagonalization of the mass matrices can be easily performed directly in \mathematica\textregistered\footnote{\mathematica\ is a registered trademark of Wolfram Research, Inc. } and \feynrules\ allows us to easily obtain the Feynman rules for the model. As already mentioned, not only the  Feynman rules of the Higgs and gauge sectors are modified with respect to the Standard Model, but also the interaction vertices in the fermionic sector change due to the mixing of the scalars and the neutral weak bosons. The vertices obtained in this way can already be used for \textit{pen \& paper} work during the model building, and to compute simple decay rates and cross sections. Since \feynrules~stores the vertices in \mathematica, it is easy to use them directly for such computations.

After this preliminary study of our model where the mass spectrum of the theory was obtained and basic sanity checks have been performed, typically the model is confronted with all relevant direct and indirect constraints coming from experiment. This is a necessary step to find areas of parameters space which are still viable. 
Once  interesting regions in parameter space are identified, the study of  the collider phenomenology of the model, \emph{e.g.}, at the LHC, 
can start with the calculation of  cross sections and decay branching ratios. Let us consider first the  the calculation of the decay widths of both SM and new particles. Using the \feynarts~implementation of the new model obtained via the \feynrules~interface, it becomes a trivial exercise to compute analytically all tree-level two-body decays for the Higgs bosons and the $Z'$ boson (alternatively, one could calculate them numerically via \eg, \calchep~or \mgme). The results for the branching ratios of the dominant decay modes are shown, for the benchmark scenario considered here, in Table~\ref{tab:hiddendecays}.
\begin{table}[!t]
\begin{center}
$\begin{array}{ccc}
\begin{tabular}{|l|r @{.} l|}
\hline
\multicolumn{3}{|c|}{Branching ratios for $h_1$} \\
\hline
$h_1\to b\,\bar b$ & 87 & 7\%\\
$h_1\to c\,\bar c$ & 8 & 1\%\\
$h_1\to \tau^+\,\tau^-$ & 4 & 2\%\\
\hline\hline
\multicolumn{3}{|c|}{Branching ratios for $h_2$}\\
\hline
$h_2\to h_1\,h_1$ & 14 & 2\%\\
$h_2\to b\bar b$ & 0 & 1\%\\
$h_2\to Z\,Z$ & 26 & 4\%\\
$h_2\to W^+\,W^-$ & 59 & 3\%\\
\hline
\end{tabular} & \phantom{aaa}
\begin{tabular}{|l|r @{.} l|}
\hline
\multicolumn{3}{|c|}{Branching ratios for $Z'$}\\
\hline
$Z'\to j\,j$ & 44 & 1\%\\
$Z'\to h_1\,h_1$ & 0 & 9\%\\
$Z'\to h_2\,h_2$ & 0 & 5\%\\
$Z'\to t\,\bar t$ & 12 & 5\%\\
$Z'\to b\,\bar b$ & 5 & 2\%\\
$Z'\to \tau^+\,\tau^-$ & 14 & 9\%\\
$Z'\to \ell^+\,\ell^-$ & 29 & 7\%\\
$Z'\to \nu\,\bar \nu$ & 9 & 5\%\\
$Z'\to W^+\,W^-$ & 2 & 0\%\\
\hline
\end{tabular}
\end{array}$
\caption{\label{tab:hiddendecays}Dominant decay channels of the new particles in the HAH model.}
\end{center}
\end{table}
Once decay widths are known, cross sections can be calculated. However, in many cases it is insufficient to have only predictions for total cross sections, as a study of differential distributions,  with possibly complicated multi-particle final states, is necessary to dig the signal out of the backgrounds. Furthermore, even a parton-level description of the events might be too simplified and additional radiation coming from the colored initial and final-state particles, as well as effects coming from hadronization and underlying events need to be accounted for. For this reason, phenomenological studies are in general performed using  generators which include (or are interfaced to) parton shower and hadronization Monte Carlo codes. The parton level events for the hard scattering can be  generated by the general purpose matrix-element (ME) program, and those events are then passed on to the parton shower codes evolving the parton-level events into physical hadronic final states. However, similar to \feynarts/\formcalc, for new models the ME generators require the form  of the new vertices, and different programs use different conventions for the vertices, making it difficult to export the implementation from one ME generator to another. To solve this issue, \feynrules\ includes interfaces to several ME generators that allow to output the interaction vertices obtained by \feynrules~directly in a format that can be read by the external codes. For the moment, such interfaces exist for \calchep/\comphep, \mgme\ and \sherpa. It should be emphasized that some of these codes have the Lorentz and/or color structures hardcoded, something that limits the range of models that can be handled by a given MC. In this respect (and others) each of MC tools has its own strengths and weaknesses: having several possibilities available maximizes the chances that at least one generator is able to efficiently deal with a given model and the case in which several MC tools can be used, as most of the examples discussed in this paper, allows for a detailed comparison and robust validation of the implementation.

\begin{figure}[htbp]
\begin{center}
\includegraphics[width=12cm]{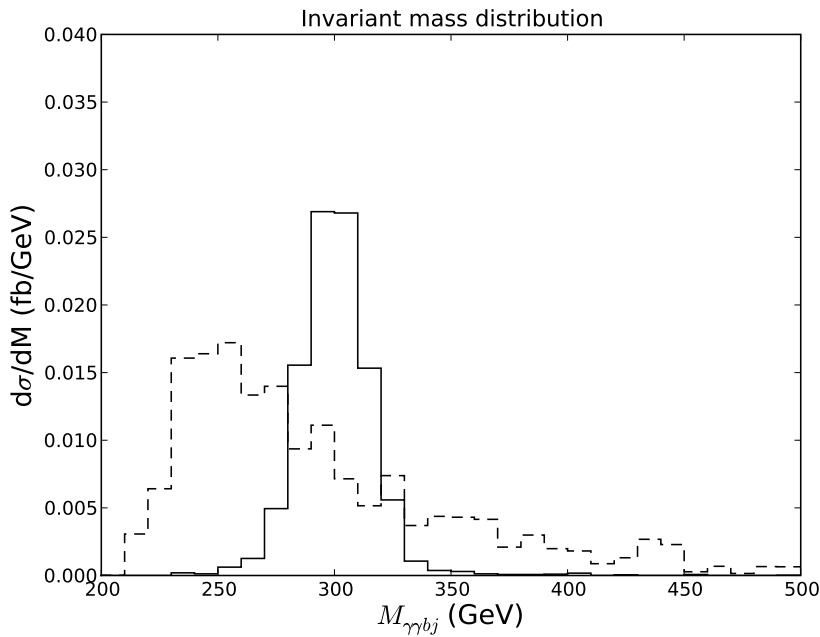}
\caption{Invariant mass distribution for the four particle final state $\gamma\gamma b\overline{b}$, both for the $gg\to h_2\to h_1 h_1\to \gamma\gamma b\overline{b}$ signal (plain) and the main SM backgrounds (dashed) events at the LHC. All simulation parameters and analysis cuts are identical to those listed in Ref.~\cite{Bowen:2007ia}.}
\label{fig:hah}
\end{center}
\end{figure}

For the sake of illustration, we used the \mgme\ implementation of the HAH model to generate signal events for the $gg\to h_2\to h_1 h_1\to \gamma\gamma b\overline{b}$ signal proposed in Ref.~\cite{Bowen:2007ia} as a signature of this model at the LHC. Using the same set of cuts, and the same smearing method as in Ref.~\cite{Bowen:2007ia}, we have been able to easily generate final state invariant mass distributions for both signal and background events. The result can be seen in Fig.~\ref{fig:hah}, which compares well to the Fig. 5 of Ref.~\cite{Bowen:2007ia}.

\subsection{Validation of New Physics Models}

In the previous subsection we discussed how the implementation of the model in \feynrules\ allows  to go all the way from model building to phenomenology without having to deal with the technicalities of  the various ME generators. In this section we argue that our approach does not only allow to exploit the strength of each ME generator, but it also has a new power in the validation of BSM models by directly comparing various ME generators among themselves. Since the various ME generators use different conventions for the interaction vertices it becomes hence possible to compare the implementations of the same model into various different matrix-element generators, and thus the different tools, among themselves. Furthermore, in many cases generator specific implementations of BSM models already exist, at least for restricted classes of models, in which case the \feynrules~model can be directly validated against the existing tool A, and then exported to any other tool B for which an interface exists. In the same spirit, any BSM model should be able to reproduce the SM results for observables which are independent of the new physics.

Let us illustrate this procedure through the example of the HAH model. We start by implementing our model into \calchep, \mgme\ and \sherpa\ by means of the corresponding interfaces. We then compute the cross sections for all interesting two-to-two processes for this model, and check that we obtain the same numbers from every ME generator. Note that, since in this case we have modified the scalar sector of the SM, we pay particular attention to the unitarity cancellations inherent to the SM in weak boson scattering, showing in this way that our implementation does not spoil these cancellations. Since the different codes used for the computation of the cross-sections all rely on different conventions for the interaction vertices, we hence demonstrated that our model is consistent not only by checking that the cancellations indeed take place in all the implementations, but we also get the same results for the total cross section after the cancellation, a strong check very rarely performed for a general BSM model implementation. We will comment on the validation of more general models in subsequent sections.

Finally, let us comment on the fact that a robust implementation of a BSM model into a code does not only require the model to be validated to a very high-level, but the implementation should also be clearly documented in order to assure its portability and reproducibility. For this reason \feynrules~includes an output to \TeX{} which allows to output the content of the \feynrules~model files in a human-readable format, including the particle content and the parameters which define the model, as well as the Lagrangian and the Feynman rules. 

\subsection{From phenomenology to experiment}

Our approach does not only apply to phenomenological studies at the theory level, but it allows in principle to continue and pass the model to an experimental collaboration for full experimental studies. In general, the experimental softwares used to simulate the detector effects have very strict requirements regarding sanity checks for new modules (\emph{e.g.}, private Monte Carlo programs) to be included in the software, and a very long and tedious validation procedure is needed. However, if we will come to the point where we have to discriminate between various competing models at the LHC, this approach becomes extremely inefficient due to the  large number of tools to be validated. In our approach this validation can be avoided, streamlining in this way the whole chain all the way from model building to the experimental studies and vice versa.

Let us again illustrate this statement through the example of the HAH model. Since this model is now implemented in \feynrules, we can easily pass it into various ME generators using the translation interfaces, and we demonstrated the validation power inherent to this approach in the previous section. These models can then be used in a ME generator in the same way as any other built-in model, without any modification to the original code, \emph{i.e.}, without creating a private version of the ME generator. If the model is validated it can easily be passed on to the experimental community, which can then read the \feynrules~output and use it in their favorite ME generator already embedded in their software framework. 

By following this procedure tedious validations for each model implementation in a given MC  are avoided. In addition, the portability and the reproducibility of the experimentally tested models is guaranteed. As at the origin of the chain is solely the \feynrules~model file, all the information is concentrated in one place, and thus everybody can reproduce all the results at any stage, from the model building to the collider signatures, by starting from the very same file. In addition, since the \feynrules~model file contains the Lagrangian of the model, it is very easy to go back to the model file and understand its physical content, a step which might be very difficult working with manually created model files for the ME generators, written in an often rather cryptic programming language hiding the essential physics. In this way it becomes very easy to use later the very same model information, and to reproduce analyses by just changing benchmark points or by adding a single new particle/interaction to the same model.

%% file: feynrules.tex
\section{\feynrules\ in a nutshell}
\label{sec:feynrules}

\feynrules\ is a \mathematica\ package which allows to derive Feynman rules directly from a Lagrangian~\cite{Christensen:2008py}. The user provides the Lagrangian for his/her model (written in \mathematica) as well as all the information about the particle and  parameter content of the model. This information uniquely defines the model, and hence is enough to derive all the interaction vertices from the Lagrangian. \feynrules\ can in principle be used with any model which fulfills basic quantum field theoretical requirements (\emph{e.g.},~Lorentz and gauge invariance), the only current limitation coming from the kinds of fields supported by \feynrules\ (see below).  In particular it can also be used to obtain Feynman rules for effective theories involving higher-dimensional operators.
In a second step, the interaction vertices obtained by \feynrules\ can be exported by the user to various matrix-element generators by means of a set of translation interfaces included in the package. In this way the user can directly obtain an implementation of his/her model into these various tools, making it straightforward to go from model building to phenomenology. Presently, interfaces to \calchep/\comphep, \feynarts/\formcalc, \mgme\ and \sherpa\ are available.
In the following we briefly describe the basic features of the package and the model files, the interfaces to the matrix-element generators being described in Section~\ref{app:Interfaces}. For more details on both the \feynrules\ package as well as the interfaces, we refer the reader to the \feynrules\ manual and to the \feynrules\ website~\cite{Christensen:2008py, FRwebpage}.

\subsection{Model description}
The \feynrules\ model definition is an extension of the \feynarts\ model file format and consists of the definitions of the particles, parameters and gauge groups that characterize the model and the Lagrangian.  This information can be placed in a text file or in a \mathematica~notebook or a combination of the two as convenient for the user.

Let us start with the particle definitions. Following the original \feynarts~convention, particles are grouped into classes describing ``multiplets'' having the same quantum numbers, but possibly different masses. Each particle class is defined in terms of a set of class properties, given as a \mathematica~replacement list. For example, the up-type quarks could be written as
\begin{center}
\begin{verbatim}
F[1] == { ClassName            -> q,
          ClassMembers         -> {u, c, t},
          SelfConjugate        -> False,
          Indices              -> {Index[Generation], Index[Colour]},
          FlavorIndex          -> Generation,
          Mass                 -> {Mq, 0, 0, {MT, 174.3}},
          Width                -> {Wq, 0, 0, {WT, 1.508}},
          QuantumNumbers       -> {Q -> 2/3},
          PDG                  -> {2, 4, 6}}
\end{verbatim}
\end{center}
This defines a Dirac fermion ({\tt F}) represented by the symbol {\tt q}. Note that the antiparticles are automatically declared and represented by the symbol {\tt qbar}. This class has three members, {\tt u}, {\tt c}, and {\tt t} ({\tt ubar}, {\tt cbar}, and {\tt tbar} for the antiparticles, respectively), distinguished by a generation index (whose range is defined at the beginning of the model definition). These fields carry an additional index labelled {\tt Colour}. The complete set of allowed particle classes is given in Table~\ref{tab:FRpartclass}. Additional information, like the mass and width of the particles, as well as the $U(1)$ quantum numbers carried by the fields can also be included. Finally, some more specific information not directly needed by \feynrules\ but required by some of the matrix-element generators (\emph{e.g.}, the Particle Data Group (PDG) codes~\cite{Amsler:2008zzb}) can also be defined. A complete description of the particle classes and properties can be found in the \feynrules\ manual.

\begin{table}[!t]
\begin{center}
\begin{tabular}{ll}
\hline\hline
{\tt S} & Scalar fields\\
{\tt F} & Fermion fields (Dirac and Majorana)\\
{\tt V} & Vector fields\\
{\tt U} & Ghost fields\\
{\tt T} & Spin two fields\\
\hline\hline
\end{tabular}
\caption{\label{tab:FRpartclass}Particle classes supported by \feynrules.}
\end{center}
\end{table}

A Lagrangian is not only defined by its particle content, but also by the local and global symmetries defining the model. \feynrules\ allows to define gauge group classes in a way similar to the particle classes. As an example, the definition of the QCD gauge group can be written
\begin{center}
\begin{verbatim}
SU3C == { Abelian              -> False,
          GaugeBoson           -> G,
          CouplingConstant     -> gs,
          StructureConstant    -> f,
          Representations      -> {T, Colour}}
\end{verbatim}
\end{center}
where the gluon field {\tt G} is defined together with the quark field during the particle declaration. The declaration of abelian gauge groups is analogous.  \feynrules\ uses this information to construct the covariant derivative and field strength tensor which the user can use in his/her Lagrangian.

The third main ingredient to define a model is the set of parameters which it depends on. In general, not all the parameters appearing in a Lagrangian are independent, but they are related through certain algebraic relations specific to each model, \emph{e.g.}, the relation $\cos\theta_w = M_W/M_Z$ relating at tree-level the masses of the weak gauge bosons to the electroweak mixing angle. \feynrules\ therefore distinguishes between \emph{external} and \emph{internal} parameters. External parameters denote independent parameters which are given as numerical inputs to the model. An example of a declaration of an external parameter reads
\begin{center}
\begin{verbatim}
aS == {ParameterType           -> External,
       Value                   ->  0.118}
\end{verbatim}
\end{center}
defining an external parameter {\tt aS} with numerical value 0.118. Several other properties representing additional information needed by matrix-element generators are also available, and we refer to the \feynrules\ manual for an extensive list of parameter class properties. Internal parameters are defined in a similar way, except that the {\tt Value} is given by an algebraic expression linking the parameter to other external and/or internal parameters. For example, the $\cos\theta_w$ parameter definition could read
\begin{center}
\begin{verbatim}
cw == {ParameterType          -> Internal,
       Value                  -> MW/MZ}
\end{verbatim}
\end{center}
Note that it is also possible to define tensors as parameters in exactly the same way, as described in more detail in the manual. 

At this point, we need to make a comment about the conventions used for the different particle and parameter names inside \feynrules. In principle, the user is free the choose the names for the gauge groups, particles and parameters at his/her convenience, without any restriction. The matrix-element generators however have certain information hardcoded (\emph{e.g.}, reference to the strong coupling constant or electroweak input parameters).  For this reason, conventions regarding the implementation of certain SM parameters has been established to ensure the proper translation to the matrix-element generator.  These are detailed in the manual and recalled in Appendix~\ref{app:FRMCconventions}.

In complicated models with large parameter spaces, it is sometimes preferable to restrict the model to certain slices of that parameter space.  This can be done as usual by adjusting the parameters to lie at that particular point in parameter space and doing the desired calculation.  However, sometimes these slices of parameter space are special in that many vertices become identically zero and including them in the Feynman diagram calculation can be very inefficient.  In order to allow both the general parameter space and a restricted parameter space, we introduce the model restriction.  A model restriction is a \mathematica~list containing replacements to the parameters which simplify the model.  For example, in the SM the CKM matrix has non-zero matrix elements, but it is sometimes useful to restrict a calculation to a purely diagonal CKM matrix.  Rather than creating a new implementation of the SM with a diagonal CKM matrix, a restriction can be created and used when desired.  The following statement restricts the SM to a diagonal CKM matrix
\begin{center}
\begin{verbatim}
CKM[i_, i_] :> 1,
CKM[i_, j_] :> 0 /; i != j,
\end{verbatim}
\end{center}
When this restriction is applied, all vertices containing off diagonal CKM elements vanish identically and are removed \emph{before} passing it on to a matrix-element generator.  The result is that these vertices never appear in Feynman diagrams and the calculation is more efficient.  Several restriction files can be created corresponding to various different slices of parameter space.  The user then selects the restriction file that they are interested in and applies it to the model before running the translation interfaces.

\subsection{Running \feynrules}
After having loaded the \feynrules\ package into \mathematica, the user can load the model and the model restrictions via the commands
\begin{center}
\begin{verbatim}
LoadModel[ < model file 1 > , < model file 2 > , ... ];

LoadRestriction[ < restriction file > ];
\end{verbatim}
\end{center}
where the model can be implemented in as many files as convenient or it can be implemented directly in the \mathematica~notebook in which case the list of files would be empty.  The restriction definitions can also be placed in a file or directly in the \mathematica~notebook.  The Lagrangian can now be entered directly into the notebook\footnote{Alternatively, the Lagrangian can also be included in the model file, in which case it is directly loaded together with the model file.} using standard \mathematica~commands, augmented by some special symbols representing specific objects like Dirac matrices. As an example, we show the QCD Lagrangian,
\begin{center}
\begin{verbatim}
L = -1/4 FS[G, mu, nu, a] FS[G, mu, nu, a]   + I qbar . Ga[mu]. DC[q, mu];
\end{verbatim}
\end{center}
where {\tt FS[G, mu, nu, a]} and {\tt DC[q, mu]} denote the $SU(3)_C$ field strength tensors and covariant derivatives automatically defined by \feynrules. At this stage, the user can perform a set of basic checks on the Lagrangian (hermiticity, normalization of kinetic terms, \ldots), or directly proceed to the derivation of the Feynman rules via the command
\begin{center}
\begin{verbatim}
verts = FeynmanRules[ L ];
\end{verbatim}
\end{center}
\feynrules\ then computes all the interaction vertices associated with the Lagrangian {\tt L} and stores them in the variable {\tt verts}. The vertices can be used for further computations within \mathematica, or they can be exported to one of the various matrix-element generators for further phenomenological studies of the model.
The translation interfaces can be directly called from within the notebook, \emph{e.g.}, for the \feynarts~interface,
\begin{center}
\begin{verbatim}
WriteFeynArtsOutput[ L ];
\end{verbatim}
\end{center}
This will produce a file formatted for use in \feynarts.  All other interfaces are called in a similar way. As already mentioned, let us note that, even if \feynrules\ is not restricted and can derive Feynman rules for any Lagrangian, the matrix-element generators usually have some information on the Lorentz and color structures hardcoded, and therefore they are much more limited in the set of vertices they can handle. Each matrix-element generator has its own strengths, and in the \feynrules\ approach the same model can be easily exported to various codes, exploiting in this way the strength of each individual tool. In practice, the interfaces check whether all the vertices are compliant with the structures supported by the corresponding matrix-element generator. If not, a warning is printed and the vertex is discarded. Each interface produces at the end a (set of) text file(s), often consistently organized in a single directory, which can be read into the matrix-element generator at runtime and allows to use the new model in a way similar to all other built-in models. For more details on the various interfaces, we refer to Section~\ref{app:Interfaces} and to the manual.

%% file: feynarts.tex
\subsection{\feynarts/\formcalc}

\feynarts~is a \mathematica~package for generating, computing and visualizing Feynman diagrams, both at tree-level and beyond~\cite{Hahn:2000kx}. 
For a given process in a specific model, \feynarts\ starts by generating all the possible topologies, taking into account the number of external legs and internal loops associated to the considered case, together with the set of constraints required by the user, such as the exclusion of one-particle reducible topologies. This stage is purely topological and does not require any physical input. Based on a pre-defined library containing topologies without any external leg for tree-level, one-loop, two-loop and three-loop calculations, the algorithm successively adds the desired number of external legs.
Then, the particles present in the model must be distributed over the obtained topologies in such a way that the resulting diagrams contain the external fields corresponding to the considered process and only vertices allowed by the model. Finally, a \mathematica~expression for the sum over all Feynman diagrams is created.

The second step in the perturbative calculation consists in the evaluation of the amplitudes generated by \feynarts. This can be handled with the help of the \mathematica~package \formcalc\ which simplifies the symbolic expressions previously obtained in such a way that the output can be directly used in a numerical program~\cite{Hahn:1998yk}. \formcalc\ first prepares an input file that is read by the program \textsc{Form} which performs most of the calculation~\cite{Vermaseren:2000nd, Fliegner:1999jq, Fliegner:2000uy, Tentyukov:2004hz, Tentyukov:2007mu}. The Lorentz indices are contracted, the fermion traces are evaluated, the color structures are simplified using the $SU(3)$ algebra, and the tensor reduction is performed\footnote{Let us note that, in order to function correctly, \formcalc\ requires the amplitude to be given in Feynman gauge.}. The results are expressed in a compact form through abbreviations and then read back into \mathematica\ where they can be used for further processing. This allows to combine the speed of \textsc{Form} with the powerful instruction set of \mathematica.

\subsubsection{Model framework}
The \feynarts~models have a very simple structure which can be easily extended to include BSM models.
In particular, the current distribution of \feynarts~contains already several models, including a complete implementation of the Standard Model, as well as a Two-Higgs-Doublet Model and a completely generic implementation of the Minimal Supersymmetric Standard Model.

The \feynarts~models are separated into two files:
\begin{itemize}
\item The generic model file: This file is not specific to any model, but it contains the expressions for the propagators and the Lorentz structures of the vertices for generic scalar, fermion and vector fields. Note that since this file is not specific to any model, different BSM models can be related to the same generic model file. 
\item The classes model file: This file is dedicated to a specific model, and contains the declarations of the particles and the analytic expressions of the couplings between the different fields. This information is stored in the two lists {\tt M\$ClassesDescription} for the particle declarations and {\tt M\$ClassesCouplings} for the couplings.
\end{itemize}
\feynarts~requires all the particles to be grouped into classes, and as a consequence also all the classes couplings must be given at the level of the particle classes. If this is the case, the number of Feynman diagrams generated at runtime is much smaller, which speeds up the code. Since the \feynrules~model files are an extension of the \feynarts~classes model files, the explicit structure of the particle class definitions is very similar to the \feynrules~particle classes discussed in Section~\ref{sec:feynrules}.

\subsubsection{\feynrules\ interface}
\feynrules~includes an interface that allows to output the interaction vertices derived from the Lagrangian as a \feynarts~model file. Note however that at the present stage only the classes model file is generated by \feynrules, the generic model file is hardcoded. The generic model file used by \feynrules~generated models,{\tt feynrules.gen}, is included in the \feynrules~distribution in the {\tt Interfaces/FeynArts} subdirectory and needs to be copied once and for all into the {\tt Models} directory of \feynarts. {\tt feynrules.gen} is based on the corresponding {\tt Lorentz.gen} file included in \feynarts, with some extensions to higher-dimensional scalar couplings as those appearing in non-linear sigma models (see Section~\ref{sec:effectiveL}). 

The \feynrules~interface to \feynarts~can be called within a \mathematica~notebook via the command
\begin{center}
\begin{verbatim}
WriteFeynArtsOutput[ L ];
\end{verbatim}
\end{center}
\feynrules~then computes all the vertices associated with the Lagrangian {\tt L}, and checks whether they all have Lorentz structures compatible with the generic couplings in the generic coupling file, and if so, it extracts the corresponding classes coupling. It not, a message is printed on the screen and the vertex is discarded. At this point we should emphasize that in order to obtain \feynarts~couplings at the level of the particle classes, it is necessary that the Lagrangian is also given completely in terms of particle classes. If the interface encounters a Lagrangian term which violates this rule, it stops and redefines all the classes such that all particles live in their own class. It then starts over and recomputes all the interaction vertices, this time for a Lagrangian where all particle classes are expanded out explicitly. In this way a consistent \feynarts~model file is obtained which can be used with \feynarts. It should however be noted that the generation of the diagrams can be considerably slower in this case, which makes it desirable to write the Lagrangian in terms of particle classes whenever possible.

The model file produced by \feynrules~has the usual \feynarts~structure. Besides the lists which contain the definitions of the particle classes and the couplings, the \feynrules~generated model files contain some more information, which can be useful at various stages during the computation:
\begin{itemize}
\item[-] {\tt M\$ClassesDescription}: This is in general a copy of the corresponding list in the original \feynrules~model file.
\item[-] {\tt M\$ClassesCouplings}: Each entry in the list represents a given interaction between the particle classes, together with the associated coupling constant, represented by an alias {\tt gcxx}, {\tt xx} being an integer. Note that at this level \feynrules~does not compute the counterterms necessary for loop calculations, but they should be added by the user by hand.
\item[-] {\tt M\$FACouplings}: A replacement list, containing the definition of the couplings {\tt gcxx} in terms of the parameters of the model.
\end{itemize}
Furthermore, several other replacement lists ({\tt M\$ExtParams}, {\tt M\$IntParams}, {\tt M\$Masses}) are included, containing the values of the parameters of the model, as well as the masses and widths of all the particles.

%% file: calchep.tex
\subsection{\calchep/\comphep }
The \calchep\  \cite{Pukhov:1999gg,Pukhov:2004ca} and \comphep\  \cite{Pukhov:1999gg,Boos:2004kh} software automate the tree-level Feynman diagram calculation and production of partonic level collider events.  Models with very general Lorentz structures are allowed and general color structures can be incorporated via auxiliary fields. Vertices with more than four particles are not supported at this time.  In this subsection, we will describe the model file structure and how the \feynrules\ interface to \calchep\  and \comphep\  works.

\subsubsection{Model framework}
Models in \calchep\  and \comphep\  are essentially comprised of four files:  
\begin{itemize}
\item \texttt{prtclsN.mdl}: a list of all the particles in the model along with information about the particles that is necessary for calculation of Feynman diagrams.
\item \texttt{varsN.mdl}: a list of the independent (external) parameters in the model along with their numerical value.  
\item  \texttt{funcN.mdl}: a list of the dependent (internal) parameters of the model along with their functional definition.  These definitions can contain any standard mathematical functions defined in the C code.
\item \texttt{lgrngN.mdl}: a list of all the vertices in the model.  It includes the specification of the particles involved in the vertex, an overall  constant to multiply the vertex with and the Lorentz form of the vertex.
\end{itemize}
Note that the letter {\tt N} in the names of the files is an integer which refers to the number of the model.

\subsubsection{\feynrules\ interface}

The \calchep/\comphep\  interface can be invoked with the command
\begin{verbatim}
WriteCHOutput[ L ]
\end{verbatim}
where {\tt L} denotes the Lagrangian.

When invoked, this interface will create the directory \texttt{M\$ModelName} with \texttt{-CH} appended if it does not already exist.  It will create the files \texttt{prtclsN.mdl}, \texttt{varsN.mdl}, \texttt{funcN.mdl} and \texttt{lgrngN.mdl}. 
It will then derive the Feynman rules with four particles or less and write them to \texttt{lgrngN.mdl}.  It will simplify the vertex list by renaming the vertex couplings as \texttt{x1}, \texttt{x2}, \texttt{x3}, etc. and write the definitions of these couplings in \texttt{funcN.mdl} along with the other internal parameters.

Although \calchep\  and \comphep\  can calculate diagrams in both Feynman and unitary gauge, they are much faster in Feynman gauge and it is highly recommended to implement a new model in Feynman gauge.  However, if a user decides to implement the model in unitary gauge, he/she should remember that according to the way \calchep/\comphep\  were written, the ghosts of the massless non-abelian gauge bosons must still  be implemented.  (In particular, the gluonic ghosts must be implemented in either gauge for this interface.)  

One major constraint of the \calchep/\comphep\  system is that the color structure is implicit.  For many vertices (\eg, quark-quark-gluon), this is not a problem.  However, for more complicated vertices, there may be an ambiguity.  For this reason, the writers of \calchep/\comphep\  chose to split them up using auxiliary fields.  Although this can be done for very general vertices, it is not yet fully automatized in \feynrules.  Currently, only the gluon four-point vertex and squark-squark-gluon-gluon vertices are automatically split up in this way.  

The model files are ready to be used and can be directly copied to the \calchep/ \comphep\  models directories.  
The default format for this interface is the \calchep\  format.  A user can direct this interface to write the files in the \comphep\  format by use of the {\tt CompHep} option.  The user who writes \comphep\  model files should note one subtlety.  If the model is written to the \comphep\  directory and if the user edits the model inside \comphep\  and tries to save it, \comphep\  will complain about any C math library functions in the model.  Nevertheless, it does understand them.  We have checked  that if the model works in \calchep, it will work in \comphep\ and give the same results.

\calchep\  has the ability to calculate the widths of the particles on the fly.  By default, this interface will write model files configured for automatic widths.  This can be turned off by setting the option {\tt CHAutoWidths} to {\tt False}.  This option is set to {\tt False} if {\tt \comphep\ } is set to {\tt True}.

This interface also contains a set of functions that read and write the external parameters from and to the \calchep\ variable files ({\tt varsN.mdl}).  After loading the model into \feynrules, the external parameters can be updated by running
\begin{verbatim}
ReadCHExtVars[ Input  ->  < file > ]
\end{verbatim}
 This function accepts all the options of the \calchep\ interface plus the option {\tt Input} which instructs \feynrules\ where to find the \calchep\ variable file.  The default is {\tt varsN.mdl} in the current working directory.  If reading a \comphep\  variable file, then the option {\tt CompHep} should be set to true.  After reading in the values of the variables in the \calchep\ file, it will update the values in \feynrules\ accordingly.

The current values of the external parameters in \feynrules\ can also be written to a \calchep\ external variable file ({\tt varsN.mdl}) using 
\begin{verbatim}
WriteCHExtVars[ Output  ->  < file >]  
\end{verbatim}
This can be done to bypass writing out the entire model if only the model parameters are changed.

%% file: madgraph.tex
\subsection{\protect\mgme}

The  {\sc MadGraph/MadEvent} {\sc v4.4} software \cite{Stelzer:1994ta,Maltoni:2002qb,Alwall:2007st,Alwall:2008pm} allows users to generate tree-level amplitudes and parton-level events for any process (with up to nine external particles). It uses the {\sc Helas} library \cite{Murayama:1992gi,Hagiwara:2008jb} to calculate matrix elements using the helicity formalism in the unitary gauge. Starting from version 4, users have the possibility to use several pre-defined BSM models, including the most generic Two-Higgs-Doublet Model and the Minimal Supersymmetric Standard Model, but can also take advantage of the {\sc \textsc{UsrMod}} interface to implement \textit{simple} Standard Model extensions.

The existing scheme for new model implementations in {\sc MadGraph/MadEvent} has two major drawbacks. First, users need to explicitly provide algebraic expressions for the coupling values used by \madgraph~to calculate amplitudes.  Second, the first version of the {\sc \textsc{UsrMod}} interface only works for models extending the existing Standard Model by \textit{adding} a limited set of new particles and/or interactions. This renders difficult any attempt to modify existing BSM models, or to generalize models previously implemented with this method.

The current version of \madgraph\ rely on a new clearly defined structure for all model libraries. New model libraries can be generated via the corresponding interface to \feynrules, which generates all the required code files automatically. Finally, a new version of the {\sc \textsc{UsrMod}} scripts exists which can be used complementary to {\sc FeynRules} for simple extensions of existing models. All these three new frameworks are introduced and described in the present section.

\subsubsection{Model framework}
\label{MGstruct}
All model libraries supported in the latest versions of {\sc MadGraph/MadEvent} now have the same structure. They are composed of a set of text and {\sc Fortran} files grouped in a single directory, stored in the {\tt Models} subdirectory of the root {\sc MadGraph/MadEvent} installation:
\begin{itemize}
\item \texttt{particles.dat}: a text file containing a list of all particles entering the model and the corresponding properties (name, spin, mass, width, color representation, PDG code, \ldots)
\item \texttt{param\_card.dat}:  a text file containing the numerical values of the necessary external parameters for a specific model. The parameter card has a format compliant with the SUSY Les Houches Accord (LHA) and is dependent on the physics model. One should pay attention to the fact that some of these parameters are related one to each other (\emph{e.g.}, the masses and the widths are generally related to more fundamental Lagrangian parameters). If possible, this file should also contain (preferably at the end) a list of Les Houches {\sc QNUMBERS} blocks describing properties of non-SM particles to facilitate the interface of matrix-element and parton-shower based generators, as proposed in Ref.~\cite{Alwall:2007mw}.
\item \texttt{intparam\_definition.inc}: a text file containing all the algebraic expressions relating internal parameters to external and/or internal parameters. There are two different kinds of internal parameters. Indeed, most of the expressions can be computed once and for all, but in some cases where the parameter depends on the scale of the process~(\emph{e.g.}, the strong coupling constant), it might be desirable to re-evaluate it at an event-by-event basis.
\item \texttt{interactions.dat}: a text file containing a list of all interactions entering the model. Each interaction is characterized by an ordered list of the involved particles, the name of the corresponding coupling, the corresponding type of coupling (for coupling order restrictions) and possible additional switches to select particular {\sc Helas} routines.
\item \texttt{couplingsXX.f} (where XX can be any integer number): these files contain the algebraic expressions for the couplings, expressed as {\sc Fortran} formulas. By convention, the file \texttt{couplings1.f} contains all expressions which should be re-evaluated when an external parameter (\emph{e.g.}, the renormalization scale) is modified on an event-by-event basis. The files \texttt{couplingsXX.f} where XX is greater than 1 contain all expressions which should be only re-evaluated if the default external parameter values are explicitly read from the LHA \texttt{param\_card.dat} parameter card. The actual number of these files may vary, but a single file should be small enough to be compiled using standard {\sc Fortran} compilers. The full list of these files should be included in the makefile through the \texttt{makeinc.inc} include file.
\item \texttt{input.inc} and \texttt{coupl.inc}: {\sc Fortran} files containing all the necessary variable declarations. All parameters and couplings can be printed on screen or written to file using the routines defined in \texttt{param\_write.inc} and \texttt{coupl\_write.inc}, respectively. If needed, the latter can also be printed in a stricter format using routines defined in \texttt{helas\_couplings.f}, so they can be used by external tools (\emph{e.g.}, {\sc Bridge} \cite{Meade:2007js}).
\end{itemize}
Additional {\sc Fortran} files, which are \textit{not} model dependent, should also be provided in order to build the full library. Most of them simply include one or more of the above files, except \texttt{lha\_read.f} which contains all the routines required to read the LHA format. A {\tt makefile} allows the user to easily compile the whole package, to produce a library or a test program called \texttt{testprog} which can be used to quickly check the library behavior by producing a standard log output.\\

\noindent{\textbf{The {\sc \textsc{UsrMod} v2} framework}}\\

The {\sc \textsc{UsrMod} v2} framework has been designed as the successor of the widely-used original {\sc \textsc{UsrMod}} template described in Ref.~\cite{Alwall:2007st}. Taking advantage of the fixed structure we just defined, it provides the user with two new possibilities. First, any pre-existing model can be used as a starting point. This of course includes all models described in the present paper and soon part of the {\sc MadGraph/MadEvent} distribution, but also all future models. This gives a natural framework for building simple extensions following a bottom-up approach, \emph{i.e.}, by adding successively new particles and new interactions and testing their implications at each step. Second, the possible modifications are no longer restricted to the \textit{addition} of new particles/interactions, but any alteration of the model content (including particle removal, modification of existing properties, \ldots) allowed in the context of {\sc MadGraph/MadEvent} is supported in an user-friendly way.

The {\sc \textsc{UsrMod} v2} approach can advantageously replace the full {\sc FeynRules} package when only minor modifications to an existing {\sc MadGraph/MadEvent} model are necessary, \emph{e.g.}, in order to study the phenomenology of a specific new particle and/or interaction, or when the use of the {\sc FeynRules} machinery is not possible, \emph{e.g.}, if a {\sc Mathematica} license is not available. However, when possible, we believe the use of {\sc FeynRules} should be favored over the {\sc \textsc{UsrMod} v2} for the consistent implementation of full new models in {\sc MadGraph/MadEvent}, especially due to the extended validation possibilities available in this context.

From the implementation point of view, the {\sc \textsc{UsrMod} v2} package consists in various {\sc Python} scripts (with one single main script) and works on any platform offering support for this programming language. The actual implementation of a new model is decomposed into four distinct phases:
\begin{enumerate}
\item \textit{Saving}: the model directory used as a starting point should be copied to a new location. The \texttt{USRMOD} script should be then run a first time to create a content archive used as a reference to identify the forthcoming modifications.
\item \textit{Modifying}: the  \texttt{particles.dat}, \texttt{interactions.dat} and  \texttt{ident\_card.dat} files can be modified to arbitrarily add, remove or modify the particle, interaction and parameter content.
\item \textit{Creating}: the \texttt{USRMOD} script should be then run a second time to actually modify all the model files to consistently reflect the changes applied in the previous phase.
\item \textit{Adjusting}: the  \texttt{couplingsXX.f} file(s) can finally be edited, if necessary, to add or modify the relevant coupling expressions. The \texttt{param\_card.dat} file can also be edited to modify default values of external parameters.
\end{enumerate}

At any time, the archive file created during the first phase can be used to restore the initial content of all model files. Several archive files can also be simultaneously saved into the same directory to reflect, for example, the successive versions of a single model. Finally, the intrinsic structure of the {\sc \textsc{UsrMod} v2} package favors various technical (not physical) consistency checks in the output files to minimize as much as possible the compilation and runtime errors.

\subsubsection{\feynrules\ interface}
The {\sc MadGraph/MadEvent} interface can be called from the {\sc FeynRules} package using the \begin{verbatim}
WriteMGOutput[ L ]
\end{verbatim}
routine described in the {\sc FeynRules} documentation\footnote{Several options are available when running the interface. We refer for a detailed review to the \feynrules\ manual.}, where \texttt{L} is the name of the model Lagrangian. Since {\sc MadGraph/MadEvent} currently only supports calculations in the unitary gauge, all the Goldstone and ghost fields are discarded in the \texttt{particles.dat} output, which is directly generated from the model description. After expanding all possible field indices (\eg, associated to flavor), an exhaustive list of non-zero vertices is generated and output as \texttt{interactions.dat}. If possible, the relevant coupling is extracted and, in case it does not already exist, stored in a new coupling variable of the form \texttt{MGVXXX} in a  \texttt{couplingsXX.f} file. All the other required model-dependent files are finally generated, including the \texttt{param\_card.dat} where the default values (which are also the default value for the reading routines in \texttt{param\_read.inc}) are set as specified in the {\sc FeynRules} model file, and where the  {\tt QNUMBERS} blocks correctly reflect the new particle content. All the produced files, together with the relevant model independent files are stored in a local directory \texttt{model\_name\_MG}, ready to be used in {\sc MadGraph/MadEvent}. As mentioned previously, the \texttt{testprog} test program can be compiled and run to check the consistency of the created library.

The two main restrictions of the {\sc MadGraph/MadEvent} interface are related to the allowed Lorentz and color structures of the vertices. As already mentioned, even though {\sc FeynRules} itself can deal with basically any interaction involving scalars, fermions, vectors and spin-two tensors, the {\sc Helas} library, used by {\sc MadGraph/MadEvent} to build and evaluate amplitudes is more restricted. In the case no correspondence is found for a specific interaction, a warning message is displayed by the interface and the corresponding vertex is discarded. If this particular vertex is required for a given application, the user has still the possibility to implement it manually following the {\sc Helas} library conventions and to slightly modify the interface files to deal with this addition. In case the vertex structure is also not present in {\sc MadGraph/MadEvent}, a more involved manual modification of the code is also required. The second limitation of the present interface comes from the fact the color factor calculations are currently hardcoded internally {\sc MadGraph}. While {\sc FeynRules}  can deal with fields in any representation of the QCD color group, {\sc MadGraph} itself is basically limited to the color representations appearing in the Standard Model and Minimal Supersymmetric Standard Model, \eg, a color sextet is not supported.

Let us mention that work to alleviate both limitations is already in progress. The {\sc FeynRules} package could, for example, be used to generate automatically missing {\sc Helas} routines, while a more open version of the {\sc MadGraph} matrix-element generator, \eg, taking advantage of a high-level programming environment, could advantageously deal with arbitrary color structures

%% file: sherpa.tex
 \subsection{\protect\sherpa}
\label{sec:sherpa}
\sherpa\ \cite{Gleisberg:2003xi,Gleisberg:2008ta} is a general-purpose Monte 
Carlo event generator aiming at the complete simulation of physical events
at lepton and hadron colliders. It is entirely written in C++ featuring 
a modular structure where dedicated modules encapsulate the simulation of
certain physical aspects of the collisions. 

\noindent
The central part is formed by the hard interaction, described using 
perturbative methods. The respective generator for matrix elements and phase-space 
integration is \amegic\ \cite{Krauss:2001iv}, which employs the spinor helicity 
formalism \cite{Kleiss:1985yh,Hagiwara:1985yu} in a fully automated approach to 
generate processes for a variety of implemented physics models, 
see Sec.~\ref{sec:sherpa_models}. Phase-space integration is accomplished using 
self-adaptive Monte Carlo integration methods 
\cite{Kleiss:1994qy,Berends:1994pv,Lepage:1980dq,Ohl:1998jn}.

\noindent
The QCD evolution of partons originating from the hard interaction down to the 
hadronization scale is simulated by the parton-shower program \apacic\ 
\cite{Krauss:2005re}. It accounts for parton emissions off all colored particles 
present in the Standard Model and the Minimal Supersymmetric Standard Model. New shower generators have recently been developed 
in the framework of \sherpa\ \cite{Schumann:2007mg,Winter:2007ye}. They account for 
QCD coherence and kinematics effects in a way consistent with NLO subtraction schemes, 
which makes them preferred over \apacic. They will be employed in future versions 
of \sherpa\ using an improved merging prescriptions for matrix elements and 
showers \cite{Hoeche:2009rj}.

\noindent
An important aspect of \sherpa\ is its implementation of a general version of the CKKW 
algorithm for merging higher-order matrix elements and parton showers
\cite{Catani:2001cc,Krauss:2002up}. It has been validated in a variety 
of processes \cite{Krauss:2004bs,Krauss:2005nu,Gleisberg:2005qq} and proved to yield 
reliable results in comparison with other generators \cite{Hoche:2006ph,Alwall:2007fs}.

\noindent
Furthermore, \sherpa\ features an implementation of the model for multiple-parton 
interactions presented in Ref.\ \cite{Sjostrand:1987su}, which was modified to allow 
for merging with hard processes of arbitrary final-state multiplicity and eventually 
including CKKW merging \cite{Alekhin:2005dx}. Furthermore \sherpa\ provides an 
implementation of a cluster-fragmentation model \cite{Winter:2003tt}, 
a hadron and tau decay package including the simulation of mixing effects for 
neutral mesons \cite{hadrons}, and an implementation of the YFS formalism
to simulate soft-photon radiation \cite{Schonherr:2008av}.
\subsubsection{Model framework}
\label{sec:sherpa_models}
\FIGURE[t]{
\centerline{
\framebox{{\sc\Large MODEL}}
\begin{picture}(0,0)(0,0)
\LongArrow(5,-10)(30, -30)
\put(45,-20){Interactions}
\LongArrow(-75,-10)(-100, -30)
\put(-160,-12){Particles}
\put(-160,-24){Parameters}
\end{picture}
}
\newline\newline\newline\newline
\centerline{
\begin{tabular}{rr}
\hspace*{2cm}
\begin{minipage}[h]{5.5cm}
\framebox{{\tt Model\_Base}}
\begin{itemize}
\item {\tt Standard\_Model}
\item {\tt MSSM}
\item ...
\item {\tt FeynRules\_Model}
\end{itemize}
\end{minipage}
&
\begin{minipage}[h]{6.5cm}
\framebox{{\tt Interaction\_Model\_Base}}
\begin{itemize}
\item {\tt Interaction\_Model\_SM}
\item {\tt Interaction\_Model\_MSSM}
\item ...
\item {\tt Interaction\_Model\_FeynRules}
\end{itemize}
\end{minipage}
\end{tabular}}
\caption{\label{fig:sherpa_models} Schematic view of the \protect\sherpa's {\tt MODEL} module, 
hosting the particle and parameter definitions of physics models as well as corresponding 
interaction vertices.}}
Physics model definitions within \sherpa\ are hosted by the module {\tt MODEL}.
Here the particle content and the parameters of any model get defined and are
made accessible for use within the \sherpa\ framework. This task is 
accomplished by instances of the basic class {\tt Model\_Base}. Furthermore
the interaction vertices of various models are defined here that in turn can
be used by \amegic\ to construct Feynman diagrams and corresponding helicity
amplitudes\footnote{Note that within \sherpa\ Feynman rules are always considered 
in unitary gauge.}. The corresponding base class from which all interaction models
are derived is called {\tt Interaction\_Model}. A schematic overview of the
{\tt MODEL} module is given in Fig.~\ref{fig:sherpa_models}.

\noindent
The list of currently implemented physics models reads: 
the Standard Model including effective couplings of the Higgs boson to 
gluons and photons \cite{Dawson:1990zj}, an extension of the SM by a general set 
of anomalous triple- and quartic gauge couplings \cite{Appelquist:1980vg,Appelquist:1993ka}, 
the extension of the SM through a single complex scalar \cite{Dedes:2008bf}, the 
extension of the Standard Model by a fourth lepton generation, the SM plus an axigluon \cite{Antunano:2007da}, 
the Two-Higgs-Doublet Model, the Minimal Supersymmetric Standard Model, and the Arkani-Hamed-Dimopoulos-Dvali (ADD) model 
of large extra dimensions \cite{ArkaniHamed:1998rs,Antoniadis:1998ig}, for details 
see Ref.~\cite{Gleisberg:2008ta}. Besides routines to set up the spectra and Feynman 
rules of the models listed above corresponding helicity-amplitude building blocks 
are provided within \amegic\ that enable the evaluation of production and decay 
processes within the supported models. In particular this 
includes all the generic three- and four-point interactions of scalar, fermionic 
and vector particles present in the SM and Minimal Supersymmetric Standard Model plus the effective operators for 
the loop-induced Higgs couplings and the anomalous gauge couplings. 
The implementation of the ADD model necessitated the extension of the helicity 
formalism to interaction vertices involving spin-two particles 
\cite{Gleisberg:2003ue}.  

\noindent
A necessary ingredient when dealing with the Minimal Supersymmetric Standard Model are specific Feynman
rules for Majorana fermions or fermion number violating interactions.  
To unambiguously fix the relative signs amongst Feynman diagrams involving 
Majorana spinors the algorithm described in~Ref.~\cite{Denner:1992vza} is used. 
Accordingly, the explicit occurrence of charge-conjugation matrices in the
Feynman rules is avoided and instead a generalized fermion flow is employed 
that assigns an orientation to complete fermion chains. This uniquely 
determines the external spinors, fermion propagators and interaction 
vertices involving fermions. 

\noindent
The implementation of new models in \sherpa\ in the traditional way is 
rather straight-forward and besides the public model implementations 
shipped with the \sherpa\ code there exist further private 
implementations that were used for phenomenological studies, 
cf.~Ref.~\cite{Agashe:2006hk,Kilic:2008ub,Meade:2009rb}. From version 
\sherpa-1.2 on \sherpa\ will support model implementations from 
\feynrules\ outputs -- facilitating the incorporation of new
models in \sherpa\ further.

\subsubsection{\feynrules\ interface}
\label{sec:sherpa_feynrules}
To generate \feynrules\ output to be read by \sherpa, the tailor-made
\feynrules\ routine
\begin{verbatim}
WriteSHOutput[ L ]
\end{verbatim}
has to be called, resulting in a
set of {\tt ASCII} files that represent the considered model through
its particle data, model parameters and interaction vertices\footnote{Note again that Feynman rules have to be considered in unitary gauge.}.  

\noindent
To allow for an on-the-flight model implementation from the \feynrules\ 
outputs, instances of the two basic classes {\tt Model\_Base} and 
{\tt Interaction\_Model\_Base} are provided dealing with the proper
initialization of all the particles and parameters, and the interaction 
vertices of the new model, respectively. The actual C++ classes for these 
tasks are called {\tt FeynRules\_Model} and 
{\tt Interaction\_Model\_FeynRules}, see Fig.~\ref{fig:sherpa_models}.

\noindent
The master switch to use a \feynrules\ generated model within \sherpa\ is

\begin{verbatim}
MODEL = FeynRules
\end{verbatim}

\noindent
to be set either in the {\tt (model)} section of the \sherpa\ run 
card or on the command line once the \sherpa\ executable is called.
Furthermore the keywords {\tt FR\_PARTICLES}, {\tt FR\_IDENTFILE}, 
{\tt FR\_PARAMCARD}, {\tt FR\_PARAMDEF} and {\tt FR\_INTERACTIONS}, 
specifying the names of corresponding input files, need to be set. The actual
format and assumed default names of these input cards will be discussed in the 
following:

\begin{itemize}
\item {\tt FR\_PARTICLES} specifies the name of the input file listing
all the particles of the theory including their SM quantum numbers
and default values for their masses and widths, default name is {\tt Particle.dat}. 
An actual particle definition, \emph{e.g.}, for the gluon, looks like
\begin{verbatim}
kf  Mass  Width 3*e Y SU(3) 2*Spin maj on stbl m_on IDName TeXName
21  0.    .0    0   0 8     2      -1  1  1    0    G      G
\end{verbatim}
Hereby {\tt kf} defines the code the particle is referred to internally
and externally, typically its PDG number \cite{Amsler:2008zzb}. 
The values for {\tt Mass} and {\tt Width} need to be given in units of $\GeV$. 
The columns {\tt 3*e} and {\tt Y} specify three times the electric charge and 
twice the weak-isospin. {\tt SU(3)} defines if the particle acts as a 
singlet ({\tt 0}), triplet ({\tt 3}) or octet ({\tt 8}) under $SU(3)_C$. 
{\tt 2*Spin} gives twice the particle's spin and {\tt maj} indicates 
if the particle is charged ({\tt 0}), self-adjoint ({\tt -1}) or a Majorana 
fermion ({\tt 1}). The flags {\tt on}, {\tt stbl} and {\tt m\_on} are internal 
basically and define if a particle is considered/excluded, 
considered stable, and if its kinematical mass is taken into account 
in the matrix-element evaluation. {\tt IDName} and {\tt TeXName} indicate 
names used for screen outputs and potential \LaTeX\ outputs, respectively.
\item In {\tt FR\_IDENTFILE} all the external parameters of the model get 
defined, default file name is {\tt ident\_card.dat}. Names and counters of 
corresponding parameter blocks to be read from {\tt FR\_PARAMCARD} are listed and 
completed by the actual variable names and their numerical types, \emph{i.e.}, real {\tt R} or complex {\tt C}. Besides, variable names for all 
particle masses and widths are defined here. To given an example, the section 
defining the electroweak inputs of the SM may look like
\begin{verbatim}
SMINPUTS 1 aEWM1 R
SMINPUTS 2 Gf    R
SMINPUTS 3 aS    R
CKMBLOCK 1 cabi  R
\end{verbatim}
\item In the file specified through {\tt FR\_PARAMCARD} the numerical 
values of all elementary parameters, particle masses and
decay widths are given, default file is {\tt param\_card.dat}. 
Following the example above the electroweak inputs of the SM can be 
set through: 
\begin{verbatim}
Block SMINPUTS
  1  1.2790000000000E+02  # aEWM1
  2  1.1663900000000E-05  # Gf
  3  1.1800000000000E-01  # aS
Block CKMBLOCK
  1  2.2773600000000E-01  # cabi
\end{verbatim}
\item {\tt FR\_PARAMDEF} gives the file name where all sorts of 
derived parameters are defined, default {\tt param\_definition.dat}. Such 
variables can be functions of the external parameters and 
subsequently other derived quantities. A few examples for the case of 
the SM again might read:
\begin{verbatim}
aEW = pow(aEWM1,-1.) R               ! Electroweak coupling constant
G = 2.*sqrt(aS)*sqrt(M_PI) R         ! Strong coupling constant
CKM11 = cos(cabi) C                  ! CKM-Matrix ( CKM11 )
\end{verbatim}
The parameter definitions get interpreted using an internal algebra interpreter, 
no additional compilation is needed for this task. All standard C++ mathematical 
functions are supported, \emph{e.g.}, {\tt sqr}, {\tt log}, {\tt exp}, {\tt abs}. For 
complex valued parameters, \emph{e.g.}, {\tt CKM11}, the real and imaginary part can be 
accessed through {\tt Real(CKM11)} and {\tt Imag(CKM11)}, the complex conjugate 
is obtained through {\tt Conjugate(CKM11)}.
\item The keyword {\tt FR\_INTERACTIONS} ({\tt Interactions.dat}) specifies the 
input file containing all the vertices of the considered model in a very simple 
format:
\begin{verbatim}
VERTEX 21 21 21  # G G G
   1 G           # right-handed coupling
   2 G           # left-handed coupling
   3 F[1,2,3]    # colour structure
   4 Gauge3      # Lorentz structure
\end{verbatim}
The keyword {\tt VERTEX} signals the start of a new Feynman rule followed
by the PDG codes of the involved particles. Note, the first particle
is always considered incoming the others outgoing. 
Counters number {\tt 1} and {\tt 2} indicate the right and
left-handed coupling of the vertex rule, the right and left-hand
projector being given by $P_{R/L} =\frac12(\1\pm \gamma_5)$, respectively.
Couplings are given in terms of the elementary and derived parameters. 
Counter number {\tt 3} explicitly gives the color structure of the 
interaction in terms of the $SU(3)_C$ structure constants or generators. 
The spin structure of the vertex is given under {\tt 4}, identified 
through a keyword used by \sherpa\ to relate a corresponding 
sub-amplitude to the correct helicity-amplitude building block.
\end{itemize}

\sherpa's interface to \feynrules\ is designed to be as general as possible,
it is, however, by construction restricted in two ways.

\noindent
The functional form of the model parameters, and respectively the couplings, 
is limited by the capabilities of the algebra interpreter that has to construe 
them. This limitation, however, might be overcome by using an external code to 
calculate the needed variables and redefining them as external giving their 
numerical values in {\tt FR\_PARAMCARD}.

\noindent
More severe limitations originate from the restricted ability of \sherpa/\amegic\ 
to handle new types of interactions. Only three and four-point functions 
can be incorporated. For the color structures only the $SU(3)_C$ objects 
$1,\;\delta_{i,j},\;\delta_{a,b},\;T^a_{ij},\;f^{abc}$ and products of those, 
\emph{e.g.}, $f^{abc}f^{cde}$, are supported. Lorentz structures not present in the 
SM or the Minimal Supersymmetric Standard Model are currently not supported by the interface. 
Furthermore, \sherpa\ cannot handle spin-$3/2$ particles. QCD parton showers 
are only invoked for the colored particles present in the SM and the Minimal Supersymmetric Standard Model. 
Hadronization of new colored states is not accomplished, they have to be decayed
before entering the stage of primary hadron generation.

%% file: sm.tex
\subsection{The Standard Model}
\label{sec:SM}

\subsubsection{Model description}

As it serves as basis to any new bottom-up implementation, we briefly describe here the Standard Model implementation.
The SM of particle physics is described by an $SU(3)_C\times SU(2)_L\times U(1)_Y$ gauge theory, where the electroweak symmetry is spontaneously broken so that the fundamental fermions and the weak gauge bosons acquire a mass.  The particle content of the SM is summarized in Table~\ref{tab:SMparts}.
\begin{table}[!t]
  \begin{center}
    \begin{tabular}{| c  | c  | c |}
      \hline  Particle & spin  &  Representations \\ 
      \hline\hline  $L^0_i = (\nu^0_{iL}, l^0_{iL})^T$ & 1/2 &  $\big(\mathbf{1}, \mathbf{2},-1/2\big)$\\ 
        $Q^0_i = (u^0_{iL}, d^0_{iL})^T$ & 1/2 &  $\big(\mathbf{3}, \mathbf{2},1/6\big)$\\ 
        $l^0_{iR}$ & 1/2 &  $\big(\mathbf{1}, \mathbf{1},-1\big)$\\       
        $u^0_{iR}$ & 1/2 &  $\big(\mathbf{3}, \mathbf{1},2/3\big)$\\ 
        $d^0_{iR}$ & 1/2 &  $\big(\mathbf{3}, \mathbf{1},-1/3\big)$\\    
      \hline  $\Phi$ & 0 &  $\big(\mathbf{1}, \mathbf{2},1/2\big)$\\   
      \hline  $B_\mu$ & 1 &  $\big(\mathbf{1}, \mathbf{1},0\big)$\\ 
        $W^i_\mu$ & 1 &  $\big(\mathbf{1}, \mathbf{3},0\big)$\\     
        $G^a_\mu$ & 1 &  $\big(\mathbf{8}, \mathbf{1},0\big)$\\ 
      \hline 
    \end{tabular}
  \end{center} 
  \caption{\label{tab:SMparts}The SM fields and their representations under the Standard Model gauge groups $SU(3)_C \times SU(2)_L \times U(1)_Y$.}
\end{table}
The Lagrangian can be written as a sum of four parts, 
\beq
\lag_{\rm SM} = \lag_{\rm SM, Gauge} + \lag_{\rm SM, Fermions} + \lag_{\rm SM, Higgs} + \lag_{\rm SM, Yukawa}.
\eeq
The pure gauge sector of the theory reads
\beq\label{SMgauge}
\lag_{\rm SM, Gauge} = -{1\over 4} \,G_{\mu\nu}^a\,G^{\mu\nu}_a -{1\over 4} \,W_{\mu\nu}^i\,W^{\mu\nu}_i -{1\over 4} \,B_{\mu\nu}\,B^{\mu\nu},
\eeq
where the SM field strength tensors are defined following the conventions introduced in Eq.~(\ref{eq:fieldstrength}).
The Lagrangian describing the matter fermions can be written as
\beq\label{SMkinferm}
\lag_{\rm SM, Fermions} = \bar Q_{i}^0\,i\slashed{D}\,Q_i^0 + \bar L_{i}^0\,i\slashed{D}\,L_i^0 + \bar u_{Ri}^0\,i\slashed{D}\,u_{Ri}^0+ \bar d_{Ri}^0\,i\slashed{D}\,d_{Ri}^0 + \bar l_{Ri}^0\,i\slashed{D}\,l_{Ri}^0,
\eeq
where $D_\mu$ denotes the $SU(3)_C\times SU(2)_L\times U(1)_Y$ covariant derivative, and we use the conventions of Eq.~(\ref{eq:covdel}). The superscript 0 refers to the gauge eigenstates. Note in particular that explicit mass terms for the matter fermions are forbidden by gauge symmetry. The Higgs field is described by the Lagrangian
\beq
\lag_{\rm SM, Higgs} = D_\mu\,\Phi^\dagger\,D^\mu\,\Phi - \mu^2 \,\Phi^\dagger\,\Phi - \lambda\,(\Phi^\dagger\,\Phi)^2.
\eeq
If $\mu^2<0$, then the Higgs field acquires a vev
that breaks the electroweak symmetry spontaneously. Expanding the Higgs field around its vev, 
\beq
\langle \Phi\rangle = {1\over\sqrt{2}}\, \doub{-i\sqrt{2}\,\phi^+}{v+H+i\phi^0},
\eeq
we generate mass terms for the Higgs boson $H$ and the electroweak gauge fields. The mass eigenstates for the gauge bosons are the $W$ and $Z$ bosons, as well as the photon, which remains massless. The relations between those fields and the original $SU(2)_L\times U(1)_Y$ gauge fields are
\beq\bsp\label{eq:weakmixing}
W_\mu^\pm &\,= {1\over \sqrt{2}}\,(W_\mu^1\mp iW^2_\mu),\\
\doub{Z_\mu}{A_\mu} &\,= \left(\begin{array}{cc} c_w & -s_w\\ s_w & c_w\end{array}\right)\,\doub{W_\mu^3}{B_\mu},
\esp\eeq
where we introduced the weak mixing angle
\beq
c_w\equiv \cos\theta_w = {M_W\over M_Z}.
\eeq
The interactions between the fermions and the Higgs field are described by the Yukawa interactions
\beq
\lag_{\rm SM, Yukawa} = - \bar u^0_{iR}\,y^u_{ij}\,Q_j^0\,\tilde\Phi -  \bar d^0_{iR}\,y^d_{ij}\,Q_j^0\,\Phi - \bar l^0_{iR}\,y^l_{ij}\,L_j^0\,\Phi +{\rm h.c.},
\eeq
where $\tilde\Phi = i\sigma^2\,\Phi^\ast$. After electroweak symmetry breaking the Yukawa interactions generate non-diagonal mass terms for the fermions that need to be diagonalized by unitary rotations on the left and right-handed fields. Since there is no right-handed neutrino, we can always rotate the leptons such that the mass matrix for the charged leptons becomes diagonal and lepton flavor is still conserved. For the quarks however, the diagonalization of the mass matrices introduces flavor mixing in the charged current interactions, described by the well-known CKM matrix. 

\subsubsection{\feynrules\ implementation}
The SM implementation in \feynrules~is divided into the four Lagrangians described in the previous section. In particular, one can use the dedicated functions for  the field strength tensors and the covariant derivative acting on the left and right-handed fermions. Matrix-element generators however need as an input the mass eigenstates of the particles, and therefore it is mandatory to rotate all the gauge eigenstates into mass eigenstates according to the prescriptions discussed in the previous section. This can be done very easily in \feynrules~by writing the Lagrangian in the gauge eigenbasis, and then letting \feynrules~perform the rotation into the mass eigenstates (note, at this point, that \feynrules~does not diagonalize the mass matrices automatically, but this information has to be provided by the user).
However, as the SM Lagrangian is the starting point for many bottom up extensions, the actual implementation was performed directly in terms of the fermion mass eigenstates. The benefit is a slight speed gain due to the rotations in the fermion sector. The default values of the external parameters in the model file are given in Table~\ref{tab:SMparams}, in Appendix \ref{app:SMvalidation}.

Three restriction files for the SM implementation are provided with the default model distribution:
\begin{itemize}
\item {\tt Massless.rst}: the electron and the muon, as well as the light quarks ($u$, $d$, $s$) are massless. 
\item {\tt DiagonalCKM.rst}: the CKM matrix is diagonal.
\item {\tt Cabibbo.rst}: the CKM matrix only contains Cabibbo mixing.
\end{itemize}

Another particularity of the SM implementation is that it was performed both in unitary and in Feynman gauge. The model file contains a switch {\tt FeynmanGauge}, which, if turned to {\tt False}, puts to zero all the terms involving ghost and Goldstone fields. The default value is {\tt False}.\\

\noindent\textbf{Possible extensions}

The SM is at the basis of almost all BSM models, and thus the number of possible extensions of the SM implementation is basically unlimited. A first extension of this model was presented in Section~\ref{sec:hidden} with the HAH model, based on the simplest possible extension of the gauge sector of the SM. Other possibilities are the addition of higher-dimensional operators compatible with the SM symmetries (see Section~\ref{sec:effectiveL}) or the inclusion of right-handed neutrinos via see-saw models.


%% file: 2hdm.tex
\subsection{The general Two-Higgs-Doublet Model}
The Two-Higgs-Doublet Model (2HDM) has been extensively studied for
more than twenty years, even though it has often been only
considered as the scalar sector of some larger model, like the Minimal Supersymmetric Standard Model or
some Little Higgs models for example. The general 2HDM considered here
already displays by itself an interesting phenomenology that justifies
its study like, for example, new sources of $CP$ violation in scalar-scalar
interactions, tree-level flavor changing neutral currents (FCNC's) due to non-diagonal Yukawa interactions, or a light pseudoscalar state and unusual Higgs decays (see Ref.~\cite{deVisscher:2009zb}).

\subsubsection{Model description}
The 2HDM considered here is based on two $SU(2)_L$ doublets $\phi_1$ and $\phi_2$ with the same hypercharge $Y=+1$. If one imposes only gauge invariance, the most general renormalizable Lagrangian is composed of four parts,
\beq
\lag_{\rm 2HDM} = \lag_{\rm 2HDM, Gauge} + \lag_{\rm 2HDM, Fermions} + \lag_{\rm 2HDM, Higgs} + \lag_{\rm 2HDM, Yukawa}.
\eeq
The gauge and fermion sectors of the model are identical to the SM,
\beq
\lag_{\rm 2HDM, Gauge} = \lag_{\rm SM, Gauge} {\rm ~~and~~} \lag_{\rm 2HDM, Fermions} = \lag_{\rm SM, Fermions}.
\eeq
The Lagrangian of the Higgs sector differs from the SM, and can be written
\beq
\lag_{\rm 2HDM, Higgs} = D_\mu\,\phi_1^\dagger\,D^\mu\,\phi_1 + D_\mu\,\phi_2^\dagger\,D^\mu\,\phi_2 - V(\phi_1,\phi_2),
\eeq
and the
scalar potential reads, in the notation of Ref.~\cite{Branco:1999fs},
\be
\label{eq:2HDMpot} \bsp
V(\phi_1,\phi_2) =&\, \mu_1 \phi_1^\dag \phi_1 +\mu_2 \phi_2^\dag \phi_2+\left(\mu_3 \phi_1^\dag \phi_2+\mathrm{h.c.}\right) +\lambda_1 \left(\phi_1^\dag \phi_1\right)^2+ \lambda_2 \left(\phi_2^\dag
  \phi_2\right)^2\\ &  + \lambda_3 \left(\phi_1^\dag
  \phi_1\right)\left(\phi_2^\dag \phi_2\right)+ \lambda_4 \left(\phi_1^\dag
  \phi_2\right)\left(\phi_2^\dag \phi_1\right)\\  & +\left[\left( \lambda_5
  \phi_1^\dag \phi_2 + \lambda_6 \phi_1^\dag \phi_1+ \lambda_7 \phi_2^\dag
  \phi_2\right)\left(\phi_1^\dag \phi_2\right)+\mathrm{h.c.}\right]~,
\esp\ee
where $\mu_{1,2}$ and $\lambda_{1,2,3,4}$ are real parameters while $\mu_3$ and $\lambda_{5,6,7}$ are \textit{a priori} complex.  We assume that the electromagnetic gauge symmetry is preserved, \ie, that the vevs of $\phi_1$ and $\phi_2$ are aligned in the $SU(2)_L$ space in such a way that a single $SU(2)_L$ gauge transformation suffices to rotate them to the neutral components,
\begin{equation}
\label{eq:2hdmvev}
\langle\phi_1\rangle = \frac{1}{\sqrt{2}}\left(
\begin{array}{c}
  0\\
  v_1
\end{array}
\right)\quad\mathrm{and}\quad 
\langle\phi_2\rangle = \frac{1}{\sqrt{2}}\left(
\begin{array}{c}
  0\\
  v_2 e^{i\theta}
\end{array}
\right),
\end{equation}
with $v_1$ and $v_2$ two real parameters such that $v_1^2+v_2^2\equiv v^2=(\sqrt{2} G_F)^{-1}$ and $v_2/v_1\equiv \tan\beta$. 

The most general form for the Yukawa interactions of the two doublets reads
\be
\label{eq:Yuk2HDM}\bsp
\lag_{\rm 2HDM, Yukawa} =&\, -\frac{\overline{Q_L}\sqrt{2}}{v}
\left[(\Delta_d \phi_1 +   \Gamma_d   \phi_2)d_R+(\Delta_u \tilde{\phi}_1
 +   \Gamma_u   \tilde{\phi}_2)u_R\right]\\
&-\frac{\overline{E_L}\sqrt{2}}{v}\left[( \Delta_e \phi_1 + 
  \Gamma_e   \phi_2)e_R\right],
\esp\ee
with $\tilde{\phi_i}\equiv i\sigma^2\phi_i^*$ and where the $3\times 3$ complex Yukawa coupling matrices $\Delta_i$ and $\Gamma_i$ are expressed in the fermion physical basis, \emph{i.e.}, in the
basis where the fermion mass matrices are diagonal. We choose as free parameters the $\Gamma_i$ matrices, while the other Yukawa couplings, the $\Delta_i$ matrices, are deduced from the matching with the observed fermion masses. Conventionally, the two indices $a$ and $b$ of the elements of the Yukawa matrices $\big(\Gamma_{i}\big)_{ab}$ and $\big(\Delta_{i}\big)_{ab}$ refer to the generations of the $SU(2)_L$ doublet and singlet, respectively.

\subsubsection{\feynrules\ implementation}
The 2HDM Lagrangian implemented in \feynrules\ is composed of Eqs.~\eqref{eq:2HDMpot}, \eqref{eq:Yuk2HDM}, together with the canonically normalized kinetic energy terms for the two doublets and the other SM terms. An important feature of this model is the freedom to redefine the two scalar fields $\phi_1$ and $\phi_2$ using arbitrary $U(2)$ transformations
\begin{equation}
\label{eq:unit2HDM}
\left(
\begin{array}{c}
  \phi_1\\
  \phi_2
\end{array}
\right)\rightarrow
\left(
\begin{array}{c}
  H_1\\
  H_2
\end{array}
\right)\equiv U\left(
\begin{array}{c}
  \phi_1\\
  \phi_2
\end{array}
\right)
\quad,\quad U\in U(2)
\end{equation}
since this transformation leaves the gauge-covariant kinetic energy terms invariant. This notion of \textit{basis invariance} has been emphasized in Ref.~\cite{Branco:1999fs} and considered in great detail more recently in Refs.~\cite{Ginzburg:2004vp,Davidson:2005cw,Haber:2006ue}. Since a given set of Lagrangian parameter values is only meaningful for a given basis, let us take advantage of this invariance property to select the \textit{Higgs basis} (by defining the additional file \texttt{HiggsBasis.fr}) where only one of the two Higgs fields acquires a non-zero vev, \ie, 
\begin{equation}
\label{eq:HB}
\langle H_1^0\rangle=\frac{v}{\sqrt{2}}\quad\mathrm{and}\quad \langle H_2^0\rangle=0\,.
\end{equation}
Let us note that the Higgs basis is not univocally defined since the reparametrization $H_2\to e^{i\alpha} H_2$ leaves the condition Eq.~\eqref{eq:HB} invariant, so that the phase of $H_2$ can be fixed in such a way that $\lambda_5$ becomes real. Other basis choices can in principle be easily implemented as different extension files for the main Lagrangian file \texttt{Lag.fr}.

The minimization conditions for potential of Eq.~\eqref{eq:2HDMpot} read, in the Higgs basis defined in Eq.~\eqref{eq:HB}:
\begin{eqnarray}
\label{eq:2HDMmin}
\mu_1&=&-\lambda_1 v^2\nonumber\,,\\
\mu_3&=&-\lambda_6 \frac{v^2}{2}\,,
\end{eqnarray}
which reduces the number of free parameters in the most general 2HDM to ten (seven real parameters, three complex ones and three minimization conditions). Besides the usual three massless would-be Goldstone bosons, the physical spectrum contains a pair of charged Higgs bosons with mass
\begin{equation}
\label{eq:chargedmass}
m_{H^\pm}^2=\frac{\lambda_3 v^2}{2}+\mu_2,
\end{equation}
and three neutral states with the squared mass matrix
\begin{equation}
\label{eq:neutralmasses}
\mathcal{M}^2=
\left(
\begin{array}{ccc}
  2\lambda_1 v^2 & \Re(\lambda_6) v^2 & -\Im(\lambda_6) v^2\\
 \Re(\lambda_6 v^2) & m_{H^\pm}^2+(\lambda_4/2+\lambda_5)v^2 & 0\\
  -\Im(\lambda_6)v^2 & 0 & m_{H^\pm}^2+(\lambda_4/2-\lambda_5)v^2
\end{array}
\right)\,.
\end{equation}
The symmetric matrix $\mathcal{M}$ is diagonalized by an orthogonal matrix $T$. The diagonalization yields masses $m_i$ for the three physical neutral scalars $S^i$ of the model (where the index $i$ refers to mass ordering),
\begin{equation}
\label{eq:Tmatrix}
\mathcal{M}^2\equiv T \mathrm{diag}(m_1^2,m_2^2,m_3^2) T^T\,.
\end{equation}
The doublet components are related to these physical states through 
\begin{equation}
\label{eq:neutralrotation}
\left(
\begin{array}{c}
  \Re(H_1^0)\\
  \Re(H_2^0)\\
  \Im(H_2^0)
\end{array}
\right)=\frac{T}{\sqrt{2}}\left(
\begin{array}{c}
  S^1\\
  S^2\\
  S^3
\end{array}
\right)
\,.
\end{equation}
The Yukawa couplings of the model are fully determined by the $\Gamma_i$ matrices in Eq.~\eqref{eq:Yuk2HDM}, since the $\Delta_i$ are, by definition, fixed to the diagonal fermion mass matrices in the Higgs basis.

In the current implementation of the 2HDM into \feynrules, the user has to provide numerical values for all the $\lambda_i$ parameters in the basis of Eq.~\eqref{eq:HB}, together with the charged Higgs mass $m_{H^\pm}$. The other parameters of the potential, such as the $\mu_i$, are then deduced using Eqs. \eqref{eq:2HDMmin} and \eqref{eq:chargedmass}. As a consequence, the orthogonal matrix $T$ must be calculated externally. This, together with the change of basis required if the user wants to provide potential parameters and Yukawa coupling values in bases different from this of Eq.~\eqref{eq:HB}, can be done using the {\sc TwoHiggsCalc}\ calculator introduced in Ref.~\cite{Alwall:2007st} which has been modified to produce a parameter file compatible with the present implementation. This calculator can also be used to calculate the required Higgs boson tree-level decay widths.



%% file: mssm1.tex
\subsection{The most general Minimal Supersymmetric Standard Model}

Most present supersymmetric models are based on the four-dimensional supersymmetric field theory of Wess and Zumino \cite{Wess:1974tw}. The simplest model is the straightforward supersymmetrization of the Standard Model, with the same gauge interactions, including $R$-parity conservation, and is called the MSSM \cite{Nilles:1983ge, Haber:1984rc}. Its main features are to link bosons with fermions and unify internal and external symmetries. Moreover, it allows for a stabilization of the gap between the Planck and the electroweak scale and for gauge coupling unification at high energies, provides the lightest supersymmetric particle as a dark matter candidate and appears naturally in string theories. However, since supersymmetric particles have not yet been discovered, supersymmetry must be broken at low energies, which makes the superpartners heavy in comparison to their Standard Model counterparts.

Supersymmetric phenomenology at colliders has been extensively investigated for basic processes at leading order \cite{Dawson:1983fw, Craigie:1983as, Craigie:1984tk, Chiappetta:1985ku, delAguila:1990yw, Baer:1993ew, Gehrmann:2004xu, Bozzi:2004qq, Bozzi:2005sy, Bozzi:2007me, Fuks:2008ab, Debove:2008nr} and next-to-leading order \cite{Beenakker:1996ch, Beenakker:1997ut, Baer:1997nh, Berger:1998kh, Berger:1999mc, Berger:2000iu, Beenakker:1999xh} of perturbative QCD. More recently, for some processes, soft-gluon radiation has been resummed to all orders in the strong coupling constant and the results have been matched with the next-to-leading order calculations \cite{Bozzi:2006fw, Bozzi:2007qr, Bozzi:2007tea, Kulesza:2008jb, Kulesza:2009kq, Debove:2009xx}. However, even if those calculations are useful for inclusive enough analyses, they are not suitable if we are interested in a proper description of the full collider environment, for which Monte Carlo event generators are needed. For a couple of years, all the multi-purpose generators already mentioned contain a built-in version of the MSSM. The model files for \feynarts/\formcalc\ are described in Ref.\ \cite{Hahn:2001rv, Hahn:2005qi}, for \calchep\ in Ref.\ \cite{Belanger:2004yn}, for \mgme\ in Ref.\ \cite{Cho:2006sx}, and for \sherpa\ in Ref.\ \cite{Hagiwara:2005wg}. The \sherpa\ and \feynarts/\formcalc\ implementations keep generic mixing in the scalar sector while the other generators rely on a simplified model with only helicity mixing for third generation sfermions.

Our MSSM implementation in \feynrules\ is the most general one in a sense that it is keeping all the flavor-violating and helicity-mixing terms in the Lagrangian and also all the possible additional $CP$-violating phases. This yields thus 105 new free parameters \cite{Dimopoulos:1995ju}, and in order to deal in a transparent way with all of those, our implementation will follow the commonly used universal set of conventions provided by the Supersymmetry Les Houches Accord (SLHA) \cite{Skands:2003cj, Allanach:2008qq}, except for some minor points. We will dedicate a complementary paper to a complete description of the model \cite{Duhr:2009xx}.

\subsubsection{Model description}
\noindent{\textbf{Field content}}
\renewcommand{\arraystretch}{1.3}
\begin{table}[!t]
  \begin{center}
    \begin{tabular}{| c || c  c | c  c | c |}
      \hline Name & Particle & spin & Superpartner & spin &  Representations \\ 
      \hline\hline (s)quarks & $(\chi_{ui} \; \; \chi_{di} )^T$ & 1/2 & $(\tilde{u}^0_{Li} \;\; \tilde{d}^0_{Li} )^T$& 0 & $\big(\mathbf{3}, \mathbf{2},1/6\big)$\\ 
       & $\chi_{\bar ui}$ & 1/2 & $\tilde{u}^{0i\dagger}_R$ &0& $\big(\bar{\mathbf{3}}, \mathbf{1},-2/3 \big)$ \\
      & $\chi_{\bar di}$ & 1/2 & $\tilde{d}^{0i\dagger}_R$ &0& $\big(\bar{\mathbf{3}}, \mathbf{1},1/3 \big)$\\ 
      \hline (s)leptons & $(\chi_{\nu i} \; \; \chi_{li})^T$ &1/2 & $(\tilde{\nu}_i^0 \; \; \tilde{l}^0_{Li})^T$ & 0 & $\big(\mathbf{1}, \mathbf{2},-1/2 \big)$ \\
      & $\chi_{\bar li}$ & 1/2 & $\tilde{l}^{0i\dagger}_R$ &0& $\big(\mathbf{1}, \mathbf{1},1 \big)$ \\ 
      \hline Higgs(inos) & $(H_u^+\;\;H_u^0 )^T$ & 0 & $(\psi_{H_u^+}\, \, \psi_{H_u^0})^T$& 1/2 & $\big(\mathbf{1}, \mathbf{2}, 1/2\big)$\\
      & $(H_d^0\;\;H_d^- )^T$ & 0 & $(\psi_{H_d^0}\, \, \psi_{H_d^-})^T$& 1/2 & $\big(\mathbf{1}, \mathbf{2}, -1/2\big)$\\ 
      \hline $B$-boson, bino & $B^0$& 1 &$\psi_B$ & 1/2 &  $\big(\mathbf{1}, \mathbf{1}, 0\big)$\\
      $W$-bosons, winos & $(W^1 \; \; W^2 \; \;  W^3)^T$ & 1  &  $(\psi_{W^1} \; \; \psi_{W^2} \; \; \psi_{W^3})^T$ & 1/2 &  $\big(\mathbf{1}, \mathbf{3}, 0\big)$\\ 
      gluon, gluino & $g$& 1 &$\psi_g $ & 1/2 &  $\big(\mathbf{8}, \mathbf{1}, 0\big)$\\
      \hline 
    \end{tabular}
  \end{center} 
  \caption{\label{tab:MSSMFieldContent}The MSSM fields and their representations under the Standard Model gauge groups $SU(3)_C \times SU(2)_L \times U(1)_Y$. The quarks and leptons are denoted in terms of two-component Weyl spinors.}
\end{table}

\noindent
Each of the Standard Model quarks and leptons is represented by a four-component Dirac spinor $f^0_i$, where $i$ stands for the generation index and the superscript $0$ denotes interaction eigenstates. It has two associated scalar superpartners, the sfermion $\tilde{f}^0_{Li}$ and the antisfermion $\tilde{f}^{0i\dagger}_R$, being related to the two-component holomorphic Weyl fermion $\chi_{fi}$ and antifermion $\chi_{\bar fi}$, respectively. Let us recall that we relate the Dirac fermion representations to the Weyl ones by
\be
   f_i^0 = \doublet{\chi_{fi}}{\bar\chi_{\bar fi}}.~
\ee For clarity, we will use in the following the left-handed component $f_{Li}^0$ and the right-handed component $f_{Ri}^0$ of the Dirac fermion $f_i^0$ and not the Weyl fermions $\chi_{fi}$ and $\chi_{\bar fi}$.
To preserve the electroweak symmetry from gauge anomaly and in order to give masses to both up-type and down-type fermions, the MSSM contains two Higgs doublets $H_i$, together with their fermionic partners, the higgsinos $\psi_{H_i}$,  
\be 
  H_u = \doublet{H_u^+}{H_u^0}, \qquad
  H_d = \doublet{H_d^0}{H_d^-}, \qquad
  \psi_{H_u} = \doublet{\psi_{H_u^+}}{\psi_{H_u^0}} {\rm ~~and~~}  
  \psi_{H_d} = \doublet{\psi_{H_d^0}}{\psi_{H_d^-}}.~
\ee 
Finally, the spin-one vector bosons of the Standard Model will be associated to Majorana fermions, the gauginos $\psi_B$, $\psi_{W^k}$ and $\psi_g$. 
The names and representations under the Standard Model gauge groups $SU(3)_C \times SU(2)_L \times U(1)_Y$ of the various fields are summarized in Table~\ref{tab:MSSMFieldContent}.\\

\noindent
The full MSSM Lagrangian can we written as
\be \label{eq:MSSM_Lag} \bsp 
\lag_{\rm MSSM} = 
  &\, \lag_{\rm MSSM, Gauge} + \lag_{\rm MSSM, Fermions} + \lag_{\rm MSSM, Yukawa} \\
  &\, + \lag_{\rm MSSM, Scalar \,\,kinetic} + \lag_{\rm MSSM, Scalar \,\, FDW}\\
&\,  + \lag_{\rm MSSM, Ino\,\, kinetic} + \lag_{\rm MSSM, Ino\, \, Yukawa} + \lag_{\rm MSSM, Ino\, \, mix} + \lag_{\rm MSSM, Soft}.~
\esp \ee Starting from the expression of the Lagrangian in the gauge-eigenstate basis of fields given above, we diagonalize the non-diagonal mass matrices arising after electroweak symmetry breaking and provide transformation rules allowing to re-express the Lagrangian in the physical basis.\\

\noindent{\textbf{ Supersymmetry-conserving Lagrangian}}
\noindent
In order to have more compact notations, we introduce the $SU(2)_L$-doublets of left-handed fermions and sfermions,
\be
  Q^0_i = \doublet{u^0_{Li}}{d^0_{Li}} {\rm ~~,~~} 
  L^0_i = \doublet{\nu^0_i}{l^0_{Li}}  {\rm ~~,~~} 
  \tilde{Q}^0_i = \doublet{\tilde{u}^0_{Li}}{\tilde{d}^0_{Li}} {\rm ~~and~~} 
  \tilde{L}^0_i = \doublet{\tilde{\nu}^0_i}{\tilde{l}^0_{Li}}.
  \ee
The pure gauge sector and the matter fermion section of the MSSM are identical to the Standard Model,
\beq
\lag_{\rm MSSM, Gauge} = \lag_{\rm SM, Gauge} {\rm ~~and~~} \lag_{\rm MSSM, Fermions} = \lag_{\rm SM, Fermions}.
\eeq
The Lagrangian for the scalar sector can be divided into one purely kinetic part,
\be \label{eq:lag_sca} \bsp
  \lag_{\rm MSSM, Scalar \,\,kinetic} = &\, 
    D_\mu H_u^\dagger\, D^\mu H_u + 
    D_\mu H_d^\dagger\, D^\mu H_d + 
    D_\mu \tilde L^{0i \dagger}   \, D^\mu \tilde L^0_i + 
    D_\mu \tilde l_R^{0i \dagger} \, D^\mu \tilde l^0_{Ri}\\
  &\, + D_\mu \tilde Q^{0i \dagger}\,D^\mu \tilde Q^0_i +
    D_\mu \tilde u_R^{0i \dagger}\,D^\mu \tilde u^0_{Ri} + 
    D_\mu \tilde d_R^{0i \dagger}\,D^\mu \tilde d^0_{Ri},
\esp\ee 
and one part derived from the $D$-terms, the $F$-terms and the superpotential $W$. Indeed, auxiliary $F$- and $D$-fields must be added to the theory in order to preserve supersymmetry when considering off-shell states, thereby keeping the total number of fermionic and bosonic degrees of freedom equal. Solving their equations of motion leads, for a set of $n$ Dirac fermions $\{\psi_n\}$ and $2 n$ associated supersymmetric scalar partners $\{\phi_{Ln}, \phi^{n\dagger}_R\}$, to 
\be  \bsp
  \lag_{\rm MSSM, Scalar \,\, FDW}=  &\, - W_L^i W_{Li}^\dagger -  W_R^i W_{Ri}^\dagger \\
  &\, - \frac{1}{2} \Big[ W^{ij}_{LL}\, \bar\psi_i^c P_L\psi_j + 2 W_{LR}^{ij}\, \bar\psi_i P_L\psi_j + W_{RR}^{ij}\, \bar\psi_i P_L\psi^c_j + \hc \Big]\\
  &\, - {1 \over 2} \Big[g\,(\phi_{L}^{i\dagger}T^a\phi_{Li}) - g\,(\phi^{i\dagger}_{R}T^a\phi_{Ri}) \Big] \Big[-g\,(\phi_{L}^{j\dagger}T^a\phi_{Lj}) +  g\,(\phi^{j\dagger}_{R}T^a\phi_{Rj}) \Big],~
\esp\ee where $\psi^c$ is the field charge-conjugated to $\psi$ and the derivatives of the superpotential $W$ are given by 
\be \bsp
   W^i_L = & \frac{\del W}{\del\phi_{Li}} {\rm ~~and~~} W^i_R = \frac{\del W}{\del\phi_{Ri}}\\
   W^{ij}_{LL} = & \frac{\del^2 W}{\del\phi_{Li} \del\phi_{Lj}} {\rm ~~,~~} W^{ij}_{LR} = \frac{\del^2 W}{\del\phi_{Li} \del\phi_{Rj}} {\rm ~~and~~} W^{ij}_{RR} = \frac{\del^2 W}{\del\phi_{Ri} \del\phi_{Rj}}.~
\esp \ee Let us note that in the case of Majorana fermions, there is only one associated scalar field. In the framework of the MSSM, the superpotential reads 
\be \label{eq:superW}
   W_{MSSM} = 
     \tilde u_R^{0i\dagger}\,({\bf y^u})_i^{~j}\,(\tilde Q_j^0 \epsilon H_u) -
     \tilde d_R^{0i\dagger}\,({\bf y^d})_i^{~j}\,(\tilde Q_j^0 \epsilon H_d) - 
     \tilde l_R^{0i\dagger}\,({\bf y^l})_i^{~j}\,(\tilde L_j^0 \epsilon H_d) + 
     \mu \, H_u\epsilon H_d, 
\ee where ${\bf y^u}$, ${\bf y^d}$ and ${\bf y^l}$ denote the $3\times3$ Yukawa matrices, $\mu$ the Higgs off-diagonal mass-mixing, and $\epsilon$ the $SU(2)_L$ invariant tensor. We could add to this superpotential other gauge-invariant and renormalizable terms, 
\be
   W_{\slashed{R}} =  {1 \over 2} \lambda^{ij}_{~~k}\,            \tilde L_i^0\, \epsilon\, \tilde L_j^0\, \tilde l_R^{0k\dagger}\,  + 
                      \lambda^{\prime ij}_{~~k}\,     \tilde L_i^0\, \epsilon\, \tilde Q_j^0\, \tilde d_R^{0k\dagger}\,  + 
		      {1 \over 2} \lambda^{\prime\prime}_{ijk}\,  \tilde u_R^{0i\dagger}\,  \tilde d_R^{0j\dagger}\, \tilde d_R^{0k\dagger}  - 
		      \kappa^i\, \tilde L_i^0\, \epsilon\, H_u,~
\ee
with the Yukawa-like couplings $\lambda$, $\lambda^\prime$ and $\lambda^{\prime\prime}$, and a slepton-Higgs off-diagonal mass term $\kappa$. Those couplings would however violate either the lepton number $L$ or the baryon number $B$, as well as the individual lepton flavors. Moreover, they allow for various $B$-violating or $L$-violating processes which have never been seen experimentally. We could just forbid these terms by postulating B and L conservation, but neither $B$ nor $L$ are fundamental symmetries of nature since they are violated by non-perturbative electroweak effects \cite{tHooft:1976up}. Therefore, an alternative symmetry is rather imposed, the $R$-parity \cite{Farrar:1978xj}, defined by 
\be 
   R =  (-1)^{3B+L+2S},~
\ee $S$ being the spin of the particle, forbidding any term different from those in Eq.\ (\ref{eq:superW}). All the Standard Model particles have thus a positive $R$-parity while the superpartners have a negative one. In this paper, we will only describe the implementation of the $R$-parity conserving MSSM. However, the $R$-parity violating extension of our implementation is straightforward, available and described in Ref.\ \cite{Duhr:2009xx}. We can now write the remaining part of the scalar Lagrangian, 
\be \bsp \label{eq:FDW}
  \lag_{\rm MSSM, Scalar \,\, FDW} = &\, -|\mu|^2\,\big(|H_u|^2+|H_d|^2\big)\\
  &\, + \bigg( \mu^\ast\, H_u^\dagger \Big[\tilde d_R^{0i\dagger}\,({\bf y^d})_i^{~j}\, \tilde Q_j^0 + \tilde l^{0i\dagger}_R\,({\bf y^l})_i^{~j}\,\tilde L_j^0 \Big]  
      + \mu^\ast\, H_d^\dagger \Big[\tilde u_R^{0i\dagger}\,({\bf y^u})_i^{~j}\, \tilde Q_j^0 \Big] +  \hc \bigg) \\  
  &\, + \big[H_u^\dagger\, \epsilon\, \tilde Q^{0i\dagger} \big] \, \big[{\bf y^u}^\dagger {\bf y^u} \big]_i^{~j} \, \big[\tilde Q_j^0\, \epsilon\, H_u\big]
      + \big[H_d^\dagger\, \epsilon\, \tilde Q^{0i\dagger} \big] \, \big[{\bf y^d}^\dagger {\bf y^d} \big]_i^{~j} \, \big[\tilde Q_j^0\, \epsilon\, H_d\big] \\
  &\, - \tilde u_R^{0i\dagger} \, \big[{\bf y^u}\, {\bf y^u}^\dagger\big]_i^{~j} \, \tilde u_{Rj}^0 \,\big|H_u\big|^2 
      - \tilde d_R^{0i\dagger} \, \big[{\bf y^d}\, {\bf y^d}^\dagger\big]_i^{~j} \, \tilde d_{Rj}^0 \,\big|H_d\big|^2\\
  &\, + \tilde u_R^{0i\dagger} \, \big[{\bf y^u}\, {\bf y^d}^\dagger\big]_i^{~j} \, \tilde d_{Rj}^0 \, H_d^\dagger H_u 
      + \tilde d_R^{0i\dagger} \, \big[{\bf y^d}\, {\bf y^u}^\dagger\big]_i^{~j} \, \tilde u_{Rj}^0\, H_u^\dagger H_d\\
  &\, + \big[H_d^\dagger\, \epsilon\, \tilde L^{0i\dagger} \big] \, \big[{\bf y^l}^\dagger {\bf y^l}\big]_i^{~j} \, \big[\tilde L^0_j\, \epsilon\, H_d\big] 
      - \tilde l_R^{0i\dagger} \, \big[{\bf y^l}\, {\bf y^l}^\dagger\big]_i^{~j} \, \tilde l_{Rj}^0\, \big|H_d\big|^2\\	
  &\, - \big[\tilde Q^{0j\dagger}\,({\bf y^u}^\dagger)_j^{~i}\,\tilde u_{Ri}^0\big] \, \big[\tilde u_R^{0k\dagger}\, ({\bf y^u})_k^{~l}\,\tilde Q_l^0\big] 
      - \big[\tilde Q^{0j\dagger}\,({\bf y^d}^\dagger)_j^{~i}\,\tilde d_{Ri}^0\big] \, \big[\tilde d_R^{0k\dagger}\, ({\bf y^d})_k^{~l}\,\tilde Q_l^0\big] \\
  &\, - \big[\tilde L^{0j\dagger}\,({\bf y^l}^\dagger)_j^{~i}\,\tilde l_{Ri}^0\big] \, \big[\tilde l_R^{0k\dagger}\, ({\bf y^l})_k^{~l}\,\tilde L_l^0\big] \\
  &\, - \big[\tilde Q^{0j\dagger}\,({\bf y^d}^\dagger)_j^{~i}\,\tilde d_{Ri}^0\big] \, \big[\tilde l_R^{0k\dagger}\,({\bf y^l})_k^{~l}\,\tilde L_l^0\big] 
      - \big[\tilde L^{0j\dagger}\,({\bf y^l}^\dagger)_j^{~i}\,\tilde l_{Ri}^0\big] \, \big[\tilde d_R^{0k\dagger}\,({\bf y^d})_k^{~l}\,\tilde Q_l^0\big] \\
  &\, -{1\over 2} \bigg[ 
         - {g^\prime \over 6} \, \tilde Q^{0i\dagger} \, \tilde Q_i^0 
	 + {g^\prime \over 2} \, \tilde L^{0i\dagger} \, \tilde L_i^0  
	 + {g^\prime \over 2} \, \tilde H_d^\dagger   \, \tilde H_d 
	 - {g^\prime \over 2} \, \tilde H_u^\dagger   \, \tilde H_u \\
  &\,\qquad 
         + {2 g^\prime \over 3}\, \tilde u_R^{0i\dagger} \, \tilde u_{Ri}^0 
	 - {g^\prime \over 3}  \, \tilde d_R^{0i\dagger} \, \tilde d_{Ri}^0 
	 - g^\prime            \, \tilde l_R^{0i\dagger} \, \tilde l_{Ri}^0 
     \bigg]^2 \\
  &\,-{1\over 2} \bigg[ 
         - {g \over 2} \, \tilde Q^{0i\dagger} \, \sigma^k \, \tilde Q_i^0 
	 - {g \over 2} \, \tilde L^{0i\dagger} \, \sigma^k \, \tilde L_i^0 
	 - {g \over 2} \, \tilde H_d^\dagger   \, \sigma^k \, \tilde H_d 
	 - {g \over 2} \, \tilde H_u^\dagger   \, \sigma^k \, \tilde H_u 
     \bigg]^2 \\
  &\, -{1\over 2} \bigg[
         - g_s \, \tilde Q^{0i\dagger}  \, T^a \, \tilde Q_i^0
	 + g_s \, \tilde u_R^{0i\dagger}\, T^a \, \tilde u_{Ri}^0
	 + g_s \, \tilde d_R^{0i\dagger}\, T^a \, \tilde d_{Ri}^0 
      \bigg]^2 .~
\esp\ee 
From the $W^{ij}$ terms of the superpotential, bilinear in the fermionic fields, we can also generate the Yukawa couplings between the matter fermions and the Higgs fields,  
\be \label{eq:lag_FD2}
  \lag_{\rm MSSM, Yukawa} = - \bar u_R^{0i} ({\bf y^u})^{~j}_i  \, Q_j^0 \, \epsilon \, H_u
    + \bar d_R^{0i} ({\bf y^d})^{~j}_i  \, Q_j^0 \, \epsilon \, H_d
    + \bar l_R^{0i} ({\bf y^l})^{~j}_i  \, L_j^0 \, \epsilon \, H_d
    + \hc 
\ee
The terms of the MSSM Lagrangian containing higgsino and gaugino fields can be divided into three parts; pure kinetic terms, 
\be \label{eq:lag_ino1}
  \lag_{\rm MSSM, Ino\,\, kinetic} = 
    {1 \over 2} \bar \psi_B    \, i \slashed{\del} \, \psi_B + 
    {1 \over 2} \bar \psi_{W^k}\, i \slashed{D}    \, \psi_{W^k} + 
    {1 \over 2} \bar \psi_{g^a}\, i \slashed{D}    \, \psi_{g^a} + 
    {1 \over 2} \bar \psi_{H_u}\, i \slashed{D}    \, \psi_{H_u} + 
    {1 \over 2} \bar \psi_{H_d}\, i \slashed{D}    \, \psi_{H_d},
\ee Yukawa interactions obtained from the superpotential terms $W_{ij}$, 
\be\bsp \label{eq:lag_FD3}
   \lag_{\rm ino\, \, Yukawa} =&\, 
     - \big[\bar u_R^{0i}\,  \epsilon\,\psi_{H_u} \big]\, ({\bf y^u})_i^{~j}\, \tilde Q^0_j \,
     + \big[\bar d_R^{0i}\,  \epsilon\,\psi_{H_d} \big]\, ({\bf y^d})_i^{~j}\, \tilde Q^0_j \,
     + \big[\bar l_R^{0i}\,  \epsilon\,\psi_{H_d} \big]\, ({\bf y^l})_i^{~j}\, \tilde L^0_j\\
   &\, - \tilde u_R^{0i\dagger}\, ({\bf y^u})_i^{~j}\, \big[ \bar Q_j^{c\,0}\, \epsilon\, \psi_{H_u} \big] \, 
       + \tilde d_R^{0i\dagger}\, ({\bf y^d})_i^{~j}\, \big[ \bar Q_j^{c\,0}\, \epsilon\, \psi_{H_d} \big] \, 
       + \tilde l_R^{0i\dagger}\, ({\bf y^l})_i^{~j}\, \big[ \bar L_j^{c\,0}\, \epsilon\, \psi_{H_d} \big]\\
   &\, - {1 \over 2} \mu\, \big( \bar \psi_{H_u} \, \epsilon\,  P_L \,\psi_{H_d} + \bar \psi_{H_d} \, \epsilon\, P_L \, \psi_{H_u} \big) + \hc,
\esp\ee 
and additional supersymmetry-conserving gauge-like interactions which have no counterpart in the Standard Model and which are not taken into account through the covariant derivatives, 
\be\bsp\label{eq:lag_ino2}
  \lag_{\rm MSSM, Ino\, \, mix} =
    &\, \sqrt{2} g^\prime \Big[ - {1 \over 6} \tilde Q^{0i\dagger}\, \big(\bar Q_i^{c\,0}\, \psi_{B}\big) 
                                + {1 \over 2} \tilde L^{0i\dagger}\, \big(\bar L_i^{c\,0}\, \psi_{B}\big) \Big] \\
    &\,+ \sqrt{2} g^\prime \Big[  {2 \over 3} \tilde u_{Ri}^0\, \big(u_R^{c\,0i}\, \psi_{B} \big) 
		   	        - {1 \over 3} \tilde d_{Ri}^0\, \big(d_R^{c\,0i}\, \psi_{B} \big)
			        -             \tilde l_{Ri}^0\, \big(l_R^{c\,0i}\, \psi_{B} \big) \Big] \\
    &\,+ \sqrt{2} g^\prime \Big[ -{1 \over 2} H_u^\dagger\, \big(\bar \psi_{H_u}^c\, \psi_{B}\big) + {1 \over 2} H_d^\dagger\, \big(\bar \psi_{H_d}^c\, \psi_{B}\big)\Big]  \\
    &\,- {g \over \sqrt{2}} \Big[   \tilde Q^{0i\dagger}\, \sigma^k\, \big(\bar Q_i^{c\,0}\, \psi_{W^k}\big) 
                                  + \tilde L^{0i\dagger}\, \sigma^k\, \big(\bar L_i^{c\,0}\, \psi_{W^k}\big) \Big] \\
    &\,- {g \over \sqrt{2}} \Big[   H_u^\dagger\, \sigma^k\, \big(\bar \psi_{H_u}^c\, \psi_{W^k}\big) 
                                  + H_d^\dagger\, \sigma^k\, \big(\bar \psi_{H_d}^c\, \psi_{W^k}\big)  \Big] \\
    &\,+ \sqrt{2} g_s \Big[ - \tilde Q^{0i\dagger}\, T^a\, \big(\bar Q_i^{c\,0}\, \psi_{g^a}\big) 
                            + \big(\bar\psi_{g^a}\, u_R^{c\,0i}\big)\, T^a\, \tilde u_{Ri}^0 
			    + \big(\bar\psi_{g^a}\, d_R^{c\,0i}\big)\, T^a\, \tilde d_{Ri}^0 \Big] \\
			    &\,+ \hc
\esp\ee

\noindent{\textbf{ Supersymmetry-breaking Lagrangian}}

As stated in the introduction, the masses of the superpartners must be considerably larger than those of the Standard Model particles. Realistic supersymmetric models must hence include supersymmetry breaking, which is expected to occur spontaneously at some high scale. The Lagrangian density then respects supersymmetry invariance, but the vacuum state does not. Moreover, in order not to introduce quadratic divergences in loop-calculations, supersymmetry has to be broken softly. In practice, since we do not know the supersymmetry-breaking mechanism and the corresponding scale, we will add all possible terms breaking supersymmetry explicitly at low-energy \cite{Girardello:1981wz},
\be \bsp \label{eq:lsoft}
   \lag_{\rm MSSM, Soft} =
   &\,- {1 \over 2} \Big[ M_1 \bar\psi_{B}\, \psi_{B} + M_2\,\bar \psi_{W^k}\, \psi_{W^k} + M_3\,\bar \psi_{g^a}\, \psi_{g^a} \Big]\\
   &\,- \,\tilde Q^{0i\dagger}\,({\bf m^2_{\tilde Q}})_i^{~j}\,\tilde Q^0_j - \,\tilde u_R^{0i\dagger}\,({\bf m^2_{\tilde U}})_i^{~j}\,\tilde u_{Rj}^0 - \,\tilde d_R^{0i\dagger}\,({\bf m^2_{\tilde D}})_i^{~j}\,\tilde d_{Rj}^0 \\
   &\,- \,\tilde L^{0i\dagger}  \, ({\bf m^2_{\tilde L}})_i^{~j}\, \tilde L^0_j - \,\tilde l_R^{0i\dagger}\, ({\bf m^2_{\tilde E}})_i^{~j}\, \tilde l_{Rj}^0 \\ 
   &\,- \, m_{H_u}^2\, H_u^\dagger\, H_u  - \, m_{H_d}^2\, H_d^\dagger\, H_d  - \, b\,      H_u\,         \epsilon\, H_d\,  - \, b^\ast\, H_u^\dagger\, \epsilon\, H_d^\dagger\\
   &\,+ \Big[ -\,\tilde u_R^{0i\dagger}\, ({\bf T_u})_i^{~j}\, \tilde Q_j^0\, \epsilon H_u 
              +\,\tilde d_R^{0i\dagger}\, ({\bf T_d})_i^{~j}\, \tilde Q_j^0\, \epsilon H_d 
	      +\,\tilde l_R^{0i\dagger}\, ({\bf T_l})_i^{~j}\, \tilde L_j^0\, \epsilon H_d \,\,+\, \hc \Big].
\esp \ee
The first line of Eq.\ (\ref{eq:lsoft}) contains gaugino mass terms, the second and third lines the sfermion mass terms, where ${\bf m^2_{\tilde Q}}$, ${\bf m^2_{\tilde L}}$, ${\bf m^2_{\tilde u}}$, ${\bf m^2_{\tilde d}}$, ${\bf m^2_{\tilde e}}$ are $3\times3$ hermitian matrices in generation space, the fourth line mass terms for the Higgs fields, and the fifth line the trilinear scalar interactions, ${\bf T_u}$, ${\bf T_d}$, and ${\bf T_e}$ being also $3\times 3$ matrices in generation space. Let us note that the additional Higgs mass terms are required in order to break electroweak symmetry spontaneously. We remind the reader that for the $R$-parity violating MSSM, additional soft supersymmetry-breaking terms must also be included,
\be \bsp
   \lag_{\rm MSSM, \slashed{R}\, Soft} = 
   &\, \bigg[ - D^i\, \tilde L_i^0\, \epsilon\, H_u\, + \, m_{L_i H}^2\, \tilde L^{0i\dagger}\, H_d + \hc \bigg]\\
   &\, + \bigg[{1 \over 2} T^{ij}_{~~k}\,            \tilde L_i^0\, \epsilon\, \tilde L_j^0\, \tilde l_R^{0k\dagger}\,  + 
      T^{\prime ij}_{~~k}\,     \tilde L_i^0\, \epsilon\, \tilde Q_j^0\, \tilde d_R^{0k\dagger}\,  + 
      {1 \over 2} T^{\prime\prime}_{ijk}\,  \tilde u_R^{0i\dagger}\,  \tilde d_R^{0j\dagger}\, \tilde d_R^{0k\dagger} + \hc \bigg].
\esp \ee\\

\noindent{\textbf{Particle mixing}}

In order to break the electroweak symmetry to electromagnetism, the classical Higgs potential, \ie \! all the Lagrangian terms quadratic or quartic in the Higgs fields, must have a non-trivial minimum. Due to gauge invariance, one of the charged Higgs vev can be rotated away, which yields a zero vev for the other charged Higgs. Again, we can use gauge-invariance to choose the remaining vevs real and positive, so that we can replace the neutral Higgs fields in $\lag_{\rm MSSM}$ by
\be
  H_u^0 \to  {v_u + h_u^0 \over \sqrt{2}} {\rm ~~and~~} 
  H_d^0 \to  {v_d + h_d^0 \over \sqrt{2}},~
\ee where $v_u$ and $v_d$ are the two vacuum expectation values of the neutral Higgses and $h_u^0$ and $h_d^0$ are complex scalar fields. Let us note that those relations are only valid in unitary gauge, which is the case we are dealing with here. We can then extract mass matrices for the gauge bosons $B^0$ and $W^k$, diagonalize them, and derive the physical mass-eigenstates, the photon $A$, the $W$ and $Z$ boson. The transformation rules relating the mass and interaction bases are given by Eq.~(\ref{eq:weakmixing}).
The weak mixing angle $\theta_w$ and the physical masses $M_Z$ and $M_W$ are defined by
\be
   \cwd = {g^2 \over g^2+g^{\prime 2}}{~~,~~}
   M_Z = {g \over 2 \cw} \sqrt{v_u^2+v_d^2} {\rm ~~and~~}
   M_W = {g \over 2} \sqrt{v_u^2+v_d^2}\,.~
\ee
In the Higgs sector, three out of the eight real degrees of freedom of the two doublets are the would-be Nambu-Goldstone bosons becoming the longitudinal modes of the weak bosons, while the five others mix to the physical Higgses, $h^0$, $H^0$, $A^0$ and $H^\pm$. The diagonalization of several mass matrices leads to the transformation rules
\be\bsp
  h_u^0 =  \ca\, h^0 + \sa\, H^0 + i \cb\, A^0 &{\rm~~and~~} h_d^0 = -\sa\, h^0 + \ca\, H^0 + i \sb\, A^0 , \\ 
  H_u^+ =  \cb\, H^+ &{\rm~~and~~} H_d^- = \sb\, H^-,~
\esp\ee where $\tb$ is the ratio of the two vevs $v_u$ and $v_d$.

Collecting the terms of $\lag_{\rm MSSM}$ bilinear in the Majorana fermions $\psi_{B}$, $\psi_{W^3}$, $\psi_{H_u^0}$ and $\psi_{H_d^0}$, we can extract the neutral gaugino-higgsino mass matrix $Y$ which can be diagonalized through a unitary matrix $N$,
\be
 N^\ast\, Y\, N^{-1} ={\rm diag}\, (m_{\tilde{\chi}^0_1},m_{\tilde{\chi}^0_2},
 m_{\tilde{\chi}^0_3},m_{\tilde{\chi}^0_4})\,.~
\ee This matrix relates the four physical neutralinos $\tilde \chi^0_i$ to the interaction-eigenstates,
\be 
  (\tilde{\chi}^0_1\,\,\, \tilde{\chi}^0_2\,\,\,  \tilde{\chi}^0_3\,\,\, \tilde{\chi}^0_4)^T = N \, (\psi_{B}\,\,\,  \psi_{W^3}\,\,\,  \psi_{H_d^0}\,\,\,  \psi_{H_u^0})^T.~
\ee Similarly, we can collect the terms yielding the charged gaugino-higgsino mass matrix $X$ and diagonalize it through the two unitary matrices $U$ and $V$,
\be
 U^\ast\, X\, V^{-1} = {\rm diag}\,(m_{\tilde{\chi}^\pm_1},m_{\tilde{\chi}^ \pm_2})\,.~
\ee Those matrices relate the interaction-eigenstates to the physical charginos $\tilde \chi^\pm_i$ according to
\be
  (\tilde{\chi}^+_1\,\,\, \tilde{\chi}^+_2)^T = V (\psi_{W^+}\, \, \, \psi_{H^+_u})^T {\rm~~and~~}  (\tilde{\chi}^-_1\,\,\, \tilde{\chi}^-_2)^T = U (\psi_{W^-}\, \,\, \psi_{H^-_d})^T.~
\ee
Let us note that the fields $\psi_{W^\pm}$ are obtained after rotating $\psi_{W^1}$ and $\psi_{W^2}$ as in Eq.\ (\ref{eq:weakmixing}) for the $W$ boson.

Within the Standard Model, the only source of flavor violation arises through the rotation of the up-type and down-type quark interaction-eigenstates basis $\{d_{Li}^0,u_{Li}^0,d_{Ri}^0,u_{Ri}^0\}$ to the basis of the physical mass eigenstates $\{d_{Li},u_{Li},d_{Ri},u_{Ri}\}$ in which the Yukawa matrices ${\bf y^u}$ and ${\bf y^d}$ are diagonal. These diagonalizations require four unitary matrices 
\be
  d_{Li}^0 = V_d\, d_{Li}\,, \quad d_{Ri}^0 = U_d\, d_R\,, \quad  u_{Li}^0 = V_u\, u_{Li}\,, \quad u_{Ri}^0 = U_u\, u_R\,, 
\ee and render the charged-current interactions proportional to the CKM matrix, 
\be
  V_{\rm CKM}=V_u^\dagger V_d~.~
\ee The leptonic sector is however diagonalized with the help of only two matrices, since the neutrinos are assumed massless,
\be
  l_{Li}^0 = V_l\, l_{Li}\,, \quad l_{Ri}^0 = U_l\, l_R\,, \quad 
  \nu_i^0 = \nu_i\,.~
\ee Let us note that the matrices $U_u$, $U_d$, $V_l$ and $U_l$ do not appear in the rotated Lagrangian. 
In the sfermion sector, we define the super-CKM basis  \cite{Hall:1985dx} as the basis in which the sfermion interaction-eigenstates undergo the same rotations as their fermionic counterparts. As a consequence, the squark charged-current interactions are also proportional to the CKM matrix. However, the fermion and sfermion fields can be misaligned due to possible off-diagonal mass terms in the supersymmetry-breaking Lagrangian $\lag_{\rm MSSM, Soft}$. The diagonalization of the four mass matrices $M_{\tilde{u}}^2$, $M_{\tilde{d}}^2$, $M_{\tilde{l}}^2$ and $M_{\tilde{\nu}}^2$ requires thus the introduction of three $6 \times 6$ matrices $R^u$, $R^d$ and $R^l$ and one $3 \times 3$ matrix $R^\nu$,
\ba
  {\rm diag}\,(m_{\tilde u_1}^2, \ldots, m_{\tilde u_6}^2) ~=~ R^u\, M_{\tilde{u}}^2\,R^{u\dag} &{\rm ,~~}& {\rm diag}\,(m_{\tilde d_1}^2, \ldots, m_{\tilde d_6}^2) ~=~ R^d\, M_{\tilde{d}}^2\, R^{d\dag},\\
  {\rm diag}\,(m_{\tilde \nu_1}^2, \ldots, m_{\tilde \nu_3}^2) ~=~ R^\nu\, M_{\tilde{\nu}}^2\,R^{\nu\dag} &{\rm ,~~}& {\rm diag}\,(m_{\tilde l_1}^2, \ldots, m_{\tilde l_6}^2) ~=~ R^l\, M_{\tilde{l}}^2\, R^{l\dag}\,.
\ea These matrices relate the physical mass-eigenstates to the interaction-eigenstates through 
\be \bsp
  (\tilde{u}_1, \tilde{u}_2, \tilde{u}_3, \tilde{u}_4, \tilde{u}_5, \tilde{u}_6)^T =&\; R^u\, (\tilde{u}_L, \tilde{c}_L, \tilde{t}_L, \tilde{u}_R, \tilde{c}_R, \tilde{t}_R)^T ,\\
  (\tilde{d}_1, \tilde{d}_2, \tilde{d}_3, \tilde{d}_4, \tilde{d}_5, \tilde{d}_6)^T =&\; R^d\, (\tilde{d}_L, \tilde{s}_L, \tilde{b}_L, \tilde{d}_R, \tilde{s}_R, \tilde{b}_R)^T , \\
  (\tilde{\nu}_1, \tilde{\nu}_2, \tilde{\nu}_3)^T =&\; R^\nu\, (\tilde{\nu}_e, \tilde{\nu}_\mu, \tilde{\nu}_\tau)^T,~\\
  (\tilde{l}_1, \tilde{l}_2, \tilde{l}_3, \tilde{l}_4, \tilde{l}_5,  \tilde{l}_6)^T =&\; R^l\, (\tilde{e}_L, \tilde{\mu}_L, \tilde{\tau}_L, \tilde{e}_R, \tilde{\mu}_R, \tilde{\tau}_R)^T.~ 
\esp\ee
The less general built-in implementation of the MSSM in \mgme\ and \calchep\ can easily be recovered by setting all the off-diagonal elements of these general mixing matrices to zero, except for the third generation flavor-conserving and helicity-mixing elements. In a framework of $R$-parity violating scenarios, additional mixings between charged leptons and charginos, neutrinos and neutralinos, Higgses, sleptons and sneutrinos can arise. Moreover, the sneutrino fields can acquire vevs, since they are not protected by lepton number conservation, unlike in the $R$-parity conserving MSSM where it is a conserved quantum number. We refer the reader to Ref.\ \cite{Duhr:2009xx} for more information.

\subsubsection{\feynrules\ implementation}
\noindent{\textbf{Current implementation}}

We describe here the implementation of the most general $R$-parity conserving MSSM in \feynrules. We divide the complete Lagrangian in six pieces, 
\be \bsp 
\lag_{\rm MSSM} = 
  &\, \lag_{\rm MSSM, Gauge} + \lag_{\rm MSSM, Fermions} + \lag_{\rm MSSM, Scalars} \\
  &\,+ \lag_{\rm MSSM, FD} + \lag_{\rm MSSM, Inos} + \lag_{\rm MSSM, Soft} \esp
~~,  \ee related to Eq.\ (\ref{eq:MSSM_Lag}) through 
\be \bsp
  \lag_{\rm MSSM, Scalars} &= \lag_{\rm MSSM, scalar \,\,kinetic}, \\
  \lag_{\rm MSSM, Inos}     &= \lag_{\rm MSSM, Ino\,\, kinetic} + \lag_{\rm MSSM, Ino\,\, mix}, \\
  \lag_{\rm MSSM, FD}      &= \lag_{\rm MSSM, Scalar \,\, FDW} + \lag_{\rm MSSM, Yukawa} + \lag_{\rm MSSM, Ino\,\, Yukawa}.~
\esp \ee
$\lag_{\rm MSSM, Scalars}$ contains the scalar kinetic terms and gauge interactions, all the $F$-terms and $D$-terms are embedded in $\lag_{\rm MSSM, FD}$, and we have grouped all the non-Yukawa terms involving neutralinos and charginos $\lag_{\rm MSSM, Inos}$. The scalar sector is implemented in terms of gauge-eigenstates while the fermion and gaugino/higgsino sectors are directly implemented in terms of mass eigenstates. For clarity and for generalization purposes, the \feynrules~implementation is splitted into 18 files,
\begin{itemize}
  \item {\tt Gauge.fr}: the definition of the gauge groups and the indices,
  \item {\tt PrmExt.fr}: the external parameters, following the SLHA conventions,
  \item {\tt PrmInt.fr}: all the internal parameters, derived from the external ones,
  \item {\tt PrmAux.fr}: auxiliary parameters, such as identity matrices,
  \item {\tt FldFer.fr}: physical fermionic states (quarks, leptons, charginos, neutralinos, gluino),
  \item {\tt FldVec.fr}: physical gauge bosons (gluon, photon, $W$ and $Z$ bosons),
  \item {\tt FldSca.fr}: physical scalar fields (sfermions and Higgses),
  \item {\tt FldAux.fr}: gauge-eigenstates (sfermions, Higgses, $SU(2)_L \times U(1)_Y$ bosons),
  \item {\tt FldGst.fr}: gluonic ghost (required in order to have \calchep\ running properly),
  \item {\tt LagGau.fr}: $\lag_{\rm Gauge}$, defined in Eq.\ (\ref{SMgauge}),
  \item {\tt LagFer.fr}: $\lag_{\rm Fermions}$, defined in Eq.\ (\ref{SMkinferm}),
  \item {\tt LagSca.fr}: $\lag_{\rm Scalars}$, defined in Eq.\ (\ref{eq:lag_sca}),
  \item {\tt LagFer.fr}: $\lag_{\rm Inos}$, defined in Eqs.\ (\ref{eq:lag_ino1}) and (\ref{eq:lag_ino2}),
  \item {\tt LagFD.fr}: the definition of the superpotential,
  \item {\tt LagFDgen.fr}: the $D$-terms and a generic routine deriving automatically the $F$-terms from the superpotential, allowing for the calculation of $\lag_{\rm FD}$,
  \item {\tt LagBrk.fr}: the soft supersymmetry-breaking Lagrangian $\lag_{\rm Soft}$, from Eq.\ (\ref{eq:lsoft}),
  \item {\tt LagGst.fr}: the ghost Lagrangian (only QCD ghosts are supported so far),
  \item {\tt SUSY.fr}: the main file collecting all the different pieces.
\end{itemize} Before running the model, the user has to provide values for the four switches {\tt FeynmanGauge}, {\tt \$CKMDiag}, {\tt \$sWScale} and {\tt \$svevScale}.
\begin{itemize}
  \item The switch {\tt FeynmanGauge} is not used in the present implementation and has to be set to {\tt False}, since the whole model is implemented in unitary gauge. However, further developments will allow for various gauges.
  \item The switch {\tt \$CKMDiag} allows for a CKM matrix different from the identity or not, depending on its {\tt True} or {\tt False} value. The CKM matrix could also be forced to the identity through a restriction file.
  \item The two switches {\tt \$sWScale} and {\tt \$svevScale} can be set to the values {\tt "weak"} or {\tt "susy"}, regarding the scale at which the electroweak parameters and the vevs of the Higgs fields will be evaluated.
\end{itemize}

\noindent{\textbf{Possible extensions}}

A large number of supersymmetric models can be built from the most general MSSM. As stated above, we can take the proper limit in the various mixing matrices and get back to the commonly studied constrained MSSM scenarios. Moreover, other gauges are planned to be implemented (\eg, Feynman gauge or non-linear gauges). Other possible extensions consist in the addition of new particles, such as an additional singlet (the Next-to-Minimal Supersymmetric Standard Model) or right-handed neutrinos (and the corresponding sneutrinos).

%% file: 3-site.tex
\subsection{The Minimal Higgsless Model}
To date, we have not detected a single fundamental scalar field and we do not know if we ever will.  Electroweak symmetry breaking may instead occur as the result of the condensation of a strongly coupled bound state as in technicolor theories~\cite{Weinberg:1979bn,Susskind:1978ms} or it may occur as the result of boundary conditions in a compactified extra dimension as in so-called Higgsless theories~\cite{Csaki:2003dt,Csaki:2003zu}.  These theories are closely related~\cite{Maldacena:1997re} and each contains a tower of vector resonances which are responsible for unitarizing $WW$ and $WZ$ scattering in the absence of a fundamental scalar field (such as the Higgs)~\cite{SekharChivukula:2001hz,Chivukula:2002ej,Chivukula:2003kq}.  Unitarity constrains the mass of the lowest of these resonances to be below $\sim1.2$ TeV, making it discoverable (or excludable) at the LHC~\cite{He:2007ge,Ohl:2008ri,He:in-prep}.

\begin{figure}
\begin{center}
\includegraphics[scale=0.5]{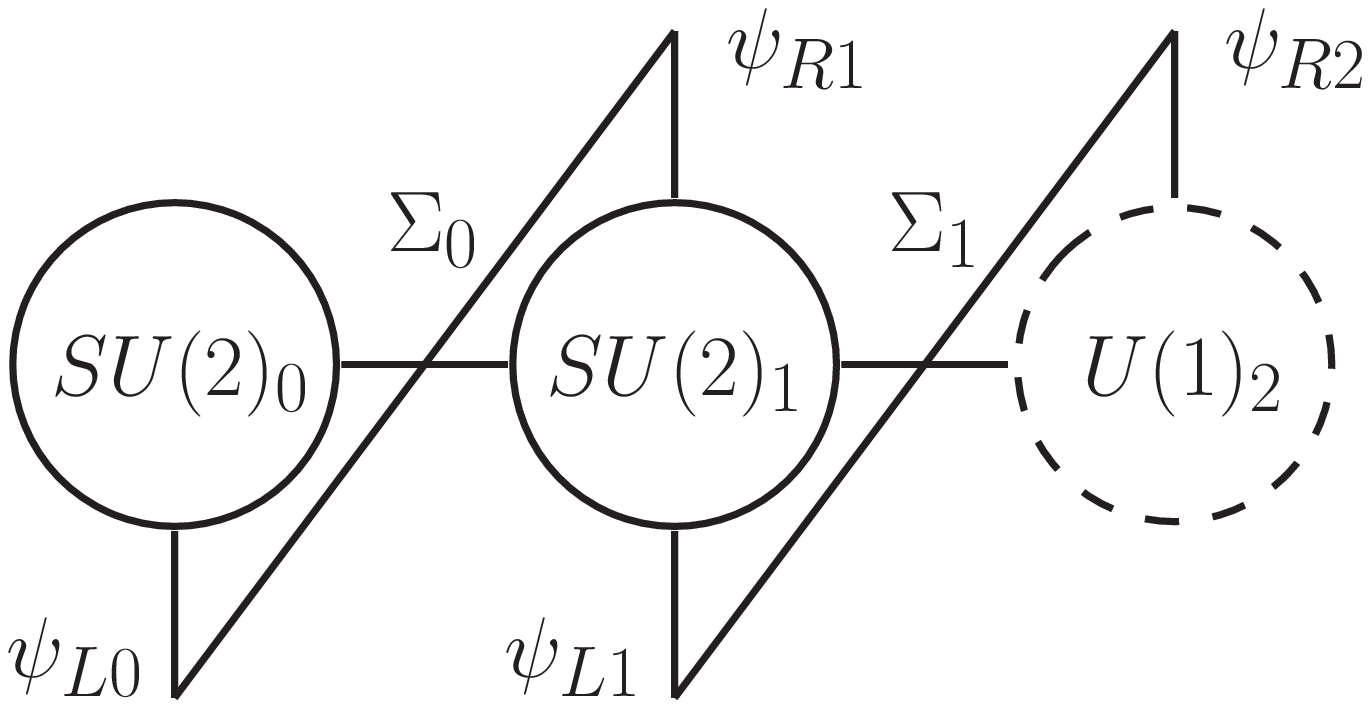}
\end{center}
\caption{\label{3-moose}A schematic ``moose'' diagram of the Three-Site model.  The circles represent gauge groups.  The two circles on the left are $SU(2)$ gauge groups while the one on the right is a $U(1)$ gauge group.  
}
\end{figure}

Although, it is impossible to implement the entire tower of resonances, it is also unnecessary as the low energy phenomenology is dominated by the lowest modes.  Deconstruction \cite{ArkaniHamed:2001ca,Hill:2000mu} gives a consistent, gauge invariant way of implementing only a subset of the resonances in a low energy effective theory.  The minimal deconstructed Higgsless model contains all the non-scalar fields of the Standard Model plus the first new resonances, the $W'$, $Z'$ and heavy partners of the fermions and is called the Minimal Higgsless Model or Three-Site Model \cite{Chivukula:2006cg}.  A schematic ``moose'' diagram of the Minimal Higgsless Model is presented in Fig.~\ref{3-moose}.

Precision electroweak constraints~\cite{Peskin:1991sw,Altarelli:1990zd,Altarelli:1991fk,Barbieri:2004qk} are satisfied in Higgsless models by allowing the fermions to delocalize into the bulk in a certain ``ideal'' way~\cite{Chivukula:2005xm,Cacciapaglia:2004rb,Foadi:2004ps}.  The consequence of this is that the heavy partners of the $W$ and $Z$, the $W'$ and $Z'$, are fermiophobic and have very small or vanishing couplings to the light SM fermions.  Nevertheless, these heavy vector resonances can be discovered (or excluded) at the LHC \cite{He:2007ge,Ohl:2008ri,He:in-prep}.

\subsubsection{\label{3-Site:model description}Model description}


\subsubsection*{\label{3-Site:gauge}Gauge Sector}

The gauge group of the Three-Site Model is
\begin{equation}
G = SU(3)_{QCD} \times SU(2)_0 \times SU(2)_1 \times U(1)_2,
\end{equation}
where $SU(2)_0$ is represented by the leftmost circle in Fig.~\ref{3-moose} and has coupling $g$, $SU(2)_1$ is represented by the center circle in  Fig.~\ref{3-moose} and has coupling $\tilde{g}$ and $U(1)_2$ is represented by the rightmost dashed circle in Fig.~\ref{3-moose} and has coupling $g'$.  We define
\begin{equation}\label{x def}
x = \frac{g}{\tilde{g}} \quad \mbox{and} \quad t = \frac{g'}{g} = \frac{s}{c},
\end{equation}
where $s^2+c^2=1$. 

The Lagrangian for the Minimal Higgsless Model can be written as a sum of six parts,
\beq
\lag_{\rm MHM} = \lag_{\rm MHM,\ Gauge} + \lag_{{\rm MHM,}\ D\Sigma} + \lag_{\rm MHM,\ GF} + \lag_{\rm MHM,\ Ghost}+ \lag_{{\rm MHM,}\ \psi} + \lag_{{\rm MHM,}\ \psi\Sigma}.
\eeq
The kinetic and self interaction terms for the gauge bosons is given by the usual gauge invariant terms:
\begin{equation}
\mathcal{L}_{\rm MHM, Gauge} = -\frac{1}{4}\,G_{\mu\nu}^a\,G^{\mu\nu}_a - \frac{1}{4}\,F_{0,\mu\nu}^i\,F_{0,i}^{\mu\nu} - \frac{1}{4}\,F_{1,\mu\nu}^i\,F_{1,i}^{\mu\nu} - \frac{1}{4} F_{2,\mu\nu}^i\,F_{2}^{\mu\nu}.
\end{equation}

The horizontal bars in Fig.~\ref{3-moose} represent non-linear sigma models $\Sigma_j$ which come from unspecified physics at a higher scale and which give mass to the six gauge bosons other than the photon.  This is encoded in the leading order effective Lagrangian term
\begin{equation}\label{L_D Sigma}
\mathcal{L}_{{\rm MHM, } D\Sigma} = \frac{f^2}{4}\mbox{Tr}\left[\left(D_\mu\Sigma_0\right)^\dagger D^\mu\Sigma_0 + \left(D_\mu\Sigma_1\right)^\dagger D^\mu\Sigma_1 \right],
\end{equation}
where
\beq\bsp
D_\mu\Sigma_0 &\,= \partial_\mu\Sigma_0 +igW_{0\mu}\Sigma_0 - i\tilde{g}\Sigma_0W_{1,\mu},\\
D_\mu\Sigma_1 &\,= \partial_\mu\Sigma_1 +i\tilde{g}W_{1\mu}\Sigma_1 - ig'\Sigma_1W_{2,\mu}.
\esp\eeq
 The non-linear sigma models can be written in exponential form
\begin{equation}
\Sigma_j = e^{i2\pi_j/f},
\end{equation}
which exposes the Goldstone bosons that become the longitudinal components of the massive gauge bosons.  $\pi_j$ and $W_j$ are written in matrix form and are
\begin{equation}
\pi_j = \left(\begin{array}{cc}
\frac{1}{2}\pi_j^0&\frac{1}{\sqrt{2}}\pi_j^+\\\frac{1}{\sqrt{2}}\pi_j^-&-\frac{1}{2}\pi_j^0
\end{array}\right)
\ \mbox{,} \ 
W_j = \left(\begin{array}{cc}
\frac{1}{2}W_j^0&\frac{1}{\sqrt{2}}W_j^+\\\frac{1}{\sqrt{2}}W_j^-&-\frac{1}{2}W_j^0
\end{array}\right)
\ \mbox{and} \
W_2 = \left(\begin{array}{cc}
\frac{1}{2}W_2^0&0\\0&-\frac{1}{2}W_2^0
\end{array}\right),
\end{equation}
where $j$ is $0$ or $1$.

The mass matrices of the gauge bosons can be obtained by going to unitary gauge ($\Sigma_j\rightarrow1$) and are,
\begin{equation}
M_\pm^2 = \frac{M_G^2}{2}\left(\begin{array}{cc}
x^2&-x\\
-x&2\\
\end{array}\right)
\quad \mbox{and} \quad
M_n^2 = \frac{M_G^2}{2}\left(\begin{array}{ccc}
x^2&-x&0\\
-x&2&-xt\\
0&-xt&x^2t^2
\end{array}\right),
\end{equation}
for the charged and neutral gauge bosons respectively where
\begin{equation}\label{M_G def}
M_G^2 = \frac{\tilde{g}^2f^2}{2},
\end{equation}
and the photon is massless.
%
After diagonalizing the gauge boson mass matrices, we find that the other masses 
are given by
\beq\bsp
M_{W}&\,=\frac{M_G}{2}\sqrt{2+x^2-\sqrt{4+x^4}}, \\
M_{W'}&\,=\frac{M_G}{2}\sqrt{2+x^2+\sqrt{4+x^4}},
\esp\eeq
for the charged gauge bosons 
and
\beq\bsp
M_{Z}&\,=\frac{M_G}{2}\sqrt{2+x^2(1+t^2)-A},\\
M_{Z'}&\,=\frac{M_G}{2}\sqrt{2+x^2(1+t^2)+A},
\esp\eeq
where
\begin{equation}
A = \sqrt{4+x^4(1-t^2)^2},
\end{equation}
for the neutral gauge bosons
.  We note that $x$ can be obtained from the ratio of the charged gauge boson masses
\begin{equation}
R_M^2 = \left(\frac{M_W}{M_{W'}}\right)^2 = \frac{2+x^2-\sqrt{4+x^4}}{2+x^2+\sqrt{4+x^4}}.
\end{equation}
%
%
%
The couplings can be determined in terms of the electric charge $e$, $x$ and $t$.
\beq\bsp
\frac{1}{e^2} &\,= \frac{1}{g^2} + \frac{1}{\tilde{g}^2} + \frac{1}{g'^2},\\
g^2 &\,= e^2\left(1+x^2+\frac{1}{t^2}\right),\\
\tilde{g}^2 &\,= e^2\left(1+\frac{1}{x^2}+\frac{1}{x^2t^2}\right),\\
(g')^2 &\,= e^2\left(1+t^2+x^2t^2\right).
\esp\eeq


\subsubsection*{Gauge Fixing Sector}
As we mentioned previously, the horizontal lines in Fig.~\ref{3-moose} represent non-linear sigma fields.  Although tree level calculations can be done in unitary gauge, there are times when a different gauge is useful.  Many calculations with gauge bosons in the external states can be computed more simply using the equivalence theorem and replacing the massive gauge bosons with the Goldstone bosons that they eat.  Another case where a gauge different from unitary gauge is advantageous is in \calchep, where the time of computation of processes is dramatically decreased when using Feynman gauge.  For this reason, we have implemented this model in both Feynman and unitary gauges.  
 
We must first determine the Goldstone bosons that are eaten by the gauge bosons.  We do this using the Lagrangian of Eq.~(\ref{L_D Sigma}).  Expanding the non-linear sigma field, we obtain the mixing terms between the gauge bosons and the Goldstone bosons.  After inserting the eigenwave functions of these fields, we obtain
\beq\bsp
\mathcal{L}_{{\rm MHM, } \pi W} &\,= \frac{1}{2}\tilde{g}f\Big(
v_\pi^0 (xv_W^0-v_W^1) + v_\pi^1 (v_W^1-\delta xtv_W^2)
\Big) \left\{ \partial_\mu \pi , W^\mu \right\}\\
&\,+\frac{1}{2}\tilde{g}f\Big(
v_{\pi'}^0 (xv_{W'}^0-v_{W'}) + v_{\pi'}^1 (v_{W'}^1-\delta xtv_{W'}^2)
\Big) \left\{ \partial_\mu \pi' , W'^\mu \right\}
\esp\eeq
where $\delta$ is $1$ if the gauge boson is neutral but $0$ otherwise.  

The gauge fixing function is constructed to fix the gauge and cancel the mixing of the Goldstone bosons and gauge bosons.  For each site, the gauge-fixing term is
\beq\bsp
G_0 &\,= \partial\cdot W_0 - \frac{\xi}{2}gf (\pi_0 ),\\
G_1 &\,= \partial\cdot W_1 - \frac{\xi}{2}\tilde{g}f (\pi_1 - \pi_0),\\
G_2 &\,= \partial\cdot W_2 - \frac{\xi}{2}g'f (-\pi_1^{ns}),
\esp\eeq
where by $\pi_1^{ns}$ we mean just the neutral sector of $\pi_1$, namely
\begin{equation}
\pi_1^{ns} = \frac{1}{2}\pi_1^0\left(\begin{array}{cc}1&0\\0&-1\end{array}\right).
\end{equation}
With this definition, the gauge fixing Lagrangian is
\begin{equation}
\mathcal{L}_{\rm MHM, GF} = -\frac{1}{\xi} \mbox{Tr}\Big( G_0^2 + G_1^2 + G_2^2 \Big).
\end{equation}


\subsubsection*{Ghost Sector}
The ghost Lagrangian terms are obtained by multiplying the BRST transformation of the gauge fixing term on the left with the antighost.  To do this, we must find the BRST transformations of the gauge fixing terms.  We begin by writing the infinitesimal BRST transformation of the fields in the gauge fixing term.
\begin{equation}
\delta_{_{BRST}}W_{\mu j} = - \Big( \partial_\mu c_j + ig_j\left[W_{\mu j} , c_j \right] \Big),
\end{equation}
for the gauge bosons where $c_j$ is the ghost for site $j$ in matrix notation
\begin{equation}
c_j = \left(\begin{array}{cc}
\frac{1}{2}c_j^0 & \frac{1}{\sqrt{2}}c_j^+ \\
\frac{1}{\sqrt{2}}c_j^- & -\frac{1}{2}c_j^0
\end{array}\right)
\quad \mbox{and} \quad 
c_2 = \left(\begin{array}{cc}
\frac{1}{2}c_2^0 & 0\\
0 & -\frac{1}{2}c_2^0
\end{array}\right),
\end{equation}
where $j$ is $0$ or $1$.
The BRST transformation to quadratic order in  the Goldstone bosons is
\begin{equation}
\delta_{_{BRST}}\pi_j = + \frac{1}{2}f\big(g_jc_j - g_{j+1}c_{j+1}\big)
+\frac{i}{2}\big[g_jc_j + g_{j+1}c_{j+1}\ ,\ \pi_j \Big]
-\frac{1}{6f}\Big[\pi_j\ ,\ \big[\pi_j\ ,\ g_jc_j - g_{j+1}c_{j+1} \big] \Big],
\end{equation}
so that
\beq\bsp
\delta_{_{BRST}}G_0 &\,= \partial\cdot \delta_{_{BRST}}W_0 - \frac{\xi}{2}gf (\delta_{_{BRST}}\pi_0\quad ),\\
\delta_{_{BRST}}G_1 &\,= \partial\cdot \delta_{_{BRST}}W_1 - \frac{\xi}{2}\tilde{g}f (\delta_{_{BRST}}\pi_1 - \delta_{_{BRST}}\pi_0),\\
\delta_{_{BRST}}G_2 &\,= \partial\cdot \delta_{_{BRST}}W_2 - \frac{\xi}{2}g'f (\quad -\delta_{_{BRST}}\pi_1^{ns}).
\esp\eeq
The ghost Lagrangian is 
\begin{equation}
\mathcal{L}_{\rm MHM, Ghost} = -\mbox{Tr}\Big( \bar{c}_0\delta_{_{BRST}}G_0 + \bar{c}_1\delta_{_{BRST}}G_1 + \bar{c}_2\delta_{_{BRST}}G_2  \Big) + \hc
\end{equation}


\subsubsection*{\label{Fermion Sector}Fermion Sector}
The vertical lines in Fig.~\ref{3-moose} represent the fermionic fields in the theory.  The vertical lines on the bottom of the circles represent the left-handed chiral fermions while the vertical lines attached to the tops of the circles are the right-handed chiral fermions.  Each fermion is a fundamental representation of the gauge group to which it is attached and a singlet under all the other gauge groups except $U(1)_2$.  The charges under $U(1)_2$ are as follows:  If the fermion is attached to an $SU(2)$ then its charge is $1/6$ for quarks and $-1/2$ for leptons.  If the fermion is attached to $U(1)_2$ its charge is the same as its electromagnetic charge: $0$ for neutrinos, $-1$ for charged leptons, $2/3$ for up type quarks and $-1/3$ for down type quarks.  The usual gauge invariant kinetic terms are used,
\begin{equation}
\mathcal{L}_{{\rm MHM, }\psi} = 
i\bar{\psi}_{L0}D\hspace{-0.1in}\slash\hspace{0.04in}\psi_{L0}
+i\bar{\psi}_{L1}D\hspace{-0.1in}\slash\hspace{0.04in}\psi_{L1}
+i\bar{\psi}_{R1}D\hspace{-0.1in}\slash\hspace{0.04in}\psi_{R1}
+i\bar{\psi}_{R2}D\hspace{-0.1in}\slash\hspace{0.04in}\psi_{R2}.
\end{equation}

The fermions attached to the internal site ($SU(2)_1$) are vectorially coupled and Dirac masses are thus allowed.  We have taken these masses to be $M_F$.  The symmetries also allow various linkings of fermions via the non-linear sigma fields.  We have assumed a very simple form, inspired by an extra dimension and represented by the diagonal lines in Fig.~\ref{3-moose}.  The left chiral field at site $j$ is linked to the right chiral field at site $j+1$ through the non-linear sigma field at link $j$.  The mass parameter for these diagonal links is taken to be $\epsilon_LM_F$ and $\epsilon_RM_F$ for the left and right links respectively.   All together, the masses of the fermions and the leading order interactions of the fermions and non-linear sigma fields are given by
\begin{equation}
\mathcal{L}_{{\rm MHM, } \psi\Sigma} = - M_F\Big[
\epsilon_L\bar{\psi}_{L0}\Sigma_0\psi_{R1} + \bar{\psi}_{L1}\psi_{R1}
+ \bar{\psi}_{L1}\epsilon_R\Sigma_1\psi_{R2}\Big],
\end{equation}
where $\epsilon_L$ is the same for all fermions but $\epsilon_R$ is a diagonal matrix which distinguishes flavors.  For example, for the top and bottom quarks we have
\begin{equation}
\epsilon_R = \left(\begin{array}{cc}\epsilon_{Rt}&0\\0&\epsilon_{Rb}\end{array}\right).
\end{equation}

The mass matrix can be obtained by going to unitary gauge and diagonalized by a biunitary transformation.  By doing this, we find the following masses,
\beq\bsp
M_{f0}&\,=\frac{M_F}{\sqrt{2}}\sqrt{1+\epsilon_L^2+\epsilon_R^2
  -C},
  \\
M_{f1}&\,=\frac{M_F}{\sqrt{2}}\sqrt{1+\epsilon_L^2+\epsilon_R^2
  +C},
\esp\eeq
where
\begin{equation}
C = \sqrt{(1+\epsilon_L^2+\epsilon_R^2)^2-4\epsilon_L^2\epsilon_R^2}.
\end{equation}

\subsubsection{\label{3-Site:FeynRules}\feynrules\ implementation}
The \feynrules\ implementation was initially based on a \lanhep\ implementation \cite{He:2007ge}.  It was translated to \feynrules\ syntax and slightly modified to fit the requirements of the \feynrules\ package and interfaces.  All symmetries, fields and parameters were implemented according to the definitions of the last subsection.  The independent variables that the user can adjust are the electromagnetic and strong couplings, the masses of the $Z$, $W$, $W'$ and SM fermions (where not set explicitly to zero) and the scale of the heavy fermions $M_F$.  The scale of the Z pole mass and Fermi constant are also implemented as required by some Monte Carlo codes, but they are not used directly in this model.  A change in these last two parameters will not affect the value of the other parameters.

Two gauges were implemented in this model file.  A variable {\tt FeynmanGauge}, similar to the SM case, was created to switch between the two.  When the switch is set to {\tt True}, Feynman gauge is chosen and the Lagrangians contain the Goldstone bosons eaten by the gauge bosons and the ghosts.  If, on the other hand, it is set to {\tt False}, all Goldstone and ghost terms are set to zero.



%% file: ued1.tex
\newcommand{\TeV}{\mathrm{\;TeV}}
\newcommand{\frmued}{FR-MUED}
\newcommand{\mued}{MUED}
\newcommand{\led}{{\sc Led}}

\subsection{Extra Dimensional models}
One popular approach to solve the hierarchy problem of the Standard Model is to extend space-time to higher dimensions \cite{ArkaniHamed:1998rs,Appelquist:2000nn,Randall:1999ee}. In this framework, the usual four-dimensional space is contained in a four-dimensional \textit{brane} embedded in a larger structure with $N$ additional dimensions, the \textit{bulk}. Moreover, gravitational and gauge interactions unify close to the only fundamental scale of the theory, the weak scale. 
In theories with Large Extra Dimensions (LED)~\cite{ArkaniHamed:1998rs, Antoniadis:1998ig, Arkani-Hamed:1998nn}, the gravitational interactions are the only ones propagating into the bulk, which dilutes their coupling strength and make it appear weaker inside the four-dimensional branes. As a consequence, the graviton field is accompanied by a tower of massive Kaluza-Klein states. In scenarios with Universal Extra Dimensions~\cite{Appelquist:2000nn}, each field of the Standard Model possesses a tower of excitations with the same quantum numbers, but different masses. Even if none of these states has been observed so far, TeV-range excitations could be detected at the present Tevatron or the future Large Hadron Collider. In a minimal Universal Extra Dimension scenario (\mued)~\cite{cheng-2002-66-2}, one has one single flat additional dimension, $y$, which is spatial and compactified on a $S_{1}/Z_{2}$ orbifold of radius $R$. Momentum conservation in the extended space-time generates a conserved quantum number, the Kaluza-Klein parity, which implies that different Kaluza-Klein modes cannot mix and that the lightest excitation could be a candidate for dark matter~\cite{Servant:2002aq, Burnell:2005hm, Kong:2005hn}.

\subsubsection{Model description}
\noindent{\textbf{Large Extra Dimension Model}}
\noindent
In a LED theory with one compact and space-like additional dimension, all the Standard Model fields are confined into a four-dimensional brane and there are only Kaluza-Klein excited states for the graviton. The Kaluza-Klein states have higher masses than the standard graviton but both behave the same, \ie,  they couple gravitationally to the Standard Model fields.  We start by specifying the generic effective Lagrangian of an unbroken gauge theory~\cite{Han:1998sg},
\begin{equation} \label{eq:lagLEDSM}
  \mathcal{L}_{\rm LED}= \mathcal{L}_{\rm LED,\ Gauge} + \mathcal{L}_{\rm LED,\ Fermions} + \mathcal{L}_{\rm LED,  \ Scalars}.
\end{equation}
The Lagrangian of each sector can be expressed as
\begin{equation} \label{eq:LLykken}
 \mathcal{L}_{{\rm LED, }\ i}=-\frac{\kappa}{2}\sum_{n}\left(h^{\mu\nu\,(n)}T^{i}_{\mu\nu}+ \sqrt{{2 \over 3 (n+2)}} \phi^{(n)}\, T^{i\,\mu}_{\quad\mu}\right) .
\end{equation}
$\kappa=\sqrt{16\pi G_{N}}$ with $G_{N}$ being the four-dimensional Newton constant,  $h^{\mu\nu\,(n)}$ the $n$-th graviton Kaluza-Klein mode in four dimensions and $\phi^{(n)}$ its scalar component in the fifth dimension. The graviton couples to the energy momentum tensor,
\beq
T_{\mu\nu}=\left.\left(-\eta_{\mu\nu}\mathcal{L}+2\frac{\delta\mathcal{L}}{\delta g^{\mu\nu}}\right)\right|_{g^{\mu\nu}=\eta^{\mu\nu}}.
\eeq
For a generic unbroken gauge theory, the various energy-momentum tensors  $T^{i}_{\mu\nu}$ for an unbroken gauge theory read\footnote{We work in unitary gauge.},
\begin{equation} \label{eq:TLykken}
\begin{split}
T_{\mu\nu}^{\rm Scalars}&\,=-\eta_{\mu\nu}\left[D^{\rho}\Phi^{\dagger}D_{\rho}\Phi-m_{\Phi}^{2}\Phi^{\dagger}\Phi\right]+\left[D_{\mu}\Phi^{\dagger}D_{\nu}\Phi+\left(\mu\leftrightarrow\nu\right)\right]  ~,~\\
T_{\mu\nu}^{\rm Fermions}&\,=-\eta_{\mu\nu}\left[\bar{\Psi}i\gamma^{\rho}D_{\rho}\Psi-m_{\Psi}\bar{\Psi}\Psi-\frac{1}{2}\partial^{\rho}\left(\bar{\Psi}i\gamma_{\rho}\Psi\right)\right]\\
&\,+\left[\frac{1}{2}\bar{\Psi}i\gamma_{\mu}D_{\nu}\Psi-\frac{1}{4}\partial_{\mu}\left(\bar{\Psi}i\gamma_{\nu}\Psi\right)+\left(\mu\leftrightarrow\nu\right)\right],~\\
T_{\mu\nu}^{\rm Gauge}&\,=-\eta_{\mu\nu}\left[-\frac{1}{4}F^{\rho\sigma}F_{\rho\sigma}+\frac{m_{A}^{2}}{2}A^{\rho}A_{\rho}\right]+\left[-F_{\mu}^{\;\rho}F_{\nu\rho}+m_{A}^{2}A_{\mu}A_{\nu}\right]~,~
\end{split} 
\end{equation}
where $\Phi$ is a (complex) scalar, $\Psi$ a fermion, and $A_{\mu}$ a vector field. $D_\mu$ and $F_{\mu\nu}$ denote the usual covariant derivatives and field strength tensors. From this general model, we can derive a realistic Large Extra Dimensional theory containing all the Standard Model fields. We choose to work in the unitary gauge also for the gravitational part, which eliminates the non-physical degrees of freedom absorbed by the massive fields, and we re-write the Lagrangian as, 
\begin{equation}
\label{eq:lagi}
 \mathcal{L}_{{\rm LED, }\ j}=-\frac{\kappa}{2}\sum_{n}\left(\mathcal{G}^{\mu\nu\,(n)}T^{j}_{\mu\nu}\right),
\end{equation}
where $\mathcal{G}^{\mu\nu\,(n)}$ is the $n$-th Kaluza-Klein mode of the physical graviton and $j$ now denotes explicitly the Standard Model sectors. The energy-momentum tensor of the Standard Model can be written as a sum of four parts,
\beq\bsp
T_{\mu\nu}&\,= T_{\mu\nu}^{\rm Fermions} + T_{\mu\nu}^{\rm Higgs}+ T_{\mu\nu}^{\rm Gauge} + T_{\mu\nu}^{\rm Yukawa}~.
\esp\eeq
The energy-momentum tensor of the fermionic sector reads,
 \begin{equation}
 \begin{split}
 &T_{\mu\nu}^{\rm Fermions}\\
\ &=\,-\eta_{\mu\nu}\left[\left(\bar{Q_{i}}\,i\,\gamma^{\rho}D_{\rho}\,Q_{i}+\bar{L_{i}}\,i\,\gamma^{\rho}D_{\rho}\,L_{i}+\bar{u}_{Ri}\,i\,\gamma^{\rho}D_{\rho}\,u_{Ri}+\bar{d}_{Ri}\,i\,\gamma^{\rho}D_{\rho}\,d_{Ri}+\bar{l}_{Ri}\,i\,\gamma^{\rho}D_{\rho}\,l_{Ri}\right)\right.\\
 \ &\,-\left.\frac{1}{2}\partial_{\rho}\left(\bar{Q_{i}}\,i\,\gamma^{\rho}\,Q_{i}+\bar{L_{i}}\,i\,\gamma^{\rho}\,L_{i}+\bar{u}_{Ri}\,i\,\gamma^{\rho}\,u_{Ri}+\bar{d}_{Ri}\,i\,\gamma^{\rho}\,d_{Ri}+\bar{l}_{Ri}\,i\,\gamma^{\rho}\,l_{Ri}\right)\right]\\
 \ &\,+\left[\frac{1}{2}\left(\bar{Q_{i}}\,i\,\gamma_{\mu}D_{\nu}\,Q_{i}+\bar{L_{i}}\,i\,\gamma_{\mu}D_{\nu}\,L_{i}+\bar{u}_{Ri}\,i\,\gamma_{\mu}D_{\nu}\,u_{Ri}+\bar{d}_{Ri}\,i\,\gamma_{\mu}D_{\nu}\,d_{Ri}+\bar{l}_{Ri}\,i\,\gamma_{\mu}D_{\nu}\,l_{Ri}\right)\right.\\
\ &\,\left.-\frac{1}{4}\partial_{\mu}\left(\bar{Q_{i}}\,i\,\gamma_{\nu}\,Q_{i}+\bar{L_{i}}\,i\,\gamma_{\nu}\,L_{i}+\bar{u}_{Ri}\,i\,\gamma_{\nu}\,u_{Ri}+\bar{d}_{Ri}\,i\,\gamma_{\nu}\,d_{Ri}+\bar{l}_{Ri}\,i\,\gamma_{\nu}\,l_{Ri}\right)+\left(\mu\rightarrow\nu\right)\right]~,
\end{split} 
 \end{equation}
where $D_{\mu}$ denotes the $SU(3)_{C}\times SU(2)_{L}\times U(1)_{Y}$ covariant derivative. Similarly, the energy-momentum tensor for the gauge sector reads,  
 \begin{equation}
 T_{\mu\nu}^{\rm Gauge}=-\eta_{\mu\nu}\left[-\frac{1}{4}B^{\rho\sigma}B_{\rho\sigma}-\frac{1}{4}W_{k}^{\rho\sigma}W_{\rho\sigma}^{k}-\frac{1}{4}G_{a}^{\rho\sigma}G_{\rho\sigma}^{a}-\right]-B_{\mu}^{\;\rho}B_{\nu\rho}-W_{\mu}^{k\,\rho}W_{\nu\rho}^{k}-G_{\mu}^{a\,\rho}G_{\nu\rho}^{a}.
 \end{equation}
Finally, the energy-momentum tensors for the sectors involving a Higgs boson read,
\beq\bsp
T_{\mu\nu}^{\rm Higgs}&\,=-\eta_{\mu\nu}\left[D_{\rho}\Phi^{\dagger}D^{\rho}\Phi+\mu^{2}\Phi^{\dagger}\Phi-\lambda\left(\Phi^{\dagger}\Phi\right)^{2}\right]+\left[D_{\mu}\Phi^{\dagger}D_{\nu}\Phi+\left(\mu\rightarrow\nu\right)\right] ,\\
T_{\mu\nu}^{\rm Yukawa}&\,=-\eta_{\mu\nu}\left[ - \bar u_{iR}\,y^u_{ij}\,Q_j\,\tilde\Phi -  \bar d_{iR}\,y^d_{ij}\,Q_j\,\Phi - \bar l_{iR}\,y^l_{ij}\,L_j\,\Phi +{\rm h.c.} \right] ,
\esp\eeq
with $\tilde\Phi = i\sigma^2\,\Phi^\ast$. \\
 
\noindent{\textbf{Minimal Universal Extra Dimension Model}}

We consider a theory in five dimensions, the fifth one being spatial and compactified on a $S_{1}/Z_{2}$ orbifold of radius $R$, \ie, the points $y$ and $-y$ are identified, where $y$ is the extra coordinate. This symmetry is essential to define chiral fermions. Unlike LED models, in UED models all fields have access to the extra dimension and depend on the fifth coordinate $y$. We split the most general Lagrangian in four pieces~\cite{Appelquist:2000nn, Matchev:2007},
\begin{equation}
 \mathcal{L}_{\rm MUED} =  \mathcal{L}_{\rm MUED,\ Gauge} +  \mathcal{L}_{\rm MUED,\ Fermions} +  \mathcal{L}_{\rm MUED,\ Higgs} +  \mathcal{L}_{\rm MUED,\ Yukawa}.
\end{equation}
The gauge sector is described by the field strength tensor terms
\begin{equation} 
\mathcal{L}_{\rm MUED, Gauge} = -\frac{1}{4}B_{MN}B^{MN} - \frac{1}{4}W_{MN}^{k}W_k^{MN} - \frac{1}{4}G_{MN}^{a}G_a^{MN}.
\end{equation}  
We define the subscript $M={(\mu,5)}$, where $\mu$ is the usual four-dimensional Lorentz index and $5$ the fifth dimension index.
The fermionic sector is decomposed into its leptonic and quark part, 
\begin{equation}
   \mathcal{L}_{\rm MUED, Fermions} =  \mathcal{L}_{\rm MUED,\ Leptons} +  \mathcal{L}_{\rm MUED,\ Quarks},~
\end{equation}
with
\begin{equation}
 \begin{split}  
     \mathcal{L}_{\rm MUED,\ Leptons} =& i\bar{L}(x^{\mu},y)\Gamma^{M}D_{M}L(x^{\mu},y) + i\bar{E}(x^{\mu},y)\Gamma^{M}D_{M}E(x^{\mu},y),\\
     \mathcal{L}_{\rm MUED,\ Quarks}=& i\bar{Q}(x^{\mu},y)\Gamma^{M}D_{M}Q(x^{\mu},y) + i\bar{U}(x^{\mu},y)\Gamma^{M}D_{M}U(x^{\mu},y) +\\ 
     & i\bar{D}(x^{\mu},y)\Gamma^{M}D_{M}D(x^{\mu},y).
  \end{split} 
\end{equation}
$Q\left(x^{\mu},y\right)$ and $L\left(x^{\mu},y\right)$ denote $SU(2)_{L}$ fermion doublets and $U\left(x^{\mu},y\right)$, $D\left(x^{\mu},y\right)$,  $E\left(x^{\mu},y\right)$ are the up-type quark, down-type quark and charged lepton singlet fields. The gamma matrices in five dimensions are defined as 
\begin{equation}
 \Gamma^{M}=\left\{ \gamma^{\mu},\,i\,\gamma^{5}\right\},~
\end{equation}
while the five-dimensional covariant derivative $D_{M}$ is
  \begin{equation}
  D_{M}=\partial_{M}-i\,Y g_{1}^{(5)} B_{M} - {1 \over 2} i\, g_{w}^{(5)} \sigma^k W^k_{M} - i\, g_{s}^{(5)} T^{a}G_{M}^{a}.
  \end{equation}
The hypercharge $Y$, the Pauli matrices $\sigma^k$ and the color matrices $T^a$ are the generators of the gauge groups, while the five-dimensional coupling constants are related to the four-dimensional ones through 
\begin{equation}
   g_{i}^{(5)} = \sqrt{\pi R} g_{i}.~
\end{equation}
Finally, the Higgs Lagrangian is given by
\begin{equation}
\begin{split}
  \mathcal{L}_{\rm MUED, Higgs} =& \left[D_{M}H(x^{\mu},y)\right]^{\dagger} \left[D^{M}H(x^{\mu},y)\right] + \mu^{2}H^{\dagger}(x^{\mu},y)\, H(x^{\mu},y)\\
   & -\lambda \left[H^{\dagger}(x^{\mu},y)\, H(x^{\mu},y)\right]^{2}, 
\end{split}
\end{equation}
and the Yukawa sector describing the interactions between the fermions and the Higgs field by
\begin{equation}
\begin{split}
  \mathcal{L}_{\rm MUED, Yukawa} =& -\lambda_{u}\bar{Q}(x^{\mu},y)U(x^{\mu},y)\,\tilde{H}(x^{\mu},y) - \lambda_{d}\bar{Q}(x^{\mu},y)D(x^{\mu},y)\, H(x^{\mu},y) \\ 
  &- \lambda_{e}\bar{Q}(x^{\mu},y)E(x^{\mu},y)\, H(x^{\mu},y),
 \end{split} 
\end{equation}
where $\tilde{H}=i\,\sigma^{2}\,H^{*}(x^{\mu},y)$, and $\lambda_{i}$ are the usual Yukawa matrices.

\subsubsection{\feynrules\ implementation}
\noindent{\textbf{Large Extra Dimension Model}}

The \feynrules\ implementation for the LED model described is based on the Lagrangian defined in Eq.~(\ref{eq:lagLEDSM}). The theory contains, besides the graviton, the full set of SM fields, together with their respective coupling to the graviton field via the stress tensors. For the graviton we restricted the implementation to the lowest mode, \ie, only the massless graviton is taken into account. An extension of this model to include Kaluza-Klein excitations of the graviton, as well as to any BSM extension of the SM, \eg, the HAH model described in Section~\ref{sec:hidden}, is straightforward.\\

\noindent{\textbf{Minimal Universal Extra Dimension Model}}

To implement the \mued\ model in \feynrules, we start with the most general five-dimensional Lagrangian described above. Then, \feynrules\ derives the effective four-dimensional Lagrangian and the corresponding Feynman rules automatically. This is achieved by expanding the five-dimensional fields and imposing the dimensional reduction by integrating out the extra coordinate $y$ \cite{Perez-Lorenzana:2005iv}, 
\begin{equation}
\begin{split}
  A_{\mu}\left(x^{\mu},y\right)&=\frac{1}{\sqrt{\pi R}}\left\{ A_{\mu}^{(0)}(x)+\sqrt{2}\sum_{n=1}^{\infty}A_{\mu}^{(n)}(x)\,\cos\left(\frac{ny}{R}\right)\right\},~ \\
  A_{5}\left(x^{\mu},y\right)&=\sqrt{\frac{2}{\pi R}}\sum_{n=1}^{\infty}A_{5}^{(n)}(x)\,\sin\left(\frac{ny}{R}\right),~\\
  H\left(x^{\mu},y\right)&=\frac{1}{\sqrt{\pi R}}\left\{ H^{(0)}(x)+\sqrt{2}\sum_{n=1}^{\infty}H^{(n)}(x)\,\cos\left(\frac{ny}{R}\right)\right\},~ \\ 
  \Psi\left(x^{\mu},y\right)&=\frac{1}{\sqrt{\pi R}}\left\{ \Psi_{L}^{(0)}(x)+\sqrt{2}\sum_{n=1}^{\infty}\left[P_{L}\Psi_{L}^{(n)}(x)\,\cos\left(\frac{ny}{R}\right)+P_{R } \Psi_{R}^{(n)}(x)\,\sin\left(\frac{ny}{R}\right)\right]\right\},~  \\
   \psi\left(x^{\mu},y\right)&=\frac{1}{\sqrt{\pi R}}\left\{ \psi_{R}^{(0)}(x)+\sqrt{2}\sum_{n=1}^{\infty}\left[P_{R}\psi_{R}^{(n)}(x)\,\cos\left(\frac{ny}{R}\right)+P_{L}\psi_{L}^{(n)}(x)\,\sin\left(\frac{ny}{R}\right)\right]\right\} ,~
\end{split}
\end{equation}
where $\left(A_{\mu}(x^{\mu},y),\,A_{5}(x^{\mu},y)\right)$, $H(x^{\mu},y)$, $\Psi(x^{\mu},y)$ and  $\psi(x^{\mu},y)$ denote a five-dimensional gauge field, a Higgs field, a fermionic doublet and a fermionic singlet field, respectively. To integrate out the extra dimensional coordinate, we follow the integration procedure described in Refs.\ \cite{Appelquist:2000nn, Matchev:2007, Perez-Lorenzana:2005iv}, using orthogonality relations. Let us note that if one would like to add other pieces to the Lagrangian, it is sufficient to deal with five-dimensional expressions, together with the appropriate definitions of the field expansions.

In our implementation we are considering Kaluza-Klein excitations only up to the first mode. The inclusion of the next Kaluza-Klein states is of course straightforward, each considered Kaluza-Klein mode has to be defined in the model file. For a given field, we identify the zeroth mode as the Standard Model particle and define as new particles the Kaluza-Klein excitations. The zeroth mode particle content is thus identical to the SM one, see Table \ref{tab:SMparts}, while the particle content of the first modes is summarized in Table \ref{KKpart}.
\begin{table}
 \begin{center}
 \begin{tabular}{ | l l | c | l | }
\hline
 \multicolumn{2}{|c|}{Particle description}  	 &			 Spin & 		Representations\\
\hline
\hline
  Lepton doublet & 	$L^{(1)}_i=\binom{\nu_{iL}^{(1)}}{L_i^{(1)}}$ & 	$1/2$ & 		$\big(\mathbf{1}, \mathbf{2},-1/2\big)$\\
  Lepton singlet & 	$\ell^{(1)}_i$ &					$1/2$& 			$\big(\mathbf{1}, \mathbf{1},-1\big)$\\
  Quark doublet & 	$Q^{(1)}_i=\binom{U_i^{(1)}}{D_i^{(1)}}$ & 		$1/2$ &			$\big(\mathbf{3}, \mathbf{2},1/6\big)$\\
  Up-quark singlet & 	$u^{(1)}_i$ &					$1/2$ &			$\big(\mathbf{3}, \mathbf{1},2/3\big)$\\
  Down-quark singlet & 	$d^{(1)}_i$ &					$1/2$ &			$\big(\mathbf{3}, \mathbf{1},-1/3\big)$\\
\hline
 		&	$B^{(1)}_{\mu}$ &				$1$ &			$\big(\mathbf{1}, \mathbf{1},0\big)$\\
  Gauge bosons 	&	$W_{\mu}^{i\,(1)}$ &				$1$ &			$\big(\mathbf{1}, \mathbf{3},0\big)$\\
		&	$G^{(1)\,a}_{\mu}$ &				$1$ &			$\big(\mathbf{8}, \mathbf{1},0\big)$\\
\hline
  Higgs 	&	$H^{(1)}$ &					$0$ &			$\big(\mathbf{1}, \mathbf{2},1/2\big)$\\
\hline
  \end{tabular}
 \end{center}
\caption{A summary of the first mode Kaluza-Klein particles.}
 \label{KKpart}
\end{table}

As for the SM particles, the new gauge-eigenstates mix to physical states 
\begin{equation}
\begin{split}
  W^{(1)\,\pm}_{\mu}&=\frac{W_{\mu}^{1\,(1)}\mp \,i W_{\mu}^{2\,(1)}} {\sqrt{2}},~\\
  Z^{(1)}_{\mu}&\approx W^{3\,(1)}_{\mu},~\\
  A^{(1)}&\approx B^{(1)}_{\mu}.~
\end{split}
\end{equation}
The first mode analog of the electroweak mixing angle is assumed to be very small. Therefore the first modes of the $Z$ boson and the photon are simply $W^{3\,(1)}_{\mu}$ and $B^{(1)}_{\mu}$.  The tree-level mass $M^{(1)}$ of the first gauge boson excitations are,
\begin{equation}
 M^{(1)} = \sqrt{\frac{1}{R^{2}} + \left(m^{(0)}\right)^{2}},
 \end{equation}  where $m^{(0)}$ is the zeroth mode mass. In the fermion sector, there might be a mixing between $SU(2)_{L}$ doublets and singlets. However, this mixing is expected to be highly suppressed. Therefore we ignore it in our implementation.  As in the SM, the zeroth mode fermions get their mass from the Yukawa couplings after the Higgs get its vev. For the Kaluza-Klein first modes, the tree-level mass terms follow the structure
\[ \left(\begin{array}{cc}
           \bar{F}^{(n)} & \bar{f}^{(n)}
           \end{array}\right)
                 \left(\begin{array}{cc}
                 \frac{n}{R} & m_{f}\\
                  m_{f} & \frac{n}{R}
                  \end{array}\right)
                      \left(\begin{array}{c}
                      F^{(n)}\\
                      f^{(n)}
                      \end{array}\right)\]
where $F^{(n)}$ and $f^{(n)}$ denote the doublet and the singlet fields, respectively. The diagonal contributions come from the kinetic terms in the fifth dimension and the off-diagonal ones are induced by the Higgs vev. After the diagonalization of the mass matrices, the masses of the physical eigenstates are obtained, 
\begin{equation}
M^{(n)}_f = \sqrt{\left(\frac{n}{R}\right)^{2}+\left(m^{(0)}_{f}\right)^{2}}.
\end{equation}

\noindent\textbf{Possible extensions}\\
As we have mentioned before, we are only considering expansions of the fields up to the first Kaluza-Klein excited states. However one could easily incorporate the next modes by expanding the fields up to the $n^{th}$ component. It is straightforward to incorporate the second mode Kaluza-Klein excitations, by following the example of the first modes.

%% file: effectivel.tex
\subsection{An access to the low energy world}
\label{sec:effectiveL}

Effective theories allow us to have predictions about the masses and the interactions at low energy without having to know everything about the fundamental theory. We only need to know the (exact or approximate) symmetries of the fundamental theory. One of the most successful examples is Chiral Perturbation Theory ($\chi$PT), the effective theory of the strong interactions. Most of the processes involving pseudoscalar mesons happen at scales at which pertubative QCD cannot be trusted anymore. Despite our resulting incapacity to compute their properties directly from the fundamental Lagrangian, the shape of the effective Lagrangian can be inferred using the global approximate chiral symmetry of QCD. The other famous example, especially in this pre-LHC era, are all the effective models developed to solve the hierarchy problem where the Higgs boson is a pseudo-Goldstone boson of a new strong sector. These models were built on the fundaments laid down by QCD. However, in this case, much less is known, neither the fundamental nor all the low energy degrees of freedom. This is probably the origin of the number of models available going from Technicolor to Extra Dimensions. However the low energy effects of many of them, called Strongly Interacting Light Higgs (SILH) Models, can be described by the same effective Lagrangian up to some coefficients~\cite{Giudice:2007fh}.

\subsubsection{Model description}

\noindent{\textbf{$\chi$PT at the lowest order}}\\
The leading effective Lagrangian invariant under the $SU(n_F)_L\times SU(n_F)_R\times U(1)_V$, where $n_F$ is the number of massless quark flavors, is \cite{Witten:1980sp}
\beq\bsp
{\cal L}_{\chi\text{PT}}&\,={\cal L}^{\left(p^0,1/N_c\right)}+{\cal L}^{\left(p^2,0\right)},\label{LchiPT}\\
{\cal L}^{\left(p^0,1/N_c\right)} &\,= \frac{f^2}{8}\frac{m_0^2}{4N_c}\left<\ln U-\ln U^\dagger\right>^2,\\
{\cal L}^{\left(p^2,0\right)} &\,= \frac{f^2}{8}\left[\left<\partial_\mu U\partial^\mu U^\dagger\right>+r\left<mU^\dagger+Um^\dagger\right>\right],
\esp\eeq
where $U$ transforms as $U\rightarrow g_LUg_R^\dagger$. The first part of the Lagrangian is due to the axial $U(1)$ breaking required by the $\eta'$ mass \cite{tHooft:1976up} and is at the source of the strong $CP$ problem. The second term is the usual non-linear sigma Lagrangian and the last one takes into account the quark masses. Only the first three  flavors are sufficiently light compared to the confinement scale of QCD of about 1 GeV to consider the chiral symmetry as approximate, so we set $n_F=3$ in the SM. At low energy, the symmetry of the strong interactions has to be broken spontaneously to its vectorial subgroup $U(n_F)_V$ in the limit where the number of colors ($N_c$) is large. The unitary matrix $U$ can thus be developed as a function of the pseudo-Goldstone bosons around its vacuum \cite{Cronin:1967jq},
\begin{equation}
U=\mathbbm{1}+\sum_{k=1}^\infty a_k\left(i\sqrt{2}\frac{\pi}{f}\right)^k\label{dev},
\end{equation}
with
\begin{equation}
\pi = \left(\begin{array}{ccc}\pi^3+\frac{1}{\sqrt{3}}\eta^8+\sqrt{\frac{2}{3}}\eta^0&\sqrt{2}\pi^+&\sqrt{2}K^+\\\sqrt{2}\pi^-&-\pi^3+\frac{1}{\sqrt{3}}\eta^8+\sqrt{\frac{2}{3}}\eta^0&\sqrt{2}K^0\\\sqrt{2}K^-&\sqrt{2}\,\overline{K^0}&-\frac{2}{\sqrt{3}}\eta^8+\sqrt{\frac{2}{3}}\eta^0
\end{array}\right).
\end{equation}
The coefficients $a_k$ are partially fixed by unitarity,
\beq\bsp
a_1&\,=1,\\
a_2&\,=\frac{1}{2},\\
a_3&\,=b,\\
a_4&\,=b-\frac{1}{8},\\
a_5&\,=c,\\
a_6&\,=c+\frac{b^2}{2}-\frac{b}{2}+\frac{1}{16},\\
\dots
\esp\eeq
The remaining free parameters, $b$, $c$, \dots, can be used to check the computation of any physical quantity which should be independent of these quantities~\cite{Callan:1969sn,Coleman:1969sm}, or they can be fixed to obtain the most suited form for $U$. For example, the common form $U=\exp\left(\frac{i\sqrt{2}\pi}{f}\right)$ requires $b=\frac{1}{6}$ and $c=\frac{1}{120}$. The mass matrix of the three neutral states, $\pi^3$, $\eta^8$ and $\eta^0$ is not diagonal. Consequently, the physical eigenstates are obtained by a rotation.\\

\noindent{\textbf{The SILH Model}}\\
The Strongly-Interacting Light Higgs Model is an effective theory of a possible strong sector responsible for the electroweak symmetry breaking (EWSB). The aim of this model is to disentangle if the EWSB is due to a strong sector or not by testing the interactions of the Higgs and the gauge bosons.

The SILH requirements are that the new strong sector has a spontaneously broken symmetry at low energy and that the Higgs boson is an exact Goldstone boson when the SM interactions are switched off. As for the mass matrix of the light quarks in QCD, the SM interactions are included order by order as small breaking parameters and induce a mass term for the Higgs boson. The little Higgses and Holographic composite Higgs models are examples of such theories.

 At scales lower than the resonances, the effects of the model are described by a set of dimension-six operators involving the Higgs boson,
\begin{eqnarray}
&&{\cal L}_{\rm SILH} = \frac{c_H}{2f^2}\partial^\mu \left( H^\dagger H \right) \partial_\mu \left( H^\dagger H \right) 
+ \frac{c_T}{2f^2}\left (H^\dagger {\overleftrightarrow { D^\mu}} H \right)  \left(   H^\dagger{\overleftrightarrow D}_\mu H\right) \nonumber \\ &&
- \frac{c_6\lambda}{f^2}\left( H^\dagger H \right)^3 
+ \left( \frac{c_yy_f}{f^2}H^\dagger H  {\bar f}_L Hf_R +{\rm h.c.}\right) \nonumber \\ &&
+\frac{ic_Wg}{2M_\rho^2}\left( H^\dagger  \sigma^i \overleftrightarrow {D^\mu} H \right )( D^\nu  W_{\mu \nu})^i
+\frac{ic_Bg'}{2M_\rho^2}\left( H^\dagger  \overleftrightarrow {D^\mu} H \right )( \partial^\nu  B_{\mu \nu})  \nonumber \\
&&
+\frac{ic_{HW} g}{16\pi^2f^2}
(D^\mu H)^\dagger \sigma^i(D^\nu H)W_{\mu \nu}^i
+\frac{ic_{HB}g^\prime}{16\pi^2f^2}
(D^\mu H)^\dagger (D^\nu H)B_{\mu \nu}
\nonumber \\
&&+\frac{c_\gamma {g'}^2}{16\pi^2f^2}\frac{g^2}{g_\rho^2}H^\dagger H B_{\mu\nu}B^{\mu\nu}+\frac{c_g g_S^2}{16\pi^2f^2}\frac{y_t^2}{g_\rho^2}H^\dagger H G_{\mu\nu}^a G^{a\mu\nu}
\label{lsilh}
\end{eqnarray}
where $H^\dagger {\overleftrightarrow D}_\mu H \equiv H^\dagger  D_\mu H- (D_\mu H^\dagger )H$, $g_\rho$ is the coupling of the new strong sector, $M_\rho$ is the mass of the new heavy states and $f=M_\rho/g_\rho$. For completeness, the model also contains the dimension-six Lagrangian for the SM vectors,
\begin{eqnarray}
{\cal L}_{\rm vect}&=
&-\frac{c_{2W}g^2}{2g_\rho^2M_\rho^2}(D^\mu W_{\mu\nu})^i (D_\rho W^{\rho\nu})^i 
-\frac{c_{2B}{g'}^2}{2g_\rho^2M_\rho^2}(\partial^\mu B_{\mu\nu})(\partial_\rho B^{\rho\nu})-\frac{c_{2g}g_3^2}{2g_\rho^2M_\rho^2}(D^\mu G_{\mu \nu})^a  (D_\rho G^{ \rho \nu})^a\nonumber \\
&&+\frac{c_{3W}g^3}{16\pi^2 M_\rho^2} \epsilon_{ijk} {W^{i}_\mu}^\nu W^{j}_{\nu\rho} W^{k\, \rho \mu}+\frac{c_{3g}g_3^3}{16\pi^2 M_\rho^2}f_{abc} {G^{a}_\mu}^\nu  G^{b}_{\nu\rho} G^{c\, \rho \mu}\, .
\label{vectors}
\end{eqnarray}
The $c_i$ are free parameters of the model and can thus be fixed to reproduce the low energy behavior of a more specific model. 

\subsubsection{\feynrules\ implementation}

\noindent{\textbf{$\chi$PT in \feynrules}}\\
$\chi$PT is implemented in \feynrules\ at the lowest order \eqref{LchiPT}, namely at the order $p^2$. In the implemented model, the Lagrangian is developed up to the $\mathscr{O}\left(\pi^6\right)$\,\footnote{All the higher order terms have been removed from the Lagrangian to save computation time.} with arbitrary coefficients $b$ and $c$. Consequently, vertices depend linearly on scalar products of the momenta and contain up to six scalars. The Lorentz structure of these vertices are not included in the default generic \feynarts\ file which contains only renormalizable structures. Let us note that at present this implementation only works with the \feynarts\ interface.

Since the isospin limit $m_u=m_d=\tilde{m}$ is taken in the \feynrules\ model file, only $\eta_8$ and $\eta_0$ mix. The mass eigenstates are given by 
\begin{equation}
\left(\begin{array}{c}\eta\\\eta'\end{array}\right)=\left(\begin{array}{cc}\cos\theta&-\sin\theta\\\sin\theta&\cos\theta 
\end{array}\right)\left(\begin{array}{c}\eta^8\\\eta^0\end{array}\right);\quad\theta\in\left[-\frac{\pi}{4},\,+\frac{\pi}{4}\right].\label{etamix}
\end{equation}
The isospin breaking can be easily added. The major modification is to extend the $2\times2$ matrix in \eqref{etamix} to a $3\times3$ mixing matrix since the three neutral states mix in this case. However, the effects are small and thus usually neglected. For the same reason, the mass matrix of the quarks is assumed to be real, but, the phase allowed by the $U(1)_A$ breaking can easily be added in the definition of the quark mass matrix.

The lowest order Lagrangian, Eq.~\eqref{LchiPT}, can reproduce the experimental data within 20\%. These discrepancies can be solved with next order corrections. The inclusion of the next order operators either $\mathscr{O}\left(p^4,0\right)$ \cite{Gerard:2004gx} or $\mathscr{O}\left(p^2,1/N_c\right)$ \cite{Leutwyler:1996sa,HerreraSiklody:1997kd} in the Lagrangian is also straightforward since $U$ is already defined in the model file. In the same way, the weak and electromagnetic interactions could also be added. 
However, the Lorentz structure of the vertices is hardcoded in the \feynarts\ interface, so the new structures generated by these new operators need to be added in the interface and in an associated \feynarts\ generic file, as it was already done for the lowest order Lagrangian.\\

\noindent{\textbf{SILH in \feynrules}}\\
The SILH implementation is based on the SM implementation but restricted to the unitary gauge. The two Lagrangians of Eqs.~\eqref{lsilh} and~\eqref{vectors} have been added to the SM one with all the $c_i$ considered as free external parameters. The first one contributes to the kinetic term of the Higgs and of the gauge bosons, so the physical fields are renormalized versions of the bare fields appearing in the Lagrangian\footnote{$(c_W+c_B)M_W^2/M_\rho^2$ is supposed to vanish to avoid non-diagonal kinetic terms.},
\beq\bsp
H &\,= H_{bare}\left(1+\xi\frac{c_H}{2}\right),\\
W^\mu_i &\,= W_{i,bare}^\mu\left(1-c_W\frac{M_W^2}{M_\rho^2}\right),\\
B^\mu &\,= B_{bare}^\mu\left(1-c_B \tan^2\theta_w\frac{M_W^2}{M_\rho^2}+\xi c_\gamma\frac{{g'}^2}{(4\pi)^2}  \frac{g^2}{g_\rho^2}\right),\\
G^\mu_a &\,= G_{a,bare}^\mu\left(1-\xi c_g \frac{g_S^2}{(4\pi)^2}\frac{y_t^2}{g_\rho}\right).
\esp\eeq
The gauge couplings have also to be redefined to obtain canonical kinetic terms for the gauge bosons,
\beq\bsp
g &\,= g_{bare}\left(1+c_W\frac{M_W^2}{M_\rho^2}\right),\\
g' &\,= g'_{bare}\left(1+c_B \tan^2\theta_w\frac{M_W^2}{M_\rho^2}-\xi c_\gamma\frac{{g'}^2}{(4\pi)^2}  \frac{g^2}{g_\rho^2}\right),\\
g_S &\,= g_{S,bare}\left(1+\xi c_g \frac{g_S^2}{(4\pi)^2}\frac{y_t^2}{g_\rho}\right).
\esp\eeq
Finally, the vev and the masses of the Higgs boson and the fermions are also corrected compared to their SM values,
\beq\bsp
v^2&\,=\frac{\mu^2}{\lambda}\left[1-\frac{3}{4}\xi c_6\right],\\
m_H^2&\,=2\mu^2\left[1-\xi\left(c_H+\frac{3}{4}c_6\right)\right],\\
m_f &\,= y_f\frac{v}{\sqrt2}\left(1-\xi\frac{c_y}{2}\right).
\esp\eeq
All of these redefinitions have been done at the first order in $\xi=\frac{v^2}{f^2}$. Any result is thus valid only at this order. It should be noted that all of these non-renormalizable interactions cannot be transferred directly to any other HEP tool for the moment.

%% file: validation.tex
\subsection{Strategy for the validation of the implemented models}

In order to validate the implementation of a model inside \feynrules, a very first natural check is to compare the obtained Feynman rules, using directly the \mathematica\ output, against the ones found in the literature. Subsequently, using the existing interfaces to symbolic tools such as \feynarts, we can go further in the validation procedure with analytical checks of some observables, confronting the results obtained with the help of the model files generated by \feynrules\ to the corresponding expressions found in the literature.

The next step in our validation procedure regards the model files generated through the different interfaces between \feynrules\ and matrix-element generators. For a given model, we calculate predictions using the set of Monte Carlo tools interfaced to \feynrules\ both with the model files generated by \feynrules\ and with the built-in (stock) implementations of the considered model, if they exist of course. After consistently fixing the set of external parameters to the values corresponding to a chosen benchmark point, results for various quantities, such as decay widths, total cross sections or (unintegrated) squared matrix elements at a given phase-space point, are computed and confronted. For a phase-space integrated observable $\sigma$, we evaluate all the possible quantities 
\be \label{eq:deltaSig}
   \Delta_{ab} = 2\,  {\big| \sigma^a - \sigma^b\big| \over \sigma^a + \sigma^b},~
\ee where $\sigma^{a,b}$ refers to predictions obtained with two specific generator and a given model file. The $\Delta_{ab}$'s quantify possible discrepancies between different implementations. For example, $\Delta_{\text{MG-FR, CH-ST}}$ represents the discrepancies between predictions obtained with \calchep, using the built-in implementation of the considered model (CH-ST), and with \madgraph, using the \feynrules-generated model files (MG-FR). For unintegrated squared matrix elements  $|M|^2$, we generalize this quantity to
\be \label{eq:deltaPS}
\Delta^{PS} = \max_{a,b} \Bigg\{ \sum_{\rm phase\,\, space\,\, points} \bigg[2\,  {\Big| |M^{a}|^2 - |M^{b}|^2\Big| \over |M^{a}|^2 + |M^{b}|^2} \bigg] \Bigg\}\,,~
\ee summing over a given number of phase-space points. 

For each model presented in the previous section, the input parameters used in the comparison and some numerical examples are given in Appendix \ref{app:val}. The complete list of results can be found on the \feynrules\ website~\cite{FRwebpage}.

\subsection{The Standard Model} 
All the Feynman rules obtained with  \feynrules\ were checked against the expressions given in the literature, while the total cross sections for a set of 35 key processes have been evaluated in \calchep, \mgme\ and \sherpa\ and compared to the existing stock versions. A selection of processes, together with the set of external parameters used for this check, is given in Appendix~\ref{app:SMvalidation}. Note in particular that \mgme\ and \sherpa\ work in unitary gauge, whereas \calchep~allows for both unitary and Feynman gauges. We therefore also demonstrated explicitly the gauge invariance of our implementation.

\subsection{The general Two-Higgs-Doublet Model}
The validation of the present implementation is done to by comparing $2\to 2$ matrix elements, for 10 different phase-space points, between the \feynrules\ unitary-gauge implementations in \mgme, \sherpa and \calchep, and the existing stock implementation in \mgme\ described in Ref.~\cite{Alwall:2007st} and distributed with the code. Since all the vertices appearing in the 2HDM are also present in various BSM models presented in this work, such as the MSSM, this limited comparison is clearly sufficient to fully validate our implementation. 

The selected benchmark point includes non-trivial values for all the $\lambda_i$ parameters of Eq.\ (\ref{eq:2HDMpot}), as well as Yukawa couplings for the second and third generation fermions leading to FCNC effects. Agreement, claimed if the quantity $\Delta^{PS}$ defined in Eq.\ \eqref{eq:deltaPS} is smaller than $0.1\%$, has been found for all the tested processes involving various combinations of scalars, fermions and gauge bosons both in the initial and final states. Furthermore, a selection of 185 $2\to 2$ cross sections have been computed using \calchep, \mgme and \sherpa, and compared to the existing \mgme\ implementation. In all cases the results agree within 1\% between the different codes. Some examples are shown in Appendix \ref{app:2hdm}. In addition to these checks, the behavior of various cross sections at high energies for processes involving new scalars has also been verified. In each case, the cross section behavior is in agreement with unitarity expectations.

\subsection{The most general Minimal Supersymmetric Standard Model}

We have compared the Feynman rules computed by \feynrules\ to those which can be found in the literature, both for the general MSSM \cite{Rosiek:1989rs, Rosiek:1995kg} and for a constraint MSSM where all the scalar mixings are neglected \cite{Haber:1984rc, Gunion:1984yn}, and we have found agreement for all the vertices.  Then, we have re-calculated all tree-level squark and gaugino hadroproduction helicity amplitudes in the case of general (and possibly complex) scalar mixing with the help of \feynarts/\formcalc\ and the model file generated by \feynrules. The results have been compared to the analytical formulas given in Refs.\ \cite{Bozzi:2007me, Fuks:2008ab} and we found complete agreement.

To validate the \feynrules-generated model files for the various Monte Carlo generators, we compared the results obtained in the very particular limit where $CP$ symmetry and flavor are assumed to be conserved within the whole model, the CKM matrix appearing in the charged-current interactions being thus neglected as well. In addition, in the scalar sector, the flavor-conserving helicity mixings are neglected for first and second generation sfermions. In this scenario, built-in implementations exist in \mgme\ and \calchep. As a benchmark scenario, we choose the typical minimal supergravity point SPS 1a \cite{Allanach:2002nj}, defined by the input parameters given in Appendix \ref{App:SUSY}. 
For \sherpa, the validation process is on-going and extensions to other programs that will be interfaced to \feynrules\ in the future are foreseen.  Let us note also that, since \calchep\ is not able to deal with files following the SLHA conventions, we had to modify the built-in files in order to correctly fix the free parameters of the model.

We start by evaluating all the 320 two-body decay widths corresponding to kinematically allowed decays in our scenario in order to check the norm of various three-point vertices. We find a complete agreement for both SM and MSSM processes. In particular, we have confronted the MG-FR implementation to the already validated MG-ST implementation \cite{Hagiwara:2005wg}, and the agreement between the two predictions was evaluated through the quantity $\Delta_{\text{MG-ST,MG-FR}}$ defined in Eq.\ (\ref{eq:deltaSig}), which has been found to be smaller than $1\%$, the differences being hence associated to a pure Monte-Carlo statistical error. Some examples can be found in the Table \ref{App:MSSMDec} of Appendix \ref{App:SUSY}.

Subsequently, we have investigated the total cross section for 636 key $2 \to 2$ processes with the help of both \mgme\ and \calchep, and with both \feynrules-generated and built-in model files. In order to properly compare the different implementations and unitary cancellations, we manually set all the widths of the particles to zero. We checked that the six quantities $\Delta_{ab}$ defined in Eq.\ (\ref{eq:deltaSig}) for any set of two predictions $a$ and $b$ are below the percent level. This check allows not only to verify the absolute value of all the three-point vertices and some four-point ones, but also to start being sensitive to the relative sign of a large part of those vertices. This is also the very first systematic comparison between predictions obtained with \mgme\ and with \calchep\ for the MSSM. We have considered the production of either a pair of any Standard Model particles, or a pair of any supersymmetric particles, from any Standard Model initial state. Even though most of these channels are not really phenomenologically relevant because most of the initial states are impossible to realize at a collider, they allow for a sensitive check of \textit{each} three-point vertex, and a large part of the four-point vertices. Let us note that the remaining untested vertices, such as the four-scalar interactions in $\lag_{\rm Scalar\, \, FDW}$ in Eq.\ (\ref{eq:FDW}), are irrelevant for $2\to 2$ leading-order calculations with a Standard Model initial state. We have considered two different cases, one where we have fixed the energies of the initial particles to 600 GeV, and one to 1 TeV. We found that the six possible $\Delta_{ab}$ quantities are below the percent level comparing any two of the four implementations, MG-FR, MG-ST, CH-FR and CH-ST, except for 
\be \label{eq:SUSYwrong} \bsp
W^+    \, W^-  \to W^+ \, W^-   {\rm ~and~~} b \, \bar t  \to Z   \, W^-
\esp \ee where the discrepancies are due to the presence of singularities in the matrix elements, which makes any of the considered Monte Carlo generators unable to correctly evaluate the total cross section, and except for about 15 processes where the built-in implementation of the MSSM in \calchep\ gives results different from those of all the other implementations due to mistakes in the model files for several triple scalar couplings. For the two processes in Eq.\ (\ref{eq:SUSYwrong}), we have evaluated the (unintegrated) squared matrix elements $|M|^2$ at given phase-space points for all models and computed the quantity $\Delta^{PS}$ of Eq.\ (\ref{eq:deltaPS}). Summing over 100 different random phase-space points, we have found that $\Delta^{PS}$ is below $1\%$, which is sufficient to conclude that the four implementations agree. Some examples of numerical results can be found in Tables \ref{App:MSSM2to2Higgs}, \ref{App:MSSM2to2SUSY1} and \ref{App:MSSM2to2SUSY2} in Appendix \ref{App:SUSY}. Let us recall that the complete list is available on the \feynrules\ webpage.

Finally, in order to test \textit{every} sign of each three-point vertex and a large part of the four-point vertices, we have evaluated squared matrix elements for 100 random phase-space points and calculated the quantity $\Delta^{PS}$ defined in Eq.\ (\ref{eq:deltaPS}). We have investigated 2708 processes, relative to the production of two supersymmetric particle plus one Standard Model particle at a center-of-mass energy of 2 TeV. Using MG-FR and MG-ST, we have found a complete agreement for each process, \ie, $\Delta^{PS}$ was below 0.1 $\%$. All the results can be found on the \feynrules\ website. The comparison with matrix elements calculated by CH-FR and CH-ST is devoted to a further study.

\subsection{The Minimal Higgsless Model} 
\label{3-Site:validation}

Our \feynrules\ implementation was compared to the \lanhep\ implementation, both in unitary and in Feynman gauge. It was run in Feynman gauge in \calchep\ and \comphep\ and in unitary gauge in \calchep, \mgme\ and \sherpa. The parameters used for this validation and the final particles cuts are given in Appendix \ref{sec:validation:3-Site}.  In all cases, the cross section was calculated, compared, and agreement to better than $1\%$ was found.

Some examples of the obtained cross sections are given in the various tables of Appendix \ref{sec:validation:3-Site}.  In Table \ref{3-Site-validation:strong}, a selection of strong processes are presented.  In Tables \ref{3-Site-validation:ffAW} and \ref{3-Site-validation:ffAA}, charged and neutral electroweak processes are presented, respectively.

\subsection{Extra Dimensional Models}

\noindent{\textbf{Large Extra Dimension Model}}\\
The Feynman rules for LED obtained by \feynrules\ were explicitly checked with those available in the literature. Unlike for the models presented above, we cannot use any matrix-element generator to compute cross sections or decay rates, because the interfaces linking \feynrules\ to Monte-Carlo generators are not yet defined to work for a theory with spin-two particles. Therefore we have chosen to validate our LED model implementation via analytical expressions. We found complete agreement both for the Feynman rules of the generic LED theory which can be found in Refs.\ \cite{Han:1998sg, Giudice:1998ck} and for those of the full LED implementation. Let us notice that from the generic LED model validated above, we can extrapolate the results of the latter to guess those of the QCD and electromagnetic part of the full LED model. This check was performed and conclusive.\\

\noindent{\textbf{Universal Extra Dimension Model}}\\
For the \mued\ model, we start by confronting the Feynman rules to those available in the literature. In a second step, we have calculated cross sections for $2 \rightarrow 2$ processes and compared the results obtained with the help of \madgraph, \sherpa\ and \calchep, and we confronted our results to those obtained with the help of an existing \calchep\ implementation~\cite{Matchev:2007}. The input parameters for the benchmark point which we have chosen for our  numerical validation are given in Appendix~\ref{App:MUED}. Let us note that the only Higgs field which we have considered is the Standard Model one, even though its first Kaluza-Klein excitation is also implemented. The comparison has been carried out for the 118 processes in total~\cite{Servant:2002aq, Burnell:2005hm, Kong:2005hn, rizzo-2001-64}. First, using \madgraph\ and considering Standard Model processes, we have calculated squared matrix elements at given phase-space points and confronted the results obtained with the \feynrules-generated model file to those obtained with the built-in Standard Model implementation. Subsequently, using the \feynrules-generated model file for \madgraph, \sherpa\ and \calchep\ and the existing \calchep\ implementation of the \mued\ model, we have compared total cross sections for the 118 chosen processes at a center-of-mass energy of 1400 GeV. We have found agreement for each of them, and some examples are shown in Tables \ref{App:muedGau} and \ref{App:muedFer} in Appendix \ref{App:MUED}.

\subsection{Low-energy effective theories}

\noindent{\textbf{$\chi$PT at lowest order}}\\
The $\chi$PT model was checked analytically. The expressions of the amplitudes computed with \textit{pen \& paper} work of Ref.~\cite{Degrande:2009ps} have been compared to the one obtained using \feynrules\ and \feynarts. The loop amplitudes have been integrated using cutoff regularization and only the terms quadratic in the cutoff $\Lambda$ have been computed. More precisely, we computed the one-loop corrections to the two-point functions $\pi-\pi$, $\eta-\eta$, $\eta'-\eta'$ and $\eta-\eta'$. The momentum independent part of each of these amplitudes, but the last one, corresponds to mass corrections and is $b$-independent as it should. For example, the renormalization of the pion wave function is 
\begin{equation}
Z = 6\left(1-8b\right)\frac{\Lambda^2}{\left(4\pi f\right)^2},\quad\text{where}\quad \pi_R \equiv \sqrt{1+Z}\pi,
\end{equation}
while its mass correction vanishes\footnote{All the other results can be found in Ref.~\cite{Degrande:2009ps}.}. The $\eta'\rightarrow\eta\pi\pi$ decay amplitude has also been computed at tree-level and at one-loop. In this last result, also the $c$-dependence cancels. Eventually, a total of about 60 diagrams were computed with both methods and perfect agreement was found.\\

\noindent{\textbf{SILH Model}}\\
The check of the SILH model consists in an analytic comparison between the decay widths computed in Ref.~\cite{Giudice:2007fh} and computations based on the vertices given by \feynrules. We used the same simplifications, \emph{i.e.}, $c_T=0$ and $(c_W+c_B)M_W^2/M_\rho^2=0$ and neglect ${\cal L}_{\rm vect}$. 
For tree-level decay widths of the Higgs into two fermions, both implementations leads to exactly the same results. For decays into a gauge boson pair, the contribution to the Feynman rules from higher-dimensional operators read, \emph{e.g.} for the $hWW$ vertex,
\begin{eqnarray}
&ig^2\frac{v}{2}(1-\xi \frac{c_H}{2})\eta^{\mu_2,\mu_3}+i\xi \frac{c_W}{2v}\frac{g^2}{g_\rho^2}\left[\eta^{\mu_2,\mu_3}\left(p_2^2+p_3^2\right)-p_2^{\mu_2}p_2^{\mu_3}-p_3^{\mu_2}p_3^{\mu_3}\right]&\nonumber\\&+i\xi\frac{c_{HW}}{2v}\frac{g^2}{(4\pi)^2}\left[\eta^{\mu_2,\mu_3}p_1^2-p_1^{\mu_2}p_1^{\mu_3}\right],&
\end{eqnarray}
where $p_1$, $p_2$ and $p_3$ denote the momenta of the Higgs boson and the two $W$ bosons respectively. 
As a consequence, the corrections are not just proportional to the SM decay widths since the vertices have a more complicated kinematic structure. We find
\begin{eqnarray}\label{eq:hWW}
\Gamma\left(h\rightarrow W^+W^-\right)&=&\Gamma\left(h\rightarrow W^+W^-\right)_{SM}\left[1-\xi\left(c_H-\frac{g^2}{g_\rho^2}c_W\right)\right]\nonumber\\&&+\xi c_{HW}\frac{g^2 }{(4\pi)^2}\frac{m_H}{16\pi v^2}\left(m_H^2+2M_W^2\right)\left(1-4\frac{M_W^2}{m_H^2}\right)^{1/2},
\end{eqnarray}
\begin{eqnarray}\label{eq:hZZ}
\Gamma\left(h\rightarrow ZZ\right)&=&\Gamma\left(h\rightarrow ZZ\right)_{SM}\left[1-\xi\left(c_H-\frac{g^2}{g_\rho^2}\left(c_W+\tan^2\theta_wc_B\right)\right)\right]\nonumber\\&&+\xi (c_{HW}+\tan^2\theta_w c_{HB})\frac{g^2 }{(4\pi)^2}\frac{m_H}{16\pi v^2}\left(m_H^2+2M_Z^2\right)\left(1-4\frac{M_Z^2}{m_H^2}\right)^{1/2},
\end{eqnarray}
in agreement with Ref.~\cite{Hagiwara:1993qt}. In the situation where we have the hierachy of couplings $g<g_\rho<4\pi$, the second term in Eqs.~(\ref{eq:hWW}) and (\ref{eq:hZZ}) is suppressed parametrically with respect to the first one and could thus be neglected, leading to just a rescaling of the SM decay widths~\cite{Giudice:2007fh}.
Let us note however that in the case where the ratio $g^2/(4\pi)^2$ is not too small, this additional term could have a numerical impact on the decay rates.

%% file: outlook.tex
\section{Outlook}
\label{sec:outlook}

We have described a new framework where BSM physics  scenarios can be developed, studied and automatically implemented in Monte Carlo or symbolic calculation tools for theoretical and experimental investigations. The main purpose of this work has been to contribute to streamlining the communication (in both directions) between the theoretical and the experimental HEP communities.

The cornerstone of our approach is the \mathematica~package \feynrules~where any perturbative quantum field theory Lagrangian, renormalizable or not, can be written in a straightforward way and the corresponding Feynman rules obtained automatically. All the relevant information can then be passed through dedicated interfaces to matrix element generators for Feynman diagram calculations at the tree level or at higher orders. The scheme itself looks very simple and is in fact not a new idea. The novelty, however, lies in several technical and design aspects which, we believe, constitute a significant improvement over the past.  

First, the use of \mathematica~as a working environment for the model development
gives all the flexibility that is needed for symbolic manipulations. The many built-in features of \mathematica, such as 
matrix diagonalization and pattern recognition functions, play an important role in building not only robust interfaces for very different codes but also to open up new possibilities. For instance, besides implementing  BSM models, new high-level functionalities/applications 
can be easily developed by the users themselves, made public and possibly included in subsequent official releases. In other words, the code is naturally very open to community contributions. A typical application that could take advantage of this open structure is the (semi-)automatic development of model calculators inside the \feynrules~package itself, including mass spectrum and decay width calculations. 

Second, the  interfaces to MC codes, all of them quite different both in philosophy, architecture and aim, offer the possibility  of testing and validating model implementations at an unprecedented level of accuracy. It also maximizes the probability that a given model might be dealt with by at least one matrix element generator. For example, purely symbolic generators such as \feynarts/\formcalc\ can be used for tree-level (or even loop) calculations which can then be compared or extended to numerical results from \calchep, \mgme~or \sherpa.  In this respect we note that one of the current major and common limitations of the matrix-element generators (but not of \feynrules) is connected to use of a fixed library of Lorentz and/or gauge structures for the vertices. These libraries are in principle extendable, but at present this is done by hand and it entails a tedious and often quite long work. The automatization of this part (and possibly the reduction of higher-dimensional operators to renormalizable ones) through \feynrules~would be the final step towards full automatization of any Lagrangian into Monte Carlo codes. Work in this direction is already in progress.

With such a framework in place, we hope that several of the current problems and drawbacks in new physics simulations
faced by the experimental groups will be alleviated if not completely solved. First the need of dedicated codes for specific models, which then call for long and tedious validations both from the physics point of view as well as in their interplay with collaboration softwares, will be greatly reduced. General purpose tools, from \herwig~and \pythia~to \sherpa, \mgme, \whizard, \calchep/\comphep\ (and potentially also \helac~and \alpgen), several of which have been successfully embedded in  frameworks such as {\sc Athena}\ (ATLAS) and {\sc CMSSW}\ (CMS), offer several ready-to-go solutions for any \feynrules~based models. Reducing the proliferation of highly dedicated tools will greatly simplify the maintenance and reliability of the software and more importantly the reproducibility of the MC samples in the mid and long terms. In addition, we believe an effort towards making new Lagrangians in \feynrules~and the corresponding benchmark points publicly available (for example by the proponents of the model themselves at the same time of the release of the corresponding publication) would certainly be a great advantage for the whole HEP community.

It is our hope that \feynrules~will effectively facilitate interactions between theorists and experimentalists. Until now there has not been a preferred way to link the two communities. Various solutions have been proposed and used, most of them plagued by significant limitations. The best available proposal so far has been to use parton-level events in the Les Houches Event File (LHEF) format \cite{Alwall:2006yp}. This is a natural place to cut a line given that theorists and phenomenologists can generate events through various private or public tools, and then pass them for showering and hadronization to codes such as \pythia~or \herwig. These codes are not only already embedded in the experimental software (for the following detector simulations), but also have been (or will be) tuned carefully to reproduce control data samples. 

While we think this is still a useful approach that should be certainly left open and supported, we are convinced the framework we propose is a promising extension. The deep reason is that, in our approach, there is no definite line between theory and experiment. On the contrary it creates a very extended region where the two overlap and work can be done in the same common framework.  This leaves much more freedom in where exactly the two ends meet and what kind of checks and information can be exchanged. As a result there are several practical advantages that come for free.  

As an example, we remind that the LHEF format only standardizes the information on the events themselves and some very basic global properties, but any information on the physical model (\ie, the explicit form of the Lagrangian) or the parameter choices is in general  absent. This is of course due to implementation differences among various codes which severely compromise any standardization attempt at this level. It is clear that, in the long run, this might lead to serious problems of traceability of the MC samples and various ambiguities in understanding experimental analyses (such as placing of exclusion limits). This problem is of course completely overcome by the approach advocated here, since models are now fully and univocally defined. In this sense \feynrules~itself offers the sought for standardization.

Another, and maybe even more striking example is that, within \feynrules, model building and/or refinement can in principle be done in realtime together with the related experimental analyses. One could imagine, for example, that if the TeV world is as rich as we hope and as data start showing hints for new particles or effects, a large number of competing Lagrangians (and not only benchmark points as used in the typical top-down SUSY analyses) could be easily and quickly implemented and readily confronted to data. This could be done in a virtuous loop, where theorists and experimentalists ``meet'' at a convenient point of the simulation chain. In other words, various top-down and bottom-up studies can fit naturally in this framework, partially addressing often reported worries about the actual possibilities to extract precise information on BSM physics from LHC data.

Finally, let us also mention a more long-term advantage of the proposed framework. Automatic NLO calculations for SM processes are now clearly in sight, and, in this context, it might be reasonable to ask whether those developments can be extended to BSM processes. We believe the answer is yes, and, since this generalization will probably rely on the simultaneous implementation of the model characteristic in different codes (\eg, dealing with different parts of the calculation like real and virtual corrections, or analytic and numerical results) our approach might also naturally play a crucial role in this context.

%% file: FRconvention.tex
\section{The \feynrules~convention for Standard Model inputs}
\label{app:FRMCconventions}

There are several parameters and particles that have special significance in Feynman diagram calculators.  Examples of these are the strong and electromagnetic couplings, the names of the fundamental representation and the structure constants of the strong gauge group and so on.  For this reason, we have chosen to fix the names of these objects at the \feynrules~level.  Adherence to these standards will increase the chances of successful translation.

The strong gauge group has special significance in many Feynman diagram calculators.  A user who implements the strong gauge group should adhere to the following rules.  The indices for the fundamental and adjoint representations of this gauge group should be called \verb+Colour+ and \verb+Gluon+ respectively.  Furthermore, the names of the QCD gauge boson, coupling constant, structure constant, totally symmetric term and fundamental representation should be given by \verb+G+, \verb+gs+, \verb+f+, \verb+dSU3+ and \verb+T+ as in the following example:
\begin{verbatim}
   SU3C == {
       Abelian           -> False,
       GaugeBoson        -> G,
       StructureConstant -> f,
       SymmetricTensor   -> dSU3,
       Representations   -> {T, Colour},
       CouplingConstant  -> gs
       }
\end{verbatim}
In addition, the strong coupling constant and its square over $4\pi$ should be declared in the parameter section in the following form:
\begin{verbatim}
   \[Alpha]S == {
       ParameterType     -> External,
        Value            -> 0.118,
        ParameterName    -> aS,
        BlockName        -> SMINPUTS,
        InteractionOrder -> {QCD, 2},
        Description      -> "Strong coupling at the Z pole."
        },
   gs == {
        ParameterType    -> Internal,
        Value            -> Sqrt[4 Pi \[Alpha]S],
        ParameterName    -> G,
        InteractionOrder -> {QCD, 1}
       }
\end{verbatim}
Note that $\alpha_S$ is given as the external parameter and $g_S$ as the internal parameter.  The description of $\alpha_S$ may be edited, but it should be remembered that, for the Monte Carlo programs that run the strong coupling constant, the value of $\alpha_S$ should be set at the Z pole.  For calculation programs that do not run the strong coupling, on the other hand, it should be set according to the scale of the interaction.  A description may also be added to the parameter $g_S$.

The electromagnetic interaction also has special significance in many Feynman diagram calculators and we outline the following standard definitions.  The electric coupling constant should be called \verb+ee+, the electric charge should be called \verb+Q+.  The declaration of the electric charge should follow the following conventions for naming:
\begin{verbatim}
   \[Alpha]EWM1 == {
        ParameterType    -> External,
        Value            -> 127.9,
        ParameterName    -> aEWM1,
        BlockName        -> SMINPUTS,
        InteractionOrder -> {QED, -2},
        Description      -> "alpha_EM inverse at the Z pole."
        },
   \[Alpha]EW == {
        ParameterType    -> Internal,
	        Value            -> 1/\[Alpha]EWM1,
	        InteractionOrder -> {QED, 2},
        ParameterName    -> aEW,
	        },
   ee == {
        ParameterType    -> Internal,
        Value            -> Sqrt[4 Pi \[Alpha]EW ],
        InteractionOrder -> {QED, 1}
        }
\end{verbatim}
As for the strong coupling, the description of $\alpha_{EW}^{-1}$ may be edited\footnote{The reason for choosing $\alpha_{EW}^{-1}$ as the external input parameter, and not $\alpha_{EW}$ itself is only to be compliant with the \emph{Les Houches Accord}.}, but it should be remembered that for calculation programs that run the electric coupling, it should be set at the Z pole.  For programs which do not run it, the electric coupling should be set at the interaction scale.  Again, a description may be added to the definition of \verb+\[Alpha]EW+ and \verb+ee+.

The Fermi constant and the Z pole mass are very precisely known and are often used in calculators to define coupling constants and the scale where couplings are run from.  They should be included in the {\tt SMINPUTS} block of the Les Houches accord and should be defined by at least the following:
\begin{verbatim}
   Gf == {
        ParameterType    -> External,
        Value            -> 1.16639 * 10^(-5),
        BlockName        -> SMINPUTS,
        InteractionOrder -> {QED, 2},
        Description      -> "Fermi constant"
        },
   ZM == {
        ParameterType -> External,
        Value         -> 91.188,
        BlockName     -> SMINPUTS,
        Description   -> "Z pole mass"
        }
\end{verbatim}

Moreover, the weak coupling constant name \verb+gw+ and the hypercharge symbol \verb+Y+ are used by some calculators and the user is encouraged to use these names where appropriate.  The masses and widths of particles should be assigned whenever possible.  If left out, \feynrules~will assign the value 1 to each.  Finally, particles are also identified by a PDG number.  The user is strongly encouraged to use existing PDG codes in their model wherever possible.  If not included, a PDG code will be automatically assigned by \feynrules~beginning at 6000001.

%% file: sm2.tex
\subsection{The Standard  Model}
\label{app:SMvalidation}

In this section we give the results for the 35 cross sections tested for the SM between \calchep, \mgme~and \sherpa~for a total center of mass energy of 550 GeV. A $p_T$ cut of 20 GeV was applied to each final state particle. The set of external parameters used for the test is given in Table~\ref{tab:SMparams}. A selection of processes is shown in Tables~\ref{tab:SMprocs1} and \ref{tab:SMprocs2}.
\begin{table}[!t]
\begin{center}
\begin{tabular}{rc l }
\hline\hline
Parameter & Symbol & Value\\
\hline
Inverse of the electromagnetic coupling  & $\alpha_{EW}^{-1}(M_Z)$ & 127.9\\
Strong coupling  & $\alpha_{s}(M_Z)$ & 0.118\\
Fermi constant & $G_F$ & 1.16639e-5 GeV$^{-2}$\\
\hline
$Z$ pole mass & $M_Z$ & 91.188\, GeV \\
$c$ quark mass & $m_c$ & 1.42 GeV\\
$b$ quark mass & $m_b$ & 4.7 GeV\\
$t$ quark mass & $m_t$ & 174.3 GeV\\
$\tau$ lepton mass & $m_\tau$ & 1.777 GeV\\
Higgs mass & $m_H$ & 120 GeV\\
Cabibbo angle & $\theta_c$ & 0.227736\\
\hline\hline
\end{tabular}
\caption{\label{tab:SMparams} Input parameters for the SM.}
\end{center}
\end{table}

\begin{table}[!h]
\begin{center}
\input{SMnumbers}

\caption{\label{tab:SMprocs2}Cross sections for a selection of SM production processes.  The built-in SM implementation in {\sc MadGraph}, {\sc CalcHep} and {\sc Sherpa} are denoted MG-ST, CH-ST and {\sc Sherpa}, respectively, while the \feynrules-generated ones MG-FR, CH-FR and SH-FR. The center-of-mass energy is fixed to 550 GeV, and a $p_T$ cut of 20 GeV is applied to each final state particle.
}
\end{table}

\clearpage

%% file: SMnumbers.tex
\renewcommand{\arraystretch}{1.3}

 \begin{tabular}{|l| r@{e}l  r@{e}l | r@{e}l  r@{e}l  | r@{e}l  r@{e}l |}
 \multicolumn{13}{c}{\bf Lepton and weak boson processes in the Standard Model}\\
\hline
\multicolumn{1}{|c|}{Process} & \multicolumn{2}{c}{MG-FR} & \multicolumn{2}{c|}{MG-ST} & \multicolumn{2}{c}{CH-FR} & \multicolumn{2}{c|}{CH-ST}  &\multicolumn{2}{c}{SH-FR} & \multicolumn{2}{c|}{SH-ST}\\
\hline \hline
$e^+ \, e^- \to e^+ \, e^-$ & $7.341$ & $+2$ & $7.343$ & $+2$ & $7.342$ & $+2$ & $7.342$ & $+2$ & $7.343$ & $+2$ & $7.343$ & $+2$\\
$e^+ \, e^- \to \mu^+ \, \mu^-$ & $3.721$ & $-1$ & $3.720$ & $-1$ & $3.719$ & $-1$ & $3.719$ & $-1$ & $3.720$ & $-1$ & $3.720$ & $-1$\\
$e^+ \, e^- \to \nu_e \, \bar \nu_e$ & $4.914$ & $+1$ & $4.913$ & $+1$ & $4.915$ & $+1$ & $4.915$ & $+1$ & $4.915$ & $+1$ & $4.915$ & $+1$\\
\hline
$\tau^+ \, \tau^- \to W^+ \, W^-$ & $5.370$ & $+0$ & $5.360$ & $+0$ & $5.368$ & $+0$ & $5.368$ & $+0$ & $5.368$ & $+0$ & $5.368$ & $+0$\\

$\tau^+ \, \tau^- \to Z \, Z$ & $3.186$ & $-1$ & $3.180$ & $-1$ & $3.182$ & $-1$ & $3.182$ & $-1$ & $3.183$ & $-1$ & $3.183$ & $-1$\\

$\tau^+ \, \tau^- \to Z \, A$ & $2.005$ & $+0$ & $2.007$ & $+0$ & $2.006$ & $+0$ & $2.006$ & $+0$ & $2.006$ & $+0$ & $2.006$ & $+0$\\

$\tau^+ \, \tau^- \to A \, A$ & $2.782$ & $+0$ & $2.780$ & $+0$ & $2.779$ & $+0$ & $2.779$ & $+0$ & $2.779$ & $+0$ & $2.779$ & $+0$\\
\hline
$Z \, Z \to Z \, Z$ & $1.960$ & $+0$ & $1.959$ & $+0$ & $1.961$ & $+0$ & $1.961$ & $+0$ & $1.961$ & $+0$ & $1.961$ & $+0$\\

$W^+ \, W^- \to Z \, Z$ & $2.726$ & $+2$ & $2.729$ & $+2$ & $2.726$ & $+2$ & $2.726$ & $+2$ & $2.726$ & $+2$ & $2.726$ & $+2$\\

$H \, H \to Z \, Z$ & $6.268$ & $+1$ & $6.266$ & $+1$ & $6.266$ & $+1$ & $6.266$ & $+1$ & $6.266$ & $+1$ & $6.266$ & $+1$\\

$H \, H \to W^+ \, W^-$ & $9.449$ & $+1$ & $9.450$ & $+1$ & $9.447$ & $+1$ & $9.447$ & $+1$ & $9.448$ & $+1$ & $9.448$ & $+1$\\
\hline
\end{tabular}

\end{center}
\caption{\label{tab:SMprocs1}Cross sections for a selection of SM production processes.  The built-in SM implementation in {\sc MadGraph} and {\sc CalcHep} and {\sc Sherpa} are denoted MG-ST, CH-ST and SH-ST, respectively, while the \feynrules-generated ones MG-FR, CH-FR, SH-FR. The center-of-mass energy is fixed to 550 GeV, and a $p_T$ cut of 20 GeV is applied to each final state particle.
}
\end{table}
\begin{table}[!h]

\begin{center}

\begin{tabular}{|l| r@{e}l  r@{e}l | r@{e}l  r@{e}l  | r@{e}l  r@{e}l |}
 \multicolumn{13}{c}{\bf Quark and gluon processes in the Standard Model}\\
\hline

\multicolumn{1}{|c|}{Process} & \multicolumn{2}{c}{MG-FR} & \multicolumn{2}{c|}{MG-ST} & \multicolumn{2}{c}{CH-FR} & \multicolumn{2}{c|}{CH-ST}   &\multicolumn{2}{c}{SH-FR} & \multicolumn{2}{c|}{SH-ST}\\

\hline \hline
$G \, G \to G \, G $ & $1.177$ & $+5$ & $1.178$ & $+5$ & $1.177$ & $+5$ & $1.177$ & $+5$ & $1.177$ & $+5$ & $1.177$ & $+5$\\
\hline
$u \, \bar u \to G \, G$ & $2.021$ & $+2$ & $2.021$ & $+2$ & $2.021$ & $+2$ & $2.021$ & $+2$ & $2.021$ & $+2$ & $2.021$ & $+2$\\
$u \, \bar u \to W^+ \, W^-$ & $1.772$ & $+0$ & $1.774$ & $+0$ & $1.774$ & $+0$ & $1.774$ & $+0$ & $1.774$ & $+0$ & $1.774$ & $+0$\\

$u \, \bar u \to Z \, Z$ & $1.936$ & $-1$ & $1.933$ & $-1$ & $1.935$ & $-1$ & $1.935$ & $-1$ & $1.935$ & $-1$ & $1.935$ & $-1$\\

$u \, \bar u \to Z \, A$ & $3.380$ & $-1$ & $3.382$ & $-1$ & $3.381$ & $-1$ & $3.381$ & $-1$ & $3.381$ & $-1$ & $3.381$ & $-1$\\

$u \, \bar u \to A \, A$ & $1.833$ & $-1$ & $1.833$ & $-1$ & $1.832$ & $-1$ & $1.832$ & $-1$ & $1.832$ & $-1$ & $1.832$ & $-1$\\
$u \, \bar u \to s \, \bar s$ & $9.864$ & $+0$ & $9.861$ & $+0$ & $9.868$ & $+0$ & $9.868$ & $+0$ & $9.869$ & $+0$ & $9.869$ & $+0$\\
$u \, \bar d \to c \, \bar s$ & $3.531$ & $-1$ & $3.531$ & $-1$ & $3.531$ & $-1$ & $3.531$ & $-1$ & $3.531$ & $-1$ & $3.532$ & $-1$\\
$u \, \bar s \to c \, \bar d$ & $1.019$ & $-3$ & $1.019$ & $-3$ & $1.019$ & $-3$ & $1.019$ & $-3$ & $1.019$ & $-3$ & $1.019$ & $-3$\\
\hline
$t \, \bar t \to G \, G$ & $6.522$ & $+1$ & $6.527$ & $+1$ & $6.528$ & $+1$ & $6.528$ & $+1$ & $6.528$ & $+1$ & $6.528$ & $+1$\\
$t \, \bar t \to Z \, A$ & $1.311$ & $+0$ & $1.311$ & $+0$ & $1.312$ & $+0$ & $1.312$ & $+0$ & $1.312$ & $+0$ & $1.312$ & $+0$\\
$t \, \bar t \to A \, A$ & $8.844$ & $-2$ & $8.846$ & $-2$ & $8.849$ & $-2$ & $8.849$ & $-2$ & $8.848$ & $-2$ & $8.848$ & $-2$\\

$t \, \bar t \to u \, \bar u$ & $1.621$ & $+1$ & $1.618$ & $+1$ & $1.619$ & $+1$ & $1.619$ & $+1$ & $1.619$ & $+1$ & $1.619$ & $+1$\\

$t \, \bar t \to W^+ \, W^-$ & $1.713$ & $+1$ & $1.713$ & $+1$ & $1.713$ & $+1$ & $1.713$ & $+1$ & $1.714$ & $+1$ & $1.714$ & $+1$\\
$t \, \bar t \to Z \, Z$ & $1.253$ & $+0$ & $1.254$ & $+0$ & $1.253$ & $+0$ & $1.253$ & $+0$ & $1.253$ & $+0$ & $1.253$ & $+0$\\

\hline

\end{tabular} \end{center}

%% file: 2hdm2.tex
\subsection{The general Two-Higgs-doublet model}
\label{app:2hdm}

In this section we give a selection of the results for the 185 cross sections tested for the 2HDM between \calchep, \mgme~and \sherpa~for a total center-of-mass energy of 800 GeV. A $p_T$ cut of 20 GeV was applied to each final state particle. The set of external parameters used for the test is given in Table~\ref{tab:2HDMparams}. A selection of processes is shown in Tables~\ref{tab:2HDMprocs1}, \ref{tab:2HDMprocs2} and \ref{tab:2HDMprocs3}.

\begin{table}[!h]
\begin{center}
\begin{tabular}{rc l }
\hline\hline
Parameter & Symbol & Value\\
\hline
Inverse of the electromagnetic coupling  & $\alpha_{EW}^{-1}(M_Z)$ & 127.934\\
Strong coupling  & $\alpha_{s}(M_Z)$ & 0.1172\\
Fermi constant & $G_F$ & 1.16637e-5 GeV$^{-2}$\\
\hline
$Z$ pole mass & $M_Z$ & 91.18876\, GeV \\
$c$ quark pole mass & $m_c$ & 1.25 GeV\\
$b$ quark pole mass & $m_b$ & 4.2 GeV\\
$t$ quark pole mass & $m_t$ & 174.3 GeV\\
$\tau$ lepton mass & $m_\tau$ & 1.777 GeV\\
Cosine of the Cabibbo angle & $\cos\theta_c$ & 0.974589144\\
\hline
$c$ quark ``Yukawa'' mass & $m_c^{Yuk}$ & 0.6 GeV\\
$b$ quark ``Yukawa'' mass & $m_b^{Yuk}$ & 3.0 GeV\\
$t$ quark ``Yukawa'' mass & $m_t^{Yuk}$ & 175 GeV\\
\hline
Potential parameters & $\lambda_{1,2,3}$ & 1.0\\
 & $\lambda_4$ & 0.5\\
 & $\lambda_5$ & 0.4 \\
 & $\Re(\lambda_6)$ & 0.3\\
 & $\Re(\lambda_7)$ & 0.2\\
Charged Higgs mass & $m_{H^\pm}$ & 300 GeV\\
\hline
Yukawa parameters (real)& $(\Gamma_d)_{2,2}$ & 0.4 GeV \\
 & $(\Gamma_d)_{2,3}$ & 0.2 GeV \\
 & $(\Gamma_d)_{3,3}$ & 5 GeV \\
  & $(\Gamma_u)_{2,2}$ & 2 GeV \\
  & $(\Gamma_u)_{2,3}$ & 1 GeV \\
  & $(\Gamma_u)_{3,3}$ & 100 GeV \\
  & $(\Gamma_l)_{2,2}$ & 0.1 GeV \\
  & $(\Gamma_l)_{2,3}$ & 0.5 GeV \\
  & $(\Gamma_l)_{3,3}$ & 3 GeV \\
\hline\hline
\end{tabular}
\caption{\label{tab:2HDMparams} Input parameters for the 2HDM. All parameters that are not quoted have a zero value.}
\end{center}
\end{table}

\begin{table}
\input{TATA}
\caption{\label{tab:2HDMprocs1}Cross sections for a selection of $\tau^+\tau^-$ initiated processes in the 2HDM.  The built-in 2HDM implementation in {\sc MadGraph} is denoted MG-ST, while the \feynrules-generated ones are MG-FR, CH-FR and SH-FR. The center-of-mass energy is fixed to 800 GeV, and a $p_T$ cut of 20 GeV is applied to each final state particle.}
\end{table}

\begin{table}
\input{TANU}
\caption{\label{tab:2HDMprocs2}Cross sections for a selection of $\tau^+\nu_\tau$ initiated processes in the 2HDM.  The built-in 2HDM implementation in {\sc MadGraph} is denoted MG-ST, while the \feynrules-generated ones are MG-FR, CH-FR and SH-FR. The center-of-mass energy is fixed to 800 GeV, and a $p_T$ cut of 20 GeV is applied to each final state particle.}
\end{table}

\begin{table}
\input{VVSS}
\caption{\label{tab:2HDMprocs3}Cross sections for a selection of processes in the 2HDM with two weak bosons in the initial state.  The built-in 2HDM implementation in {\sc MadGraph} is denoted MG-ST, while the \feynrules-generated ones are MG-FR, CH-FR  and SH-FR. The center-of-mass energy is fixed to 800 GeV, and a $p_T$ cut of 20 GeV is applied to each final state particle.}
\end{table}

\clearpage

%% file: TATA.tex
\renewcommand{\arraystretch}{1.3}
\newcommand{\FRHDMtab}[9]{$#1$ & $#2$ & $#3$ & $#4$ & $#5$ & $#6$ & $#7$ & $#8$ & $#9$ }

\begin{center} \begin{tabular}{|l| r@{e}l  r@{e}l | r@{e}l  |  r@{e}l  | }
\multicolumn{9}{c}{\bf Lepton processes in the Two-Higgs-Doublet Model}\\
\hline
\multicolumn{1}{|c|}{Process} & \multicolumn{2}{c}{MG-FR} & \multicolumn{2}{c|}{MG-ST} & \multicolumn{2}{c|}{CH-FR} & \multicolumn{2}{c|}{SH-FR} \\
\hline \hline
\FRHDMtab{\tau^+ \, \tau^- \to h_1 \, h_1}{1.917}{-5}{1.916}{-5}{1.916}{-5}{1.916}{-5}\\
\FRHDMtab{\tau^+ \, \tau^- \to h_1 \, h_2}{2.043}{-3}{2.043}{-3}{2.043}{-3}{2.043}{-3}\\
\FRHDMtab{\tau^+ \, \tau^- \to h_1 \, h_3}{2.043}{-3}{2.041}{-3}{2.042}{-3}{2.043}{-3}\\
\FRHDMtab{\tau^+ \, \tau^- \to h_2 \, h_2}{2.349}{-5}{2.349}{-5}{2.349}{-5}{2.348}{-5}\\
\FRHDMtab{\tau^+ \, \tau^- \to h_2 \, h_3}{2.260}{-4}{2.259}{-4}{2.259}{-4}{2.259}{-4}\\
\FRHDMtab{\tau^+ \, \tau^- \to h_3 \, h_3}{7.242}{-4}{7.240}{-4}{7.241}{-4}{7.240}{-4}\\
\FRHDMtab{\tau^+ \, \tau^- \to h^+ \, h^-}{1.345}{-2}{1.345}{-2}{1.345}{-2}{1.345}{-2}\\
\hline
\FRHDMtab{\tau^+ \, \tau^- \to Z \, h_1}{5.818}{-4}{5.824}{-4}{5.824}{-4}{5.821}{-4}\\
\FRHDMtab{\tau^+ \, \tau^- \to Z \, h_2}{8.012}{-3}{8.016}{-3}{8.012}{-3}{8.011}{-3}\\
\FRHDMtab{\tau^+ \, \tau^- \to Z \, h_3}{4.844}{-3}{4.842}{-3}{4.842}{-3}{4.842}{-3}\\
\FRHDMtab{\tau^+ \, \tau^- \to \gamma \, h_1}{2.915}{-3}{2.915}{-3}{2.914}{-3}{2.913}{-3}\\
\FRHDMtab{\tau^+ \, \tau^- \to \gamma \, h_2}{8.059}{-5}{8.059}{-5}{8.060}{-5}{8.058}{-5}\\
\FRHDMtab{\tau^+ \, \tau^- \to \gamma \, h_3}{4.347}{-3}{4.348}{-3}{4.347}{-3}{4.346}{-3}\\
\FRHDMtab{\tau^+ \, \tau^- \to W^- \, h^+}{2.038}{-3}{2.037}{-3}{2.037}{-3}{2.037}{-3}\\
\FRHDMtab{\tau^+ \, \tau^- \to W^+ \, h^-}{2.037}{-3}{2.039}{-3}{2.037}{-3}{2.037}{-3}\\
\hline
\FRHDMtab{\tau^+ \, \tau^- \to Z \, Z}{1.782}{-1}{1.784}{-1}{1.783}{-1}{1.784}{-1}\\
\FRHDMtab{\tau^+ \, \tau^- \to W^+ \, W^-}{3.017}{+0}{3.018}{+0}{3.018}{+0}{3.018}{+0}\\
\hline
\FRHDMtab{\tau^+ \, \tau^- \to \mu^+ \, \mu^-}{1.756}{-1}{1.753}{-1}{1.755}{-1}{1.755}{-1}\\
\FRHDMtab{\tau^+ \, \tau^- \to \mu^+ \, \tau^-}{1.453}{-8}{1.453}{-8}{1.453}{-8}{1.452}{-8}\\
\FRHDMtab{\tau^+ \, \tau^- \to \tau^+ \, \mu^-}{1.452}{-8}{1.453}{-8}{1.453}{-8}{1.452}{-8}\\
\FRHDMtab{\tau^+ \, \tau^- \to \tau^+ \, \tau^-}{7.423}{+2}{7.421}{+2}{7.421}{+2}{7.422}{+2}\\
\hline
\FRHDMtab{\tau^+ \, \tau^- \to c \, \bar t}{1.189}{-7}{1.189}{-7}{1.189}{-7}{1.189}{-7}\\
\FRHDMtab{\tau^+ \, \tau^- \to t \, \bar c}{1.189}{-7}{1.189}{-7}{1.189}{-7}{1.189}{-7}\\
\FRHDMtab{\tau^+ \, \tau^- \to t \, \bar t}{2.690}{-1}{2.686}{-1}{2.687}{-1}{2.688}{-1}\\
\FRHDMtab{\tau^+ \, \tau^- \to s \, \bar s}{1.431}{-1}{1.432}{-1}{1.431}{-1}{1.433}{-1}\\
\FRHDMtab{\tau^+ \, \tau^- \to s \, \bar b}{5.252}{+0}{5.252}{-9}{5.249}{-9}{5.242}{-9}\\
\FRHDMtab{\tau^+ \, \tau^- \to b \, \bar s}{5.248}{+0}{5.249}{-9}{5.249}{-9}{5.242}{-9}\\
\FRHDMtab{\tau^+ \, \tau^- \to b \, \bar b}{1.431}{-1}{1.431}{-1}{1.431}{-1}{1.431}{-1}\\
\hline
\end{tabular} \end{center}

%% file: TANU.tex
\renewcommand{\arraystretch}{1.3}
\newcommand{\FRHDMtab}[9]{$#1$ & $#2$ & $#3$ & $#4$ & $#5$ & $#6$ & $#7$& $#8$ & $#9$ }

\begin{center} \begin{tabular}{|l| r@{e}l  r@{e}l | r@{e}l  |r@{e}l  | }
 \multicolumn{9}{c}{\bf Neutrino processes in the Two-Higgs-Doublet Model}\\
 \hline
\multicolumn{1}{|c|}{Process} & \multicolumn{2}{c}{MG-FR} & \multicolumn{2}{c|}{MG-ST} & \multicolumn{2}{c|}{CH-FR} & \multicolumn{2}{c|}{SH-FR} \\
\hline \hline
\FRHDMtab{\tau^+ \, \nu_\tau \to h^+ \, h_1}{1.440}{-2}{1.439}{-2}{1.440}{-2}{1.441}{-2}\\
\FRHDMtab{\tau^+ \, \nu_\tau \to h^+ \, h_2}{4.277}{-3}{4.277}{-3}{4.278}{-3}{4.277}{-3}\\
\FRHDMtab{\tau^+ \, \nu_\tau \to h^+ \, h_3}{4.063}{-3}{4.066}{-3}{4.064}{-3}{4.065}{-3}\\
\FRHDMtab{\tau^+ \, \nu_\tau \to h^+ \, Z}{1.731}{-3}{1.731}{-3}{1.731}{-3}{1.731}{-3}\\
\FRHDMtab{\tau^+ \, \nu_\tau \to h^+ \, \gamma}{1.271}{-3}{1.271}{-3}{1.271}{-3}{1.271}{-3}\\
\hline
\FRHDMtab{\tau^+ \, \nu_\tau \to W^+ \, h_1}{9.814}{-4}{9.803}{-4}{9.808}{-4}{9.807}{-4}\\
\FRHDMtab{\tau^+ \, \nu_\tau \to W^+ \, h_2}{1.802}{-2}{1.801}{-2}{1.802}{-2}{1.802}{-2}\\
\FRHDMtab{\tau^+ \, \nu_\tau \to W^+ \, h_3}{1.076}{-2}{1.076}{-2}{1.077}{-2}{1.077}{-2}\\
\FRHDMtab{\tau^+ \, \nu_\tau \to W^+ \, Z}{2.250}{+0}{2.251}{+0}{2.251}{+0}{2.250}{+0}\\
\FRHDMtab{\tau^+ \, \nu_\tau \to W^+ \, \gamma}{1.439}{+0}{1.439}{+0}{1.439}{+0}{1.440}{+0}\\
\hline
\FRHDMtab{\tau^+ \, \nu_\tau \to e^+ \, \nu_e}{1.816}{-1}{1.816}{-1}{1.816}{-1}{1.816}{-1}\\
\FRHDMtab{\tau^+ \, \nu_\tau \to \mu^+ \, \nu_\mu}{1.816}{-1}{1.816}{-1}{1.816}{-1}{1.816}{-1}\\
\FRHDMtab{\tau^+ \, \nu_\tau \to \mu^+ \, \nu_\tau}{1.002}{-8}{1.002}{-8}{1.002}{-8}{1.002}{-8}\\
\FRHDMtab{\tau^+ \, \nu_\tau \to \tau^+ \, \nu_\tau}{9.179}{+0}{9.180}{+0}{9.180}{+0}{9.180}{+0}\\
\hline
\FRHDMtab{\tau^+ \, \nu_\tau \to u \, \bar d}{5.174}{-1}{5.175}{-1}{5.174}{-1}{5.175}{-1}\\
\FRHDMtab{\tau^+ \, \nu_\tau \to u \, \bar s}{2.733}{-2}{2.733}{-2}{2.733}{-2}{2.734}{-2}\\
\FRHDMtab{\tau^+ \, \nu_\tau \to c \, \bar d}{2.733}{-2}{2.734}{-2}{2.733}{-2}{2.734}{-2}\\
\FRHDMtab{\tau^+ \, \nu_\tau \to c \, \bar s}{5.175}{-1}{5.174}{-1}{5.174}{-1}{5.184}{-1}\\
\FRHDMtab{\tau^+ \, \nu_\tau \to c \, \bar b}{1.202}{-7}{1.202}{-7}{1.202}{-7}{1.202}{-7}\\ 
\FRHDMtab{\tau^+ \, \nu_\tau \to t \, \bar s}{4.363}{-9}{4.364}{-9}{4.364}{-9}{4.364}{-9}\\
\FRHDMtab{\tau^+ \, \nu_\tau \to t \, \bar b}{5.097}{-1}{5.097}{-1}{5.098}{-1}{5.099}{-1}\\
\hline
\end{tabular} \end{center}

%% file: VVSS.tex
\renewcommand{\arraystretch}{1.3}
\newcommand{\FRHDMtab}[9]{$#1$ & $#2$ & $#3$ & $#4$ & $#5$ & $#6$ & $#7$& $#8$ & $#9$ }

\begin{center} \begin{tabular}{|l| r@{e}l  r@{e}l | r@{e}l  | r@{e}l  | }
 \multicolumn{9}{c}{\bf Weak boson processes in the Two-Higgs-Doublet Model}\\
\hline
\multicolumn{1}{|c|}{Process} & \multicolumn{2}{c}{MG-FR} & \multicolumn{2}{c|}{MG-ST} & \multicolumn{2}{c|}{CH-FR}& \multicolumn{2}{c|}{SH-FR} \\
\hline \hline
\FRHDMtab{W^+ \, W^- \to W^+ \, W^-}{1.347}{+3}{1.346}{+3}{1.347}{+3}{1.347}{+3}\\
\FRHDMtab{W^+ \, W^- \to Z \, Z}{2.787}{+2}{2.774}{+2}{2.782}{+2}{2.782}{+2}\\
\FRHDMtab{W^+ \, W^- \to Z \, \gamma}{1.510}{+2}{1.513}{+2}{1.511}{+2}{1.509}{+2}\\
\FRHDMtab{Z \, Z \to Z \, Z}{1.616}{+1}{1.615}{+1}{1.616}{+1}{1.616}{+1}\\
\hline
\FRHDMtab{W^+ \, W^- \to W^+ \, h^-}{3.589}{-1}{3.586}{-1}{3.588}{-1}{3.587}{-1}\\
\FRHDMtab{W^+ \, W^- \to W^- \, h^+}{3.586}{-1}{3.589}{-1}{3.588}{-1}{3.587}{-1}\\
\FRHDMtab{W^+ \, W^- \to Z \, h_1}{1.788}{-1}{1.787}{-1}{1.787}{-1}{1.788}{-1}\\
\FRHDMtab{W^+ \, W^- \to Z \, h_2}{3.680}{+1}{3.685}{+1}{3.685}{+1}{3.684}{+1}\\
\FRHDMtab{W^+ \, W^- \to Z \, h_3}{2.130}{+1}{2.131}{+1}{2.130}{+1}{2.129}{+1}\\
\FRHDMtab{Z \, Z \to Z \, h_1}{5.934}{-1}{5.943}{-1}{5.940}{-1}{5.939}{-1}\\
\hline
\FRHDMtab{W^+ \, W^- \to h_1 \, h_1}{6.617}{-1}{6.620}{-1}{6.618}{-1}{6.617}{-1}\\
\FRHDMtab{W^+ \, W^- \to h_2 \, h_2}{4.455}{-1}{4.449}{-1}{4.453}{-1}{4.452}{-1}\\
\FRHDMtab{W^+ \, W^- \to h_2 \, h_3}{1.077}{+0}{1.076}{+0}{1.076}{+0}{1.076}{+0}\\
\FRHDMtab{W^+ \, W^- \to h^+ \, h^-}{2.446}{+0}{2.444}{+0}{2.443}{+0}{2.443}{+0}\\
\FRHDMtab{W^+ \, Z \to h^+ \, h_1}{2.888}{-1}{2.888}{-1}{2.887}{-1}{2.887}{-1}\\
\FRHDMtab{W^+ \, Z \to h^+ \, h_2}{1.485}{-2}{1.486}{-2}{1.485}{-2}{1.485}{-2}\\
\FRHDMtab{W^+ \, Z \to h^+ \, h_3}{3.014}{-2}{3.013}{-2}{3.012}{-2}{3.011}{-2}\\
\FRHDMtab{W^+ \, \gamma \to h^+ \, h_1}{1.970}{-2}{1.969}{-2}{1.969}{-2}{1.969}{-2}\\
\FRHDMtab{W^+ \, \gamma \to h^+ \, h_2}{7.925}{-3}{7.927}{-3}{7.926}{-3}{7.925}{-3}\\
\FRHDMtab{W^+ \, \gamma \to h^+ \, h_3}{1.692}{-2}{1.693}{-2}{1.693}{-2}{1.693}{-2}\\
\FRHDMtab{Z \, Z \to h_1 \, h_1}{2.333}{+0}{2.333}{+0}{2.335}{+0}{2.334}{+0}\\
\FRHDMtab{Z \, Z \to h_2 \, h_2}{6.636}{-1}{6.627}{-1}{6.630}{-1}{6.629}{-1}\\
\FRHDMtab{Z \, Z \to h_2 \, h_3}{1.065}{+0}{1.066}{+0}{1.066}{+0}{1.065}{+0}\\
\FRHDMtab{Z \, Z \to h^+ \, h^-}{1.356}{+0}{1.356}{+0}{1.356}{+0}{1.356}{+0}\\
\hline
\FRHDMtab{Z \, h_1 \to h_2 \, h_3}{2.929}{+0}{2.928}{+0}{2.930}{+0}{2.929}{+0}\\
\FRHDMtab{W^+ \, h_1 \to h^+ \, h_2}{4.732}{+0}{4.727}{+0}{4.731}{+0}{4.729}{+0}\\
\FRHDMtab{W^+ \, h_2 \to h^+ \, h_2}{2.998}{+0}{2.997}{+0}{2.999}{+0}{2.997}{+0}\\
\FRHDMtab{W^+ \, h^- \to h_2 \, h_3}{2.294}{+0}{2.292}{+0}{2.293}{+0}{2.292}{+0}\\
\hline
\end{tabular} \end{center}

%% file: mssm2.tex
\subsection{The most general Minimal Supersymmetric Standard Model}\label{App:SUSY}

In order to fully determine the MSSM Lagrangian at low energy scale, it is sufficient to fix the SM sector
 and the supersymmetry-breaking scenario. We have chosen the typical minimal supergravity point SPS 1a \cite{Allanach:2002nj} which is completely defined once we fix the values of four parameters at the gauge coupling unification scale and one sign. The complete set of input parameters is shown in Table~\ref{tab:MSSMparams}. We then use the numerical program {\sc SOFTSUSY}~\cite{Allanach:2001kg} to solve the renormalization group equations linking this restricted set of supersymmetry-breaking parameters at high-energy scale to the complete set of masses, mixing matrices and parameters appearing in $\lag_{\rm MSSM}$ at the weak scale. The output is encoded in a file following the SLHA conventions, readable by \feynrules\ after the use of an additional translation interface taking into account the small differences between the SLHA2 format and the one of our implementation. This interface is available on the \feynrules\ website, as well as an the corresponding SLHA2 output file. 

In Tables \ref{App:MSSMDec}, \ref{App:MSSM2to2Higgs}, \ref{App:MSSM2to2SUSY1} and \ref{App:MSSM2to2SUSY2}, we give some examples of the numerical checks which we have performed in order to validate the implementation of the MSSM in \feynrules. We recall that the built-in MSSM implementation in \madgraph\ and \calchep\ are denoted MG-ST and CH-ST, respectively, while the \feynrules-generated ones MG-FR and CH-FR. A $p_T$ cut of 20 GeV is applied to all final-state charged leptons, photons, and jets (including $b$-jets). The full list of results can be found on the \feynrules\ webpage.

\begin{table}[!ht]
\begin{center}
\begin{tabular}{rc l }
\hline\hline
Parameter & Symbol & Value\\
\hline
Inverse of the electromagnetic coupling  & $\alpha_{EW}^{-1}(M_Z)$ & 127.934\\
Strong coupling  & $\alpha_{s}(M_Z)$ & 0.1172\\
Fermi constant & $G_F$ & 1.16637e-5 GeV$^{-2}$\\
\hline
$Z$ mass & $M_Z$ & 91.1876\, GeV \\
$b$ quark mass & $m_b(m_b)$ & 4.25 GeV\\
$t$ quark mass & $m_t$ & 175 GeV\\
$\tau$ lepton mass & $m_\tau$ & 1.777 GeV\\
\hline
Universal scalar mass & $m_0$ & 100 GeV\\
Universal gaugino mass & $m_{1/2}$ & 250 GeV\\
Universal trilinear coupling & $A_0$ & -100 GeV\\
Ratio of the two vevs & $\tb$ & 10\\
Off-diagonal Higgs mixing parameter & ${\rm sign}(\mu)$ & +\\
\hline\hline
\end{tabular}
\caption{\label{tab:MSSMparams} Input parameters for the SPS 1a benchmark point for the MSSM.}
\end{center}
\end{table}

\newcommand{\tabbenj}[6]{$#1$ & #2 & #3 & $#4$ & #5 & #6\\}
\renewcommand{\arraystretch}{1.1}
\begin{table}[!h]
\begin{center}\begin{tabular}{|l|c c| |l|c c|}
\multicolumn{6}{c}{\bf Decay widths in the MSSM}\\
  \hline
  Process & MG-FR & MG-ST & Process & MG-FR & MG-ST\\
  \hline \hline
  \tabbenj{t   \,\to\, W^+ \,b   }{1.5608   }{1.5561   }{Z   \,\to\, d \,\bar d}{3.6625e-1}{3.6712e-1}
  \tabbenj{W^+ \,\to\, c \,\bar s}{6.6807e-1}{6.6830e-1}{h^0 \,\to\, b \,\bar b}{1.6268e-3}{1.6268e-3}
  \tabbenj{A^0 \,\to\, b \,\bar b}{4.5479e-1}{4.5479e-1}{H^+ \,\to\, t \,\bar b}{4.3833e-1}{4.3833e-1}
  \hline
  \tabbenj{\tilde u_1   \,\to\, \tilde\chi_1^+ \,b       }{1.3661   }{1.3661   }{\tilde u_2   \,\to\, \tilde\chi_1^0 \,u      }{1.1373   }{1.1373   }
  \tabbenj{\tilde d_4   \,\to\, \tilde\chi_1^0 \,d       }{2.8196e-1}{2.8196e-1}{\tilde d_5   \,\to\, \tilde\chi_1^- \,u      }{3.2220   }{3.2220   }
  \tabbenj{\tilde \nu_1 \,\to\, \tilde\chi_1^0 \,\nu_\tau}{1.4545e-1}{1.4545e-1}{\tilde l_3   \,\to\, \tilde\chi_1^0 \,\mu^-  }{2.1612e-1}{2.1612e-1}
  \hline
  \tabbenj{\tilde g        \,\to\, \tilde\d_1^\dagger \,b       }{5.5408e-1}{5.5370e-1}{\tilde \chi_2^0 \,\to\, \tilde l_1^-    \,\tau^+}{9.1581e-3}{9.1507e-3}
  \tabbenj{\tilde \chi_3^0 \,\to\, \tilde\chi_1^+     \,W^-     }{5.6624e-1}{5.6460e-1}{\tilde \chi_4^0 \,\to\, \tilde\chi_1^+  \,W^-   }{6.4519e-1}{6.4576e-1}
  \tabbenj{\tilde \chi_1^+ \,\to\, \tilde l_1^+       \,\nu_\tau}{1.5768e-2}{1.5748e-2}{\tilde \chi_2^+ \,\to\, \tilde\chi_2^0  \,W^+   }{7.2945e-1}{7.2998e-1}
  \hline
\end{tabular}\end{center}
\caption{\label{App:MSSMDec}Widths (in GeV) of some of the allowed decay channels in the SPS 1a scenario. MG-FR amd MG-ST denote the \feynrules-generated and built-in \madgraph\ implementations.}
\end{table}

\newcommand{\FRtab}[9]{$#1$ & $#2$ & $#3$ & $#4$ & $#5$ & $#6$ & $#7$ & $#8$ & $#9$ }
\begin{table}
\begin{center} \begin{tabular}{|l| r@{e}l  r@{e}l | r@{e}l  r@{e}l | }
\multicolumn{9}{c}{\bf Higgs production in the MSSM}\\
\hline
\multicolumn{1}{|c|}{Process} & \multicolumn{2}{c}{MG-FR} & \multicolumn{2}{c|}{MG-ST} & \multicolumn{2}{c}{CH-FR} & \multicolumn{2}{c|}{CH-ST} \\
\hline \hline
\FRtab{e^+ \, e^- \to Z \, h^0}{8.787}{-3}{8.788}{-3}{8.787}{-3}{8.787}{-3}\\
\FRtab{e^+ \, e^- \to H^+ \, H^-}{8.121}{-3}{8.121}{-3}{8.119}{-3}{8.119}{-3}\\
\FRtab{\tau^+ \, \tau^- \to h^0 \, H^0}{1.610}{-5}{1.610}{-5}{1.610}{-5}{1.610}{-5}\\
\FRtab{\tau^+ \, \tau^- \to A^0 \, h^0}{1.741}{-5}{1.741}{-5}{1.741}{-5}{1.741}{-5}\\
\FRtab{\tau^- \, \bar \nu_\tau \to H^- \, h^0}{6.245}{-6}{6.245}{-6}{6.244}{-6}{6.243}{-6}\\
\FRtab{\tau^- \, \bar \nu_\tau \to W^- \, A^0}{1.810}{-2}{1.811}{-2}{1.810}{-2}{1.810}{-2}\\
\FRtab{\tau^- \, \bar \nu_\tau \to Z \, H^-}{3.125}{-2}{3.123}{-2}{3.124}{-2}{3.124}{-2}\\
\hline
\FRtab{u \, \bar u \to Z \, h^0}{3.331}{-3}{3.331}{-3}{3.331}{-3}{3.331}{-3}\\
\FRtab{d \, \bar d \to Z \, H^0}{4.213}{-7}{4.215}{-7}{4.214}{-7}{4.214}{-7}\\
\FRtab{b \, \bar b \to A^0 \, A^0}{7.221}{-5}{7.218}{-5}{7.220}{-5}{7.214}{-5}\\
\FRtab{b \, \bar b \to H^+ \, H^-}{9.240}{-4}{9.237}{-4}{9.237}{-4}{9.237}{-4}\\
\FRtab{b \, \bar t \to H^- \, H^0}{2.070}{-3}{2.070}{-3}{2.069}{-3}{2.069}{-3}\\
\FRtab{b \, \bar t \to Z \, H^-}{2.590}{-2}{2.587}{-2}{2.592}{-2}{2.592}{-2}\\
\hline
\FRtab{W^+ \, W^- \to h^0 \, H^0}{1.109}{-3}{1.110}{-3}{1.110}{-3}{1.110}{-3}\\
\FRtab{W^+ \, W^- \to Z \, h^0}{8.217}{+1}{8.216}{+1}{8.213}{+1}{8.213}{+1}\\
\FRtab{W^+ \, W^- \to H^+ \, H^-}{3.689}{-2}{3.686}{-2}{3.685}{-2}{3.685}{-2}\\
\FRtab{Z \, Z \to h^0 \, h^0}{7.829}{+0}{7.827}{+0}{7.827}{+0}{7.827}{+0}\\
\FRtab{Z \, \gamma \to H^+ \, H^-}{1.126}{-2}{1.124}{-2}{1.124}{-2}{1.124}{-2}\\
\FRtab{W^- \, Z \to W^- \, A^0}{4.375}{-4}{4.387}{-4}{4.378}{-4}{4.378}{-4}\\
\FRtab{W^- \, \gamma \to W^- \, h^0}{1.588}{+1}{1.589}{+1}{1.589}{+1}{1.589}{+1}\\
\hline
\end{tabular}
\end{center}
\caption{\label{App:MSSM2to2Higgs}Cross sections for a selection of Higgs production processes in the MSSM scenario SPS 1a. The built-in MSSM implementation in \madgraph\ and \calchep\ are denoted MG-ST and CH-ST, respectively, while the \feynrules-generated ones are MG-FR and CH-FR. The center-of-mass energy is fixed to 1200 GeV.}
\end{table}

\begin{table}
\begin{center} \begin{tabular}{|l| r@{e}l  r@{e}l | r@{e}l  r@{e}l | }
\multicolumn{9}{c}{\bf Supersymmetric particle production from fermions}\\
\hline
\multicolumn{1}{|c|}{Process} & \multicolumn{2}{c}{MG-FR} & \multicolumn{2}{c|}{MG-ST} & \multicolumn{2}{c}{CH-FR} & \multicolumn{2}{c|}{CH-ST} \\
\hline \hline
\FRtab{e^+ \, e^- \to \tilde{l}_2^- \, \tilde{l}_2^+}{1.944}{-1}{1.944}{-1}{1.944}{-1}{1.944}{-1}\\
\FRtab{e^+ \, e^- \to \tilde{\nu}_3 \, \tilde{\nu}_3^\ast}{4.862}{-1}{4.863}{-1}{4.862}{-1}{4.863}{-1}\\
\FRtab{e^+ \, e^- \to \tilde{u}_4 \, \tilde{u}_4^\ast}{1.664}{-3}{1.662}{-3}{1.663}{-3}{1.663}{-3}\\
\FRtab{e^+ \, e^- \to \tilde{d}_1 \, \tilde{d}_2^\ast}{2.858}{-4}{2.857}{-4}{2.858}{-4}{2.858}{-4}\\
\FRtab{\tau^+ \, \tau^- \to \tilde{l}_1^+ \, \tilde{l}_6^-}{4.332}{-2}{4.326}{-2}{4.329}{-2}{4.382}{-2}\\
\FRtab{\tau^- \, \bar \nu_\tau \to \tilde{l}_6^- \, \tilde{\nu}_1^\ast}{1.206}{-1}{1.206}{-1}{1.206}{-1}{1.206}{-1}\\
\hline
\FRtab{e^+ \, e^- \to \tilde{\chi}^0_2 \, \tilde{\chi}^0_2}{4.334}{-2}{4.334}{-2}{4.330}{-2}{4.330}{-2}\\
\FRtab{e^+ \, e^- \to \tilde{\chi}^+_1 \, \tilde{\chi}^-_1}{1.023}{-1}{1.023}{-1}{1.023}{-1}{1.023}{-1}\\
\FRtab{\tau^+ \, \tau^- \to \tilde{\chi}^0_3 \, \tilde{\chi}^0_3}{9.280}{-5}{9.267}{-5}{9.272}{-5}{9.271}{-5}\\
\FRtab{\tau^+ \, \tau^- \to \tilde{\chi}^+_1 \, \tilde{\chi}^-_2}{1.169}{-2}{1.170}{-2}{1.170}{-2}{1.170}{-2}\\
\FRtab{e^- \, \bar \nu_e \to \tilde{\chi}^-_1 \, \tilde{\chi}^0_1}{5.445}{-2}{5.444}{-2}{5.446}{-2}{5.446}{-2}\\
\FRtab{\tau^- \, \bar \nu_\tau \to \tilde{\chi}^-_2 \, \tilde{\chi}^0_4}{5.306}{-2}{5.313}{-2}{5.306}{-2}{5.306}{-2}\\
\hline
\FRtab{u \, \bar u \to \tilde{l}_3^- \, \tilde{l}_3^+}{2.123}{-3}{2.123}{-3}{2.123}{-3}{2.123}{-3}\\
\FRtab{u \, \bar u \to \tilde{u}_5 \, \tilde{u}_2^\ast}{6.141}{-1}{6.142}{-1}{6.141}{-1}{6.141}{-1}\\
\FRtab{d \, \bar d \to \tilde{\nu}_2 \, \tilde{\nu}_2^\ast}{3.607}{-3}{3.606}{-3}{3.607}{-3}{3.607}{-3}\\
\FRtab{d \, \bar d \to \tilde{d}_4 \, \tilde{d}_4^\ast}{1.306}{-1}{1.307}{-1}{1.307}{-1}{1.307}{-1}\\
\FRtab{b \, \bar b \to \tilde{d}_1 \, \tilde{d}_1^\ast}{3.403}{-1}{3.401}{-1}{3.401}{-1}{3.401}{-1}\\
\FRtab{b \, \bar b \to \tilde{u}_6 \, \tilde{u}_6^\ast}{5.913}{-3}{5.915}{-3}{5.912}{-3}{5.270}{-3}\\
\FRtab{b \, \bar t \to \tilde{l}_6^- \, \tilde{\nu}_1^\ast}{1.114}{-2}{1.114}{-2}{1.115}{-2}{1.115}{-2}\\
\FRtab{b \, \bar t \to \tilde{d}_2 \, \tilde{u}_6^\ast}{4.417}{-1}{4.419}{-1}{4.420}{-1}{4.420}{-1}\\
\hline
\FRtab{u \, \bar u \to \tilde{\chi}^0_1 \, \tilde{\chi}^0_4}{1.272}{-4}{1.272}{-4}{1.272}{-4}{1.272}{-4}\\
\FRtab{u \, \bar u \to \tilde{\chi}^+_2 \, \tilde{\chi}^-_2}{1.321}{-2}{1.317}{-2}{1.319}{-2}{1.320}{-2}\\
\FRtab{b \, \bar b \to \tilde{\chi}^0_3 \, \tilde{\chi}^0_4}{1.087}{-2}{1.087}{-2}{1.085}{-2}{1.085}{-2}\\
\FRtab{b \, \bar b \to \tilde{\chi}^+_2 \, \tilde{\chi}^-_2}{9.556}{-2}{9.545}{-2}{9.556}{-2}{9.556}{-2}\\
\FRtab{b \, \bar t \to \tilde{\chi}^-_1 \, \tilde{\chi}^0_2}{1.556}{-2}{1.557}{-2}{1.557}{-2}{1.557}{-2}\\
\FRtab{b \, \bar t \to \tilde{\chi}^-_2 \, \tilde{\chi}^0_3}{3.981}{-2}{3.971}{-2}{3.977}{-2}{3.977}{-2}\\
\hline
\end{tabular} \end{center}
\caption{\label{App:MSSM2to2SUSY1}Cross sections for a selection of supersymmetric particle pair production processes in the MSSM scenario SPS 1a. The built-in MSSM implementation in \madgraph\ and \calchep\ are denoted MG-ST and CH-ST, respectively, while the \feynrules-generated ones are MG-FR and CH-FR. The center-of-mass energy is fixed to 1200 GeV.}
\end{table}

\begin{table}
\begin{center} \begin{tabular}{|l| r@{e}l  r@{e}l | r@{e}l  r@{e}l | }
\multicolumn{9}{c}{\bf Supersymmetric particle production from gauge bosons}\\
\hline
\multicolumn{1}{|c|}{Process} & \multicolumn{2}{c}{MG-FR} & \multicolumn{2}{c|}{MG-ST} & \multicolumn{2}{c}{CH-FR} & \multicolumn{2}{c|}{CH-ST} \\
\hline \hline
\FRtab{W^+ \, W^- \to \tilde{l}_1^- \, \tilde{l}_1^+}{2.289}{-3}{2.290}{-3}{2.289}{-3}{1.682}{-2}\\
\FRtab{W^+ \, W^- \to \tilde{\nu}_3 \, \tilde{\nu}_3^\ast}{5.561}{-2}{5.564}{-2}{5.562}{-2}{5.562}{-2}\\
\FRtab{W^+ \, W^- \to \tilde{u}_1 \, \tilde{u}_6^\ast}{5.338}{-2}{5.349}{-2}{5.344}{-2}{9.183}{-1}\\
\FRtab{Z \, Z \to \tilde{\nu}_1 \, \tilde{\nu}_1^\ast}{7.678}{-2}{7.686}{-2}{7.686}{-2}{7.686}{-2}\\
\FRtab{Z \, Z \to \tilde{d}_5 \, \tilde{d}_5^\ast}{4.693}{-2}{4.695}{-2}{4.693}{-2}{4.693}{-2}\\
\FRtab{Z \, \gamma \to \tilde{u}_1 \, \tilde{u}_6^\ast}{3.283}{-2}{3.285}{-2}{3.286}{-2}{3.286}{-2}\\
\FRtab{Z \, \gamma \to \tilde{l}_2^- \, \tilde{l}_2^+}{1.712}{-2}{1.711}{-2}{1.712}{-2}{1.712}{-2}\\
\FRtab{W^- \, Z \to \tilde{l}_5^- \, \tilde{\nu}_3^\ast}{3.952}{-2}{3.941}{-2}{3.950}{-2}{3.950}{-2}\\
\FRtab{W^- \, Z \to \tilde{d}_6 \, \tilde{u}_4^\ast}{2.690}{-2}{2.689}{-2}{2.690}{-2}{2.690}{-2}\\
\FRtab{W^- \, \gamma \to \tilde{d}_2 \, \tilde{u}_6^\ast}{3.618}{-4}{3.618}{-4}{3.618}{-4}{3.618}{-4}\\
\FRtab{g \, \gamma \to \tilde{u}_1 \, \tilde{u}_1^\ast}{1.129}{-1}{1.129}{-1}{1.129}{-1}{1.129}{-1}\\
\FRtab{g \, Z \to \tilde{d}_4 \, \tilde{d}_4^\ast}{4.637}{-3}{4.633}{-3}{4.634}{-3}{4.634}{-3}\\
\FRtab{g \, W^- \to \tilde{d}_5 \, \tilde{u}_5^\ast}{2.569}{-1}{2.566}{-1}{2.566}{-1}{2.566}{-1}\\
\FRtab{g \, W^+ \to \tilde{d}_2^\ast \, \tilde{u}_6}{2.208}{-2}{2.206}{-2}{2.206}{-2}{2.206}{-2}\\
\FRtab{g \, g \to \tilde{u}_3 \, \tilde{u}_3^\ast}{1.865}{-1}{1.865}{-1}{1.866}{-1}{1.866}{-1}\\
\hline
\FRtab{W^+ \, W^- \to \tilde{\chi}^0_2 \, \tilde{\chi}^0_3}{6.514}{-1}{6.515}{-1}{6.515}{-1}{6.515}{-1}\\
\FRtab{W^+ \, W^- \to \tilde{\chi}^+_1 \, \tilde{\chi}^-_1}{1.836}{+0}{1.837}{+0}{1.835}{+0}{1.835}{+0}\\
\FRtab{\gamma \, \gamma \to \tilde{\chi}^+_1 \, \tilde{\chi}^-_1}{6.250}{-1}{6.263}{-1}{6.257}{-1}{6.257}{-1}\\
\FRtab{W^- \, Z \to \tilde{\chi}^0_4 \, \tilde{\chi}^-_1}{4.738}{-1}{4.751}{-1}{4.746}{-1}{4.746}{-1}\\
\FRtab{W^- \, \gamma \to \tilde{\chi}^0_2 \, \tilde{\chi}^-_2}{5.235}{-2}{5.235}{-2}{5.236}{-2}{5.236}{-2}\\
\hline
\end{tabular}
\end{center}
\caption{\label{App:MSSM2to2SUSY2}Cross sections for a selection of supersymmetric particle pair production processes in the MSSM scenario SPS 1a. The built-in MSSM implementation in \madgraph\ and \calchep\ are denoted MG-ST and CH-ST, respectively, while the \feynrules-generated ones are MG-FR and CH-FR. The center-of-mass energy is fixed to 1200 GeV.}
\end{table}

\clearpage

%% file: 3-site2.tex
\subsection{\label{sec:validation:3-Site}The Minimal Higgsless Model}

The external parameters used for the validation of the Minimal Higgsless Model are shown in Table~\ref{tab:MHMparams}, and all the widths are set to zero. We employ a center-of-mass energy of 600 GeV and a $p_T$ cut of 20 GeV if only SM particles are present, a center-of-mass energy of 1200 GeV and a $p_T$ cut of 200 GeV if heavy vector bosons are present but heavy fermions not, and a center-of-mass energy of 10000 GeV and a $p_T$ cut of 2000 GeV if heavy fermions are present.

\begin{table}[!t]
\begin{center}
\begin{tabular}{rc l }
\hline\hline
Parameter & Symbol & Value\\
\hline
Inverse of the electromagnetic coupling  & $\alpha_{EW}^{-1}(M_Z)$ & 127.9\\
Strong coupling  & $\alpha_{s}(M_Z)$ & 0.1172\\
Fermi constant & $G_F$ & 1.16637e-5 GeV$^{-2}$\\
\hline
$Z$ mass & $M_Z$ & 91.1876\, GeV \\
$W$ mass & $M_W$ & 80.398 GeV\\
$W'$ mass & $M_{W'}$ & 500 GeV\\
Heavy fermion mass & $M_F$ & 4 TeV\\
\hline\hline
\end{tabular}
\caption{\label{tab:MHMparams} Input parameters for the MHM.}
\end{center}
\end{table}

\begin{table}
\begin{center} 
\begin{tabular}{|c|l|ll|l|l|}
\multicolumn{6}{c}{\bf Strong Processes in the Minimal Higgsless Model}\\
\hline
 Process & CH-LH-F & CH-FR-F & CH-FR-U & SH-FR-U & MG-FR-U \\
\hline\hline
$G,G\to G,G$ & 1.143e+5 & 1.143e+5 & 1.143e+5 & 1.143e+5 & 1.143e+5 \\
\hline
$u,\bar{u}\to G,G$ & 1.705e+2 & 1.705e+2 & 1.705e+2 & 1.705e+2 & 1.706e+2 \\
$U,\bar{u}\to G,G$ & 0.000e-1 & 0.000e-1 & 0.000e-1 & 0.000e-1 & 0.000e-1 \\
$U,\bar{U}\to G,G$ & 8.696e-2 & 8.696e-2 & 8.696e-2 & 8.696e-2 & 8.690e-2 \\
\hline
$u,\bar{u}\to t,\bar{t}$ & 8.116e+0 & 8.116e+0 & 8.116e+0 & 8.117e+0 & 8.100e+0 \\
$U,\bar{u}\to t,\bar{t}$ & 6.464e-2 & 6.464e-2 & 6.464e-2 & 6.467e-2 & 6.466e-2 \\
$U,\bar{u}\to T,\bar{t}$ & 6.592e-1 & 6.592e-1 & 6.592e-1 & 6.594e-1 & 6.597e-1 \\
$U,\bar{u}\to t,\bar{T}$ & 6.592e-1 & 6.592e-1 & 6.592e-1 & 6.594e-1 & 6.596e-1 \\
$U,\bar{U}\to t,\bar{t}$ & 5.033e-2 & 5.033e-2 & 5.033e-2 & 5.035e-2 & 5.044e-2 \\
$U,\bar{U}\to T,\bar{t}$ & 1.366e-1 & 1.366e-1 & 1.366e-1 & 1.367e-1 & 1.365e-1 \\
$U,\bar{U}\to T,\bar{T}$ & 1.589e-1 & 1.589e-1 & 1.589e-1 & 1.589e-1 & 1.589e-1 \\
\hline
\end{tabular}
\end{center}
\caption{\label{3-Site-validation:strong}Cross sections for a selection of strong processes in the Minimal Higgsless Model. 
The \lanhep-generated MHM implementation in \calchep\ is denoted CH-LH, while the \feynrules-generated ones are MG-FR, CH-FR and SH-FR. F means the calculation was done in Feynman gauge while U means it was done in unitary gauge The center-of-mass energy is fixed to 600 GeV and a $p_T$ cut of 20 GeV is applied to each final state particle, if only SM particles are present. A center-of-mass energy of 1200 GeV and a $p_T$ cut of 200 GeV is used if heavy vector bosons are present but heavy fermions not, and a center-of-mass energy of 10000 GeV and a $p_T$ cut of 2000 GeV if heavy fermions are present.}
\end{table}


\begin{table}
\begin{center} 
\begin{tabular}{|c|l|ll|l|l|}
\multicolumn{6}{c}{\bf Charged Electroweak Processes in the Minimal Higgsless Model}\\
\hline
Process & CH-LH-F & CH-FR-F & CH-FR-U & SH-FR-U & MG-FR-U \\
\hline\hline
$Z,W^{+}\to W^{+},Z$ & 3.121e+2 & 3.121e+2 & 3.121e+2 & 3.119e+2 & 3.118e+2 \\
$W^{'+},Z\to W^{+},Z$ & 4.084e+0 & 4.084e+0 & 4.084e+0 & 4.077e+0 & 4.080e+0 \\
$Z^{'},W^{+}\to W^{+},Z$ & 4.214e+0 & 4.214e+0 & 4.214e+0 & 4.210e+0 & 4.214e+0 \\
$Z^{'},W^{'+}\to W^{+},Z$ & 3.072e+1 & 3.072e+1 & 3.072e+1 & 3.071e+1 & 3.074e+1 \\
$W^{'+},Z\to W^{'+},Z$ & 2.289e+1 & 2.289e+1 & 2.289e+1 & 2.289e+1 & 2.288e+1 \\
$W^{'+},Z\to W^{+},Z^{'}$ & 1.303e+2 & 1.303e+2 & 1.303e+2 & 1.302e+2 & 1.304e+2 \\
$Z^{'},W^{'+}\to W^{'+},Z$ & 8.098e+0 & 8.098e+0 & 8.098e+0 & 8.095e+0 & 8.111e+0 \\
$Z^{'},W^{'+}\to W^{+},Z^{'}$ & 1.914e+1 & 1.914e+1 & 1.914e+1 & 1.913e+1 & 1.915e+1 \\
$Z^{'},W^{'+}\to W^{'+},Z^{'}$ & 6.967e+2 & 6.967e+2 & 6.967e+2 & 6.963e+2 & 6.965e+2 \\
\hline
$u,\bar{d}\to Z,W+$ & 1.112e+0 & 1.112e+0 & 1.112e+0 & 1.110e+0 & 1.112e+0 \\
$u,\bar{d}\to Z,W^{'+}$ & 8.639e-2 & 8.639e-2 & 8.639e-2 & 8.638e-2 & 8.647e-2 \\
$u,\bar{d}\to Z^{'},W+$ & 8.463e-2 & 8.463e-2 & 8.463e-2 & 8.460e-2 & 8.456e-2 \\
$U,\bar{d}\to Z,W+$ & 7.987e-2 & 7.987e-2 & 7.987e-2 & 7.986e-2 & 7.996e-2 \\
$U,\bar{D}\to Z,W+$ & 1.794e+0 & 1.794e+0 & 1.794e+0 & 1.793e+0 & 1.793e+0 \\
$U,\bar{d}\to Z^{'},W+$ & 1.464e-1 & 1.464e-1 & 1.464e-1 & 1.464e-1 & 1.463e-1 \\
$U,\bar{d}\to Z,W^{'+}$ & 1.528e-1 & 1.528e-1 & 1.528e-1 & 1.529e-1 & 1.529e-1 \\
$U,\bar{D}\to Z^{'},W+$ & 2.499e+0 & 2.499e+0 & 2.499e+0 & 2.499e+0 & 2.499e+0 \\
$U,\bar{D}\to Z,W^{'+}$ & 2.564e+0 & 2.564e+0 & 2.564e+0 & 2.563e+0 & 2.562e+0 \\
$U,\bar{d}\to Z^{'},W^{'+}$ & 2.512e-1 & 2.512e-1 & 2.512e-1 & 2.512e-1 & 2.513e-1 \\
$U,\bar{D}\to Z^{'},W^{'+}$ & 2.287e+0 & 2.287e+0 & 2.287e+0 & 2.286e+0 & 2.286e+0 \\
\hline
$e,\bar{\nu}_1\to b,\bar{t}$ & 9.514e-1 & 9.514e-1 & 9.514e-1 & 9.515e-1 & 9.514e-1 \\
$e,\bar{\nu}_1\to B,\bar{t}$ & 5.346e-5 & 5.346e-5 & 5.346e-5 & 5.347e-5 & 5.339e-5 \\
$E^-,\bar{\nu}_1\to B,\bar{t}$ & 8.287e+0 & 8.287e+0 & 8.287e+0 & 8.289e+0 & 8.287e+0 \\
$e,\bar{\nu}_1\to B,\bar{T}$ & 1.106e-3 & 1.106e-3 & 1.106e-3 & 1.106e-3 & 1.107e-3 \\
$E^-,\bar{N}_1\to b,\bar{t}$ & 8.071e-2 & 8.071e-2 & 8.071e-2 & 8.075e-2 & 8.054e-2 \\
$E^-,\bar{\nu}_1\to B,\bar{T}$ & 5.904e-1 & 5.904e-1 & 5.904e-1 & 5.906e-1 & 5.905e-1 \\
$E^-,\bar{N}_1\to B,\bar{t}$ & 1.717e+0 & 1.717e+0 & 1.717e+0 & 1.717e+0 & 1.715e+0 \\
$E^-,\bar{N}_1\to B,\bar{T}$ & 1.732e+0 & 1.732e+0 & 1.732e+0 & 1.733e+0 & 1.733e+0 \\
\hline
\end{tabular}
\end{center}
\caption{\label{3-Site-validation:ffAW}Cross sections for a selection of charged electroweak processes in the Minimal Higgsless Model. 
The \lanhep-generated MHM implementation in \calchep\ is denoted CH-LH, while the \feynrules-generated ones are MG-FR, CH-FR and SH-FR. F means the calculation was done in Feynman gauge while U means it was done in unitary gauge The center-of-mass energy is fixed to 600 GeV and a $p_T$ cut of 20 GeV is applied to each final state particle, if only SM particles are present. A center-of-mass energy of 1200 GeV and a $p_T$ cut of 200 GeV is used if heavy vector bosons are present but heavy fermions not, and a center-of-mass energy of 10000 GeV and a $p_T$ cut of 2000 GeV if heavy fermions are present.}
\end{table}


\begin{table}
\begin{center} 
\begin{tabular}{|c|l|ll|l|l|}
\multicolumn{6}{c}{\bf Neutral Electroweak Processes in the Minimal Higgsless Model}\\
\hline
Process & CH-LH-F & CH-FR-F & CH-FR-U & SH-FR-U & MG-FR-U \\
\hline\hline
$W^{+},W^{-}\to W^{+},W^{-}$ & 1.407e+3 & 1.407e+3 & 1.407e+3 & 1.407e+3 & 1.407e+3 \\
$W^{+},W^{-}\to W^{+},W^{'-}$ & 2.917e+0 & 2.917e+0 & 2.917e+0 & 2.917e+0 & 2.918e+0 \\
$W^{+},W^{'-}\to W^{+},W^{-}$ & 5.313e+0 & 5.313e+0 & 5.313e+0 & 5.313e+0 & 5.316e+0 \\
$W^{+},W^{-}\to W^{'+},W^{'-}$ & 7.078e+0 & 7.078e+0 & 7.078e+0 & 7.074e+0 & 7.078e+0 \\
$W^{+},W^{'-}\to W^{+},W^{'-}$ & 1.406e+2 & 1.406e+2 & 1.406e+2 & 1.405e+2 & 1.404e+2 \\
$W^{'+},W^{'-}\to W^{+},W^{-}$ & 3.123e+1 & 3.123e+1 & 3.123e+1 & 3.123e+1 & 3.126e+1 \\
$W^{'+},W^{'-}\to W^{+},W^{'-}$ & 1.948e+1 & 1.948e+1 & 1.948e+1 & 1.947e+1 & 1.944e+1 \\
$W^{+},W^{'-}\to W^{'+},W^{'-}$ & 6.991e+0 & 6.991e+0 & 6.991e+0 & 6.995e+0 & 6.974e+0 \\
$W^{'+},W^{'-}\to W^{'+},W^{'-}$ & 7.245e+2 & 7.245e+2 & 7.245e+2 & 7.241e+2 & 7.256e+2 \\
\hline
$u,\bar{u}\to Z,Z$ & 1.957e-1 & 1.957e-1 & 1.957e-1 & 1.958e-1 & 1.955e-1 \\
$u,\bar{u}\to Z,Z^{'}$ & 9.367e-5 & 9.367e-5 & 9.367e-5 & 9.362e-5 & 9.368e-5 \\
$U,\bar{u}\to Z,Z$ & 4.497e+0 & 4.497e+0 & 4.497e+0 & 4.497e+0 & 4.494e+0 \\
$U,\bar{U}\to Z,Z$ & 6.302e-1 & 6.302e-1 & 6.302e-1 & 6.302e-1 & 6.304e-1 \\
$U,\bar{u}\to Z,Z^{'}$ & 9.733e+0 & 9.733e+0 & 9.733e+0 & 9.732e+0 & 9.744e+0 \\
$u,\bar{u}\to Z^{'},Z^{'}$ & 1.196e-5 & 1.196e-5 & 1.196e-5 & 1.196e-5 & 1.196e-5 \\
$U,\bar{U}\to Z,Z^{'}$ & 1.371e+0 & 1.371e+0 & 1.371e+0 & 1.371e+0 & 1.371e+0 \\
$U,\bar{u}\to Z^{'},Z^{'}$ & 5.268e+0 & 5.268e+0 & 5.268e+0 & 5.268e+0 & 5.266e+0 \\
$U,\bar{U}\to Z^{'},Z^{'}$ & 8.832e-1 & 8.832e-1 & 8.832e-1 & 8.832e-1 & 8.828e-1 \\
\hline
$e,\bar{e}\to t,\bar{t}$ & 4.480e-1 & 4.480e-1 & 4.480e-1 & 4.480e-1 & 4.478e-1 \\
$e,\bar{e}\to T,\bar{t}$ & 1.388e-5 & 1.388e-5 & 1.388e-5 & 1.388e-5 & 1.389e-5 \\
$E^-,\bar{e}\to T,\bar{t}$ & 1.978e+0 & 1.978e+0 & 1.978e+0  & 1.978e+0 & 1.978e+0 \\
$e,\bar{e}\to T,\bar{T}$ & 4.093e-4 & 4.093e-4 & 4.093e-4 & 4.093e-4 & 4.093e-4 \\
$E^-,\bar{E}\to t,\bar{t}$ & 4.155e-2 & 4.155e-2 & 4.155e-2 & 4.157e-2 & 4.153e-2 \\
$E^-,\bar{e}\to T,\bar{T}$ & 1.359e-1 & 1.359e-1 & 1.359e-1 & 1.359e-1 & 1.359e-1 \\
$E^-,\bar{E}\to T,\bar{t}$ & 4.097e-1 & 4.097e-1 & 4.097e-1 & 4.098e-1 & 4.095e-1 \\
$E^-,\bar{E}\to T,\bar{T}$ & 4.207e-1 & 4.207e-1 & 4.207e-1 & 4.207e-1 & 4.202e-1 \\
\hline
\end{tabular}
\end{center}
\caption{\label{3-Site-validation:ffAA}Cross sections for a selection of neutral electroweak processes in the Minimal Higgsless Model. 
The \lanhep-generated MHM implementation in \calchep\ is denoted CH-LH, while the \feynrules-generated ones are MG-FR, CH-FR and SH-FR. F means the calculation was done in Feynman gauge while U means it was done in unitary gauge The center-of-mass energy is fixed to 600 GeV and a $p_T$ cut of 20 GeV is applied to each final state particle, if only SM particles are present. A center-of-mass energy of 1200 GeV and a $p_T$ cut of 200 GeV is used if heavy vector bosons are present but heavy fermions not, and a center-of-mass energy of 10000 GeV and a $p_T$ cut of 2000 GeV if heavy fermions are present.}
\end{table}

\clearpage

%% file: ued2.tex
\subsection{Extra Dimensional models}
\label{App:MUED}

\begin{table}
\begin{center}
\begin{tabular}{rc l }
\hline\hline
Parameter & Symbol & Value\\
\hline
Inverse of the electromagnetic coupling 		& $\alpha_{EW}^{-1}(M_Z)$ & 128 \\
Strong coupling 						& $\alpha_{s}(M_Z)$ 	& 0.1172\\
Fermi constant 							& $G_F$ 			        & 1.16639e-5 GeV$^{-2}$\\
\hline
$Z$ pole mass				 			& $M_Z$ 			        & 91.1876\, GeV \\
$c$ quark mass 						& $m_c$ 		         	& 1.42 GeV\\
$b$ quark mass 						& $m_b$ 		         	& 4.2 GeV\\
$t$ quark mass 						& $m_t$ 		         	& 175 GeV\\
$\tau$ lepton mass 						& $m_\tau$ 			& 1.777 GeV\\
Higgs mass                                                         & $m_H$ 				& 120 GeV\\
Sine of the electroweak mixing angle		 	&  $s_{w}$  	         	& 0.48076\\
Parameter of the CKM matrix				&  $s_{12}$			& 0.221\\
Parameter of the CKM matrix				&  $s_{23}$			& 0.041\\
Parameter of the CKM matrix				&  $s_{13}$			& 0.0035\\
\hline
Extra dimensional radius					& $R$				& 0.002 $\GeV^{-1}$\\
Cutoff scale							& $\Lambda$			& 10  $\TeV$\\
\hline \hline
\end{tabular}
\caption{\label{tab:muedparams} Input parameters for our \mued\ benchmark scenario.}
\end{center}
\end{table}

The benchmark point which we have used for our numerical comparison of the \mued\ implementation in \feynrules\ to the existing one in \calchep\ is defined through the various external parameters shown in Table \ref{tab:muedparams}. The masses of the first Kaluza-Klein excitations are computed via one-loop calculations \cite{Matchev:2007, cheng-2002-66-1}. We obtain for the excitations of the quarks, 
\be \bsp 
m_{u_D^1}= m_{d_D^1} = m_{c_D^1}= m_{s_D^1} = 573.3793 {\rm~GeV,~}\\
m_{t_D^1} = 560.4622 {\rm~GeV~and~}
m_{b_D^1} = 558.9203  {\rm~GeV,~}\\
m_{u_S^1} = m_{c_S^1} = 562.0523 {\rm~GeV,~}\\
m_{d_S^1} = m_{s_S^1} = 560.2356 {\rm~GeV,~}\\
m_{t_S^1} = 586.2638 {\rm~GeV~and ~}
m_{b_S^1} = 560.2514 {\rm~GeV,~}
\esp\ee for those of the leptons
\be\bsp
m_{e_D^{1-}} = m_{\mu_D^{1-}} = m_{\tau_D^{1-}}= m_{\nu_e^1} = m_{\nu_\mu^1} = m_{\nu_\tau^1} = 514.9604 {\rm~GeV,~}\\
m_{e_S^{1-}} = m_{\mu_S^{1-}} = m_{\tau_S^{1-}} = 505.4502 {\rm~GeV,~}\\
\esp\ee and for those of the gauge bosons
\be\bsp
m_{G^1}     =&\, 603.3141 {\rm~GeV,~}\\
m_{Z^1}     =&\, 535.4923 {\rm~GeV,~}
m_{W^{1\pm}}= 500.8931 {\rm~GeV~and~}
m_{B^1}     = 500.8931 {\rm~GeV.}\\
\esp\ee
Some examples of the obtained results for the calculation of cross sections of some $2\to 2$ processes relative to the production of two Standard Model particles or two Kaluza-Klein excitations are shown in Tables \ref{App:muedGau} and \ref{App:muedFer}. We set the center-of-mass energy to 1400 GeV and we applied a $p_T$ of 20 GeV on each final state particle.\\

\renewcommand{\FRtab}[9]{$#1$ & $#2$ & $#3$ & $#4$ & $#5$ & $#6$ & $#7$ & $#8$ & $#9$ }
\begin{table}
\begin{center} \begin{tabular}{|l| r@{e}l | r@{e}l  r@{e}l |   r@{e}l | }
\multicolumn{9}{c}{\bf MUED processes with gauge boson excitations}\\
\hline
\multicolumn{1}{|c|}{Process} & \multicolumn{2}{c|}{MG-FR} & \multicolumn{2}{c}{CH-FR} & \multicolumn{2}{c|}{CH-ST} & \multicolumn{2}{c|}{SH-FR} \\
\hline \hline
\FRtab{Z^1 \, Z^1 \to W^{-} \, W^{+}}{2.856}{+1}{2.854}{+1}{2.855}{+1}{2.855}{+1}\\
\FRtab{W^{1+} \, W^{1-} \to Z \, Z}{8.400}{+0}{8.408}{+0}{8.408}{+0}{8.408}{+0}\\
\FRtab{W^{1+} \, W^{1-} \to Z \, \gamma}{5.077}{+0}{5.074}{+0}{5.074}{+0}{5.074}{+0}\\
\FRtab{W^{1+} \, W^{1-} \to W^{+} \, W^{-}}{8.707}{+0}{8.714}{+0}{8.714}{+0}{8.713}{+0}\\
\FRtab{W^{1+} \, W^{1-} \to \gamma \, \gamma}{7.648}{-1}{7.656}{-1}{7.656}{-1}{7.657}{-1}\\
\FRtab{\gamma^{1} \, \gamma^{1} \to t \, \bar t}{8.988}{-2}{8.985}{-2}{8.985}{-2}{8.985}{-2}\\
\FRtab{Z^1 \, Z^1 \to d \, \bar d}{3.553}{-1}{3.556}{-1}{3.556}{-1}{3.557}{-1}\\
\FRtab{W^{1+} \, W^{1-} \to e^- \, e^+}{1.560}{-1}{1.557}{-1}{1.557}{-1}{1.557}{-1}\\
$G^1 \, G^1 \to G \, G$ & $7.890$ & $+1$ & \multicolumn{2}{c}{-} & $7.858$ & $+1$ & $7.854$ &$+1$\\
\FRtab{G^1 \, Z^1 \to c \, \bar c}{6.890}{-1}{6.881}{-1}{6.882}{-1}{6.882}{-1}\\
\FRtab{G^1 \, \gamma^{1} \to b \, \bar b}{1.096}{-1}{1.098}{-1}{1.098}{-1}{1.098}{-1}\\
\FRtab{G^1 \, G^1 \to G^1 \, G^1}{1.459}{+5}{1.462}{+5}{1.462}{+5}{1.462}{+5}\\
\FRtab{G^1 \, G \to G^1 \, G}{9.534}{+4}{9.542}{+4}{9.539}{+4}{9.535}{+4}\\
\FRtab{Z \, \gamma \to W^{1+} \, W^{1-}}{3.185}{+0}{3.185}{+0}{3.185}{+0}{3.185}{+0}\\
\FRtab{W^{+} \, W^{1-} \to Z^1 \, \gamma}{1.279}{+2}{1.278}{+2}{1.278}{+2}{1.280}{+2}\\
\FRtab{Z \, Z^1 \to W^{+} \, W^{1-}}{4.944}{+2}{4.944}{+2}{4.944}{+2}{4.952}{+2}\\
\FRtab{W^{+} \, W^{1-} \to Z \, Z^1}{4.930}{+2}{4.926}{+2}{4.926}{+2}{4.932}{+2}\\
\FRtab{W^{1+} \, W^{1-} \to W^{1+} \, W^{1-}}{2.199}{+3}{2.202}{+3}{2.202}{+3}{2.203}{+3}\\
\FRtab{Z^1 \, Z^1 \to W^{1+} \, W^{1-}}{1.905}{+3}{1.905}{+3}{1.905}{+3}{1.904}{+3}\\
\hline
\end{tabular} \end{center}
\caption{\label{App:muedGau}Cross sections for a selection of processes in MUED with two gauge bosons in the initial state.  The existing \mued\ implementation in \calchep\ is denoted CH-ST, while the \feynrules-generated ones in \madgraph\, \calchep\ and \sherpa\ are denoted MG-FR, CH-FR and SH-FR. The center-of-mass energy is fixed to 1400 GeV, and a $p_T$ cut of 20 GeV is applied to each final state particle.}
\end{table}

\begin{table}
\begin{center} \begin{tabular}{|l| r@{e}l | r@{e}l  r@{e}l |   r@{e}l | }
\multicolumn{9}{c}{\bf MUED processes with fermion excitations}\\
\hline
\multicolumn{1}{|c|}{Process} & \multicolumn{2}{c|}{MG-FR} & \multicolumn{2}{c}{CH-FR} & \multicolumn{2}{c|}{CH-ST} & \multicolumn{2}{c|}{SH-FR} \\
\hline \hline
\FRtab{e^{1-}_S \, e^{1+}_S \to u \, \bar u}{1.109}{-1}{1.109}{-1}{1.109}{-1}{1.110}{-1}\\
\FRtab{e^{1-}_S \, e^{1-}_S \to e^- \, e^-}{1.071}{+0}{1.071}{+0}{1.071}{+0}{1.071}{+0}\\
\FRtab{e^{1-}_S \, \mu^{1+}_S \to e^- \, \mu^+}{4.766}{-1}{4.768}{-1}{4.768}{-1}{4.768}{-1}\\
\FRtab{e^{1-}_S \, e^{1+}_S \to \gamma \, \gamma}{2.078}{-1}{2.079}{-1}{2.079}{-1}{2.079}{-1}\\
\FRtab{\nu_e^1 \, \nu_e^1~ \to u \, \bar u}{1.635}{-1}{1.635}{-1}{1.635}{-1}{1.636}{-1}\\
\FRtab{\nu_e^1 \, \nu_e^1~ \to W^{+} \, W^{-}}{5.905}{-1}{5.901}{-1}{5.901}{-1}{5.897}{-1}\\
\FRtab{e^{1-}_D \, e^{1+}_D \to u \, \bar u}{2.298}{-1}{2.298}{-1}{2.298}{-1}{2.299}{-1}\\
\FRtab{e^{1-}_D \, e^{1+}_D \to e^- \, e^+}{2.496}{-1}{2.498}{-1}{2.498}{-1}{2.498}{-1}\\
\FRtab{e^{1-}_D \, \nu_e1~ \to d \, \bar u}{6.399}{-1}{6.400}{-1}{6.400}{-1}{6.403}{-1}\\
\FRtab{e^{1-}_D \, \nu_e1 \to e^- \, \nu_e}{1.052}{+0}{1.052}{+0}{1.052}{+0}{1.052}{+0}\\
\FRtab{\tau^{1-}_S \, \tau^{1+}_S \to u \, \bar u}{1.110}{-1}{1.109}{-1}{1.109}{-1}{1.110}{-1}\\
\FRtab{\tau^{1-}_S \, \tau^{1+}_S \to \tau^- \, \tau^+}{2.553}{-1}{2.554}{-1}{2.554}{-1}{2.553}{-1}\\
\FRtab{\tau^{1-}_S \, \mu^{1-}_S \to \tau^- \, \mu^-}{6.585}{-1}{6.582}{-1}{6.582}{-1}{6.582}{-1}\\
\FRtab{e^{1-}_S \, \tau^{1+}_S \to e^- \, \tau^+}{4.765}{-1}{4.768}{-1}{4.768}{-1}{4.768}{-1}\\
\FRtab{\nu_\tau^1 \, \nu_\tau^1~ \to t \, \bar t}{1.502}{-1}{1.504}{-1}{1.504}{-1}{1.504}{-1}\\
\FRtab{\nu_\mu^1 \, \nu_\mu^1~ \to c \, \bar c}{1.634}{-1}{1.635}{-1}{1.635}{-1}{1.636}{-1}\\
\FRtab{\nu_\tau^1 \, \nu_\tau^1~ \to Z \, Z}{4.141}{-1}{4.135}{-1}{4.135}{-1}{4.133}{-1}\\
\FRtab{\tau^{1-}_D \, \tau^{1+}_D \to b \, \bar b}{1.426}{-1}{1.427}{-1}{1.427}{-1}{1.428}{-1}\\
\FRtab{\tau^{1-}_D \, \nu_\tau^1~ \to b \, \bar t}{6.557}{-1}{6.560}{-1}{6.560}{-1}{6.563}{-1}\\
\FRtab{\tau^{1-}_D \, \nu_\tau^1 \to \tau^- \, \nu_\tau}{1.053}{+0}{1.052}{+0}{1.052}{+0}{1.052}{+0}\\
\FRtab{\gamma^{1} \, \tau^{1-}_S \to \gamma \, \tau^-}{1.638}{-1}{1.639}{-1}{1.639}{-1}{1.639}{-1}\\
\hline
\FRtab{u^{1}_D \, Z^1 \to G \, u}{2.065}{+0}{2.068}{+0}{2.068}{+0}{2.072}{+0}\\
\FRtab{d^{1}_D \, d^{1}_D \to d \, d}{9.147}{+0}{9.136}{+0}{9.139}{+0}{9.143}{+0}\\
\FRtab{u^{1}_D \, \bar u^{1}_D \to u \, \bar u}{7.976}{+0}{7.986}{+0}{7.989}{+0}{7.992}{+0}\\
\FRtab{u^{1}_S \, u^{1}_S \to u \, u}{7.149}{+0}{7.147}{+0}{7.150}{+0}{7.152}{+0}\\
\FRtab{d^{1}_S \, \bar d^{1}_S \to d \, \bar d}{9.098}{+0}{9.100}{+0}{9.103}{+0}{9.107}{+0}\\
\FRtab{t^{1}_D \, t^{1}_D \to t \, t}{8.471}{+0}{8.474}{+0}{8.477}{+0}{8.474}{+0}\\
\FRtab{t^{1}_D \, \bar t^{1}_D \to t \, \bar t}{7.502}{+0}{7.492}{+0}{7.495}{+0}{7.492}{+0}\\
\FRtab{t^{1}_S \, t^{1}_S \to t \, t}{7.672}{+0}{7.674}{+0}{7.677}{+0}{7.674}{+0}\\
\FRtab{b^{1}_S \, b^{1}_S \to b \, b}{5.853}{+0}{5.858}{+0}{5.860}{+0}{5.862}{+0}\\
\FRtab{t^{1}_S \, \bar t^{1}_S \to t \, \bar t}{9.208}{+0}{9.211}{+0}{9.215}{+0}{9.211}{+0}\\
\FRtab{b^{1}_S \, \bar b^{1}_S \to b \, \bar b}{9.092}{+0}{9.100}{+0}{9.104}{+0}{9.107}{+0}\\
\hline
\end{tabular} \end{center}
\caption{\label{App:muedFer}Cross sections for a selection of fermionic processes in MUED.  The existing \mued\ implementation in \calchep\ is denoted CH-ST, while the \feynrules-generated ones in \madgraph\, \calchep\ and \sherpa\ are denoted MG-FR, CH-FR and SH-FR. The center-of-mass energy is fixed to 1400 GeV, and a $p_T$ cut of 20 GeV is applied to each final state particle.}
\end{table}